\documentclass[11pt]{article} 
\usepackage[english]{babel}
\usepackage[a4paper,margin=1in]{geometry} 
\usepackage{authblk} 
\usepackage[colorlinks=true, linkcolor=black, citecolor=black, urlcolor=blue, filecolor=blue,breaklinks=true]{hyperref} 
\usepackage{setspace} 
\usepackage{lineno} 
\usepackage{csquotes} 
\usepackage{ragged2e}
\usepackage{array} 

\usepackage[no-math]{fontspec}
\usepackage{amsmath}

\usepackage[capitalise, noabbrev, nameinlink]{cleveref}
\usepackage[labelsep=period]{caption} 
\crefdefaultlabelformat{#2#1#3}
\Crefname{figure}{Figure}{Figures}
\Crefname{table}{Table}{Tables}
\Crefname{section}{\textbf{Section}}{\textbf{Sections}}

\usepackage{graphicx}
\graphicspath{{./main_figs/}{./supplement/suppl_figs/}} 
\DeclareGraphicsExtensions{.pdf,.jpeg,.JPG,.png,.PNG, .eps, .tiff}
\usepackage{subcaption} 
\DeclareCaptionLabelFormat{bold}{{(#2)}} 
\newcommand{\labelphantom}[1]{
  \parbox{0pt}{\phantomsubcaption\label{#1}}%
}
\usepackage{pgffor} 
\usepackage{alphalph} 
\usepackage[labelfont=bf, textfont=bf, singlelinecheck=off, textfont=footnotesize]{caption} 
\usepackage{booktabs}
\usepackage{multirow}
\usepackage{rotating}
\usepackage{adjustbox}
\usepackage{siunitx}

\newcommand{\generateFigSubpanels}[6][0]{ 
     \expandafter\newcommand\csname#2\endcsname{ 
      \begin{figure}[htbp]
        \foreach [count=\i] \x in #6{%
            \labelphantom{fig:#2:\AlphAlph{\i}}
        }
        \centering
        \includegraphics[width=#4\linewidth,angle=#1]{#3}
        \caption{\textbf{\color{sectioncolor}\normalsize #5}}\label{fig:#2}
        \footnotesize
        \justifying
        \foreach [count=\i] \x in #6{%
            \noindent\subref*{fig:#2:\AlphAlph{\i}}~\x\space\\ 
        }
      \end{figure}
    }
}

\newcommand{\generateFig}[6][0]{ 
     \expandafter\newcommand\csname#2\endcsname{
      \begin{figure}[htbp]
        \centering
        \includegraphics[width=#4\linewidth,angle=#1]{#3}
        \caption{\textbf{\color{sectioncolor}\normalsize #5}\label{fig:#2}
        \footnotesize
        \normalfont
        #6
        }
        
      \end{figure}
    }
}

\newcommand{\generateSidewaysFigSubpanels}[6][0]{
     \expandafter\newcommand\csname#1\endcsname{ 
      \begin{figure}[htbp]
        \foreach [count=\i] \x in #5{%
            \labelphantom{fig:#2:\AlphAlph{\i}}
        }
        \centering
        \includegraphics[width=#3\linewidth,angle=#1]{#3}
        \caption{\textbf{\color{sectioncolor}\normalsize #5}}\label{fig:#2}
        \footnotesize
        \justifying
        \foreach [count=\i] \x in #6{%
            \noindent\subref*{fig:#2:\AlphAlph{\i}}~\x\space\\ 
        }
      \end{figure}
    }
}

\newcommand{\generateSidewaysFig}[6][0]{
     \expandafter\newcommand\csname#2\endcsname{ 
      \begin{sidewaysfigure}[htbp]
        \centering
        \includegraphics[width=#4\linewidth,angle=#1]{#3}
        \caption{\textbf{\color{sectioncolor}\normalsize #5}\label{fig:#2}
        \footnotesize
        \normalfont
        #6
        }
        
      \end{sidewaysfigure}
    }
}

\newcommand{\generateTab}[6][]{
     \expandafter\newcommand\csname#2\endcsname{ \begin{table}[ht]
          \centering
                \resizebox{#4\textwidth}{!}{
                \input{#3}
            }
            \caption{\textbf{\normalsize #5}\label{tab:#2}
            \footnotesize
            \normalfont
            #6
            }
            
        \end{table}
    }
}

\newcommand{\generateSidewaysTab}[6][]{
     \expandafter\newcommand\csname#2\endcsname{ \begin{sidewaystable}[ht]
          \centering
                \resizebox{#4\textwidth}{!}{
                \input{#3}
            }
            \caption{\textbf{\normalsize #5}\label{tab:#2}
            \footnotesize
            \normalfont
            #6
            }
            
        \end{sidewaystable}
    }
}

\usepackage[dvipsnames]{xcolor} 
\usepackage{changepage} 
\usepackage{enumitem} 

\usepackage{xurl}
\usepackage{ulem}
\usepackage{xeCJK}

\usepackage[backend=biber, citestyle=numeric-comp, bibstyle=authoryear, sorting=none, natbib=true, url=false, minbibnames=3, maxbibnames=3, uniquename=false, uniquelist=false, giveninits=true, eprint=false]{biblatex} 

\defbibenvironment{bibliography}
  {\list
    {\printtext[labelnumberwidth]{%
       \printfield{labelprefix}%
       \printfield{labelnumber}}}
    {\setlength{\labelwidth}{\labelnumberwidth}%
     \setlength{\leftmargin}{\labelwidth}%
     \setlength{\labelsep}{\biblabelsep}%
     \addtolength{\leftmargin}{\labelsep}%
     \setlength{\itemsep}{\bibitemsep}%
     \setlength{\parsep}{\bibparsep}}%
     }
  {\endlist}
  {\item}

\DeclareFieldFormat{labelnumberwidth}{#1.}

\DeclareNameAlias{author}{family-given}

\AtEveryBibitem{\clearfield{month}}
\AtEveryBibitem{\clearfield{number}}
\AtEveryBibitem{\clearfield{issn}}

\DeclareFieldFormat[article]{title}{#1}

\DeclareFieldFormat{journaltitle}{#1\isdot}

\renewbibmacro{in:}{}

\renewbibmacro*{volume+number+eid}{%
  \printfield{volume}%
}
\DeclareFieldFormat[article]{volume}{\textbf{#1}\addcolon}

\DeclareFieldFormat{doi}{DOI: \href{https://doi.org/#1}{\nolinkurl{#1}}}

\DefineBibliographyStrings{english}{%
  page             = {},
  pages            = {},
} 

\usepackage{xcolor}

\definecolor{sectioncolor}{HTML}{9c4847} 
\definecolor{black}{HTML}{000000}
\definecolor{linkcolor}{HTML}{008ebe}

\usepackage{titlesec}
\titleformat{\section}
{\color{sectioncolor}\normalfont\Large\bfseries}
{\thesection}{1em}{}

\titleformat{\subsection}
{\color{sectioncolor}\normalfont\large\bfseries}
{\thesubsection}{1em}{}

\DeclareCaptionFont{figuretitlecolor}{\color{sectioncolor}}
\DeclareCaptionFont{figurecaptioncolor}{\color{black}}
\DeclareCaptionFont{tabletitlecolor}{\color{sectioncolor}}
\DeclareCaptionFont{tablecaptioncolor}{\color{black}\footnotesize}
\hypersetup{
    colorlinks=true,
    linkcolor=linkcolor,
    urlcolor=linkcolor,
    citecolor=linkcolor
}

\makeatletter
\renewcommand\maketitle{\par
  \begingroup
    \flushleft
    \Large{\@title}\par
  \endgroup
  \vspace{0.5em}
  \begin{flushleft}
    \@author
  \end{flushleft}
}
\makeatother

\renewenvironment{abstract}{
  \begin{flushleft}
    \textbf{\abstractname}
  \end{flushleft}
  \vspace{-1.5em}
  \par
  \noindent\justify
}{
  \par 
}

\title{Flare-driven habitability:  \\ Expanding life's potential around low-mass stars}

\author[1,2,6]{Dong-Yang Gao}
\author[1,4,6,*]{Hui-Gen Liu}
\author[3,5,6,*]{Ming Yang}
\author[1,4]{Ji-Lin Zhou}
\affil[1]{School of Astronomy and Space Science, Nanjing University, Nanjing, 210023, China}
\affil[2]{Shandong Key Laboratory of Space Environment and Exploration Technology, Institute of Space Sciences, School of Space Science and Technology, Shandong University, Weihai, 264209, China}
\affil[3]{College of Surveying and Geo-Informatics, Tongji University, Shanghai, 200092, China}
\affil[4]{Key Laboratory of Modern Astronomy and Astrophysics, Ministry of Education, Nanjing, 210023, China}
\affil[5]{Shanghai Key Laboratory for Planetary Mapping and Remote Sensing for Deep Space Exploration, Shanghai, 200092, China} 
\affil[6]{These authors contributed equally to this work}
\affil[*]{Correspondence: \href{mailto:huigen@nju.edu.cn}{huigen@nju.edu.cn} (H.-G.L.); \href{mailto:myang@tongji.edu.cn}{myang@tongji.edu.cn} (M.Y.)}
\affil[ ]{ } 
\date{} 

\newcommand{\makeAbstract}{
\renewcommand{\abstractname}{\color{sectioncolor}\colorbox{yellow}{ABSTRACT}}
\begin{abstract}
\noindent \color{linkcolor}The traditional definition of the circumstellar habitable zone (HZ) focuses on liquid water, but neglects the crucial role of ultraviolet (UV) radiation in prebiotic chemistry. Low-mass stars typically emit insufficient UV radiation for photochemistry throughout the liquid water HZs during quiescent states. However, frequent flares can provide substantial UV fluxes, potentially fostering habitable conditions. We refine the concept of a UV habitable zone (UV-HZ) by incorporating a temperature-dependent model for RNA precursor synthesis. Furthermore, we explore a parameterized spectral energy distribution model and adopt an empirical flare frequency distribution for flares on different stars to quantify their UV contribution. Applying this framework to different flaring stars, we find the UV-HZ around low-mass stars can extend to inner regions, and overlap with the traditional HZ in wide ranges. Apply the analysis to 9 planets around \textit{Kepler} flaring stars, three planets are located within both the refined UV-HZ and liquid water habitable zone (LW-HZ) without causing ozone depletion. Our findings highlight the significant role of flares in expanding the potential for life around low-mass stars, offering a revised perspective on exoplanet habitability criteria.
\end{abstract}
}

\begin{document}
  
    \setstretch{1.15} 

\newenvironment{tips} 
    {
        \footnotesize\color{CadetBlue}
        \begin{adjustwidth}{1cm}{1cm}
        \begin{singlespace}
        \vspace{.5em}
        \nolinenumbers
    }
    {
        \end{singlespace}
        \end{adjustwidth}
        \vspace{.5em}
    }

\newcommand{\titleTips}
    {
    \begin{tips}
    \centering
        State a result with the title if possible. Aim for fewer than 10 words.
    \end{tips}
    }
    
\newcommand{\abstractTips}
    {
    \begin{tips}
        \noindent $<$ 250 words. Shorter is better. Check journal limits, structured or unstructured, and so on. \medskip
        
        \noindent Make a compelling elevator pitch for the paper that includes a brief intro, methods, results, conclusions. It can be split into sections or not. Always start with sections to provide a framework, but remove if needed. Revisit the abstract as you write the paper (even if you have written abstracts for conferences previously), because you will find better ways to summarize the paper as it evolves! 
        
        First sentence should make a broad and general statement that sets up to the importance of the topic. Avoid definitions in the first sentence. For example, don’t say \textit{TADs are 3D structures.} It is boring and does not convey importance or a gap. The first sentence can be the same as the first sentence of the Intro but often is more specific/less broad. Overall, it should establish the gap in knowledge as briefly as possible and avoid too much background. Put your main question/gap early (2\textsuperscript{nd} or 3\textsuperscript{rd} sentence). End the abstract by making the significance and implications for the field (and maybe next steps) clear. Be as broad as possible.  No citations and minimize explicit references to other work in the abstract (meta-discourse like \textit{Previous work that investigated X showed Y}); instead, just talk about Y and the gap that you will fill.
    \end{tips}
    }
    
\newcommand{\introTips}
    {
    \begin{tips}
        $<$ 1 page, approx. 3 paragraphs \medskip

        \noindent Begin by framing the broad field in which you are working very briefly. You do not need to provide a full review of this field. Strategically cite a few reviews. Just identify the main relevant work and cite review articles for those who need more context. Get to the critical gap/problem as quickly as possible. 
        
        Then elaborate on the specific problem that you are working on and explain why solving it is important (if you haven’t already). Presumably, other folks have also looked at this problem. Briefly review what they have found and/or the relevant hypotheses. Set up why there is a gap/problem (e.g., \textit{We need to test this new hypothesis, there are better machine learning methods, etc.}).
        
        Then explicitly detail what questions/hypotheses/etc you are trying to answer/test. Feel free to use a list. Describe your innovative approach---you may introduce such an approach in the paragraph(s) directly before this one. Note major predictions here and reemphasize the significance. Someone should be able to read only this paragraph and have a good sense of what the study/you/the team has tried to accomplish. (You can bullet point or list out the aims/problems/questions if you really want to draw attention to them).
    \end{tips}
    }

\newcommand{\resultsTips}
    {
    \begin{tips}
        As long as needed, but usually 4-7 subsections \medskip
        
        \noindent Divide this section into subsections with a logical progression from one analysis to another. The subsection titles should provide an outline of the results that enables skimming. Thus, try to make each results subsection title state a brief result. Do not feel constrained to follow the order in which you did the analyses. Tell the most logical and easy to follow story that highlights the most exciting parts of your findings early and ends with a finding that leads most towards future directions (or is a preliminary finding toward the new direction).  Make sure to reference and use all data presented in figures and tables. Negative results are useful. Actions should be written in past-tense, while statements/results are in the present-tense. E.g., (\textit{We TESTED the correlation of X vs. Y, and we OBSERVED that they ARE correlated}).
        
        Each results section should contain the following: context, approach, result, details, conclusion (see below for template). Repeat as needed. Depending on the relatedness of results, it is possible to have multiple of these ‘modules’ under a single heading. Conversely, you may have more headings than figures if, for example, multiple sections are needed to describe a multi-panel figure. (In this case, consider if the figure should be split up or not.)
    \end{tips}
    }
    
\newcommand{\captionTips}
    {
    \begin{tips}
        \noindent Tips for captions and figures: All figures and tables should be inserted into the document as soon as possible after the first mention in the text. Make sure to number them sequentially as they are referenced in the text and provide a succinct descriptive caption. If there are multiple parts to a figure, give clear capital letter labels (A, B, C, etc.) to each panel. Make sure each figure and sub-figure are referenced somewhere in the text! The images should be saved in vector formats (such as PDF or EPS) to maintain high resolution. The exceptions to this are: 1) if you are including a photograph or 2) if you are plotting a file with thousands of overlapping individual data points that would each be represented in the vector file (use a bitmap format here). The title of the figure caption should state the overarching result. (This will often be similar to a results section title.) Captions should interpret the results briefly. The figures and captions should be able to be ``stand alone'' from the main text with enough context and methods for interpretation. (Many readers will only look at the figures and captions.) Subplots should contribute to one major overarching scientific point. If you can’t come up with one conclusion/title for a figure, consider if it should be two figures. If two figures show the same conclusion, one should probably go to the supplement (e.g., demonstrating that a results replicates across cell types or with a different control).

    \end{tips}
    }
    
\newcommand{\discussionTips}
    {
    \begin{tips}
        Approx 1 page \medskip
        
        \noindent The discussion should be a broad overview of the significance of your findings to the community you are addressing. You can mirror the results section, but you don’t have to. You should start with a brief summary of the main findings and then a paragraph for the accompanying context and interpretation of each of the big picture conclusions or contributions. Then you should address high-level limitations of the study (not just small limitations that you might address in your results/methods) Finally, end with future directions on a positive note. How can people use your model/framework results? This positive-critical-positive is a compliment sandwich.

        Everything discussed should be within the frame of reference of your paper’s major conclusions or contributions. You will likely want to address a shortlist of the key papers your paper speaks to, but this is not a literature review. Even if you don’t talk extensively about a paper, you can still cite it if it is relevant to your conclusions. (There is little cost to being free with your citations!) 
        
        Contrary to popular belief, you can bring up a new result in the discussion, especially if it is preliminary or a response to reviewer critiques. Save this for findings that are in service to the discussion, rather than just a main finding of a paper. You can also use sub-sections if it is helpful in framing but this is somewhat non-standard. 

        Optional: Consider proposing a model or framework in your discussion that integrates or contextualizes your results. This can be accompanied by a schematic or ``model'' figure. This is a great way to emphasize the contribution of your study and guide future work.

    \end{tips}
    }
    
\newcommand{\methodsTips}
    {
    \begin{tips}
        No length limit \medskip
        
        \noindent \underline{Goal}: Communicate the data and methods used to produce the results. Start writing this as you are doing your analyses. \underline{Objective}: (1) To write clearly how you produced all parts of the results and (2) promote reproducibility of the findings. \underline{Style}: Overall, the methods section should be clearly and thoroughly written. In theory, any knowledgeable reader (e.g., a graduate student working in your area) would be able to reproduce any of the findings just by reading your methods. \underline{Organization}: Write subsections with clear subheadings for each method. These should be descriptive. Ideally, subsections will be structured so that the reader can “look up” the details for any subsection of the results. \underline{Ordering subsections}: Subsection order should generally match the order of the results. However, if you repeat the same type of analyses in two separate parts of the paper, you can describe this once and refer to the sections those methods address. \underline{Voice}: Write in the active past tense (E.g. “We computed…, we intersected…, we quantified…”). \underline{Note on publicly available datasets}: Include a citation or url link to datasets used, along with the last download date. \underline{Citations}: if you use it, cite it. There is no ambiguity here.
        
        In some journals, the methods section precedes the results section, while in others it follows the Discussion section. If it comes before, then the text should provide some orientation and motivation. If it comes after, then it can be more of a list. Unless you know it will come first, write the methods section as if it were following the discussion section and then add orienting text later if necessary. 
        
        \underline{Other tips}: Think of the methods section is a cookbook, and each subsection is its own chapter. Some chapters will discuss where to get the finest ingredients, some will discuss specific cooking techniques, and the rest will discuss the recipes that combine ingredients and technique. For each recipe (subsection), you must be clear on which ingredients are needed and the order that the ingredients are assembled.
        
        You do not need to divulge every quotidian analytical detail (e.g., \textit{I first loaded .tsv into a data frame using pandas (v14.0) function pd.read\_csv(filename) and pivoted it into a table}). But you should provide every analytical detail relevant to reproducing the final data analyzed (e.g. \textit{We removed all loci mapping to sex chromosomes using the subtract function from BEDTools (v2.2.1; Quinlan and Hall 2010).})
    \end{tips}
    }
    
\newcommand{\methodsDataTips}
    {
    \begin{tips}
        \begin{itemize}[noitemsep]
            \item For publicly available data: Dataset name, genome build, url/citation, last-download date, sample size, etc.
            \item For private data: Describe samples, sample collection, IRBs, names of kits and reagents used to process samples, sample size. 
            \item Include any details about the dataset critical for analyses and generation of results (I.e. controls, selection criteria, covariates, ancestry, age, sex, status, cell type, tissue source, assay details). 
            \item If applicable, explain how data were processed (i.e. excluded sex chromosomes, removed centromeric and repeat elements, etc.) 
            \item Ask someone else in the lab to look over the draft to make sure you have not left out any essential details.
        \end{itemize}
    \end{tips}
    }
    
\newcommand{\methodsAnalysisTips}{
    \begin{tips}
        \begin{itemize}[noitemsep]
            \item State the data inputs. 
            \item State your controls.  
            \item State sample sizes.
            \item State any software package, its version, and arguments used to run the analysis. 
            \item Explain any processing steps, the order, and the rationale of those steps relevant for producing the result figure. 
        \end{itemize}
    \end{tips}
    }
    
\newcommand{\generalTips}
    {
    \begin{tips}
        \begin{itemize}[noitemsep]
            \item Formatting/scientific corrections 
                \begin{itemize}[noitemsep]
                    \item The first time you report a p-value, give the test used. 
                    \item Report p values as ($P = 0.\#$), be careful with ($P = 0$) (use $< \#$ when $P$ is very small)
                    \item Change “mutation” to “variant”...maybe change “SNP” or “SNV” to “variant(s)” if specificity is not important (i.e., don’t forget that indels and CNVs/SVs are also types of genetic variants)
                    \item For genomic coordinates/distances, the format is \# kb or \# Mb (space between number and unit, not \# Kb or \# mb)
                    \item Use American spelling conventions.
                \end{itemize}
            \item Organization
                \begin{itemize}[noitemsep]
                    \item Put answers to the “why” question in the discussion
                    \item Bring important findings to the first sentence of a paragraph/section
                    \item With section headers, figure titles, and paper titles, try to state a result if possible (rather than a method, e.g. “Neanderthals were bald” rather than “Investigating hair patterns of Neanderthals”)
                \end{itemize}
            \item Wordsmithing
                \begin{itemize}[noitemsep]
                    \item Where you can avoid being vague, be specific:  instead of “establishes the relationship” say the relationship that was established. Instead of “Is incompletely understood” say what is the specific thing that is incompletely understood
                    \item Get rid of “metadiscourse”. E.g., Change “Our findings XY and Z support previous findings, including studies by Smith et al  and Johnson et al, which reported that Neanderthals might have been bald at one point” to “XY and Z support findings that Neanderthals may have been bald (citation)”.
                    \item Don’t claim you’re the first ever (or save that for the cover later)
                    \item Generally speaking, avoid passive voice. If you can add “by zombies” after the verb and it still makes sense...rethink your phrasing
                \end{itemize}
            \item Word choice
                \begin{itemize}[noitemsep]
                    \item Change utilize to use 
                    \item Change Impacted to influenced
                    \item Delete uses of “those/these” in reference to previous sentence (make sure the subject of the sentence is clear)
                    \item “Characterize” connotes that you don’t really have a hypothesis (try quantify?)
                    \item Some favorite transition words: thus, nonetheless, furthermore, indeed, therefore
                    \item Fewer (discrete quantities) vs less (continuous quantities)
                    \item (e.g.,\dots) and (i.e.,\dots) can be helpful
                    \item “Data” is the plural of “datum”.
                    \item Avoid personifying inanimate objects (“the gene wants to...”)
                    \item Splitting infinitives is ok.
                    \item Contractions are not acceptable in academic writing.
                    \item Remove: Interestingly, surprisingly
                \end{itemize}
            \item Punctuation
                \begin{itemize}[noitemsep]
                    \item Parenthetical punctuation: clauses and sentences
                    \item Use an Oxford comma to separate the last and second-to-last elements in a list.
                    \item Beware of consistency between Hyphens(-), n-dash(--), m-dash(---)
                    \item With reporting intervals be consistent in how you use dashes and spaces (1-2 vs 1 - 2 vs 1 – 2, etc)
                    \item Use a comma between independent clauses joined by a conjunction.
                    \item Use "that" for restrictive clauses (no commas). Use "which" for non-restrictive clauses (use commas). 
                    \item Pluralization rules for abbreviations.
                \end{itemize}
        \end{itemize}
    \end{tips}
    }

\begingroup
  \flushleft
  {\LARGE Flare-driven habitability:\newline Expanding life's potential around low-mass stars}
  \endgroup
  \vspace{0.5em}
\newline
Dong-Yang~Gao$^{1,2,6}$,
Hui-Gen~Liu$^{1,4,6,\ast}$,
Ming~Yang$^{3,5,6,\ast}$, and
Ji-Lin~Zhou$^{1,4}$\\ 
\small$^6$These authors contributed equally to this work.\\
\small$^\ast$Corresponding author. Email: \href{mailto:huigen@nju.edu.cn}{huigen@nju.edu.cn} (H.-G.L.); \href{mailto:myang@tongji.edu.cn}{myang@tongji.edu.cn} (M.Y.)

\section{GRAPHICAL ABSTRACT}

\begin{figure}[h]
\centering
\includegraphics[width=1.0\linewidth]{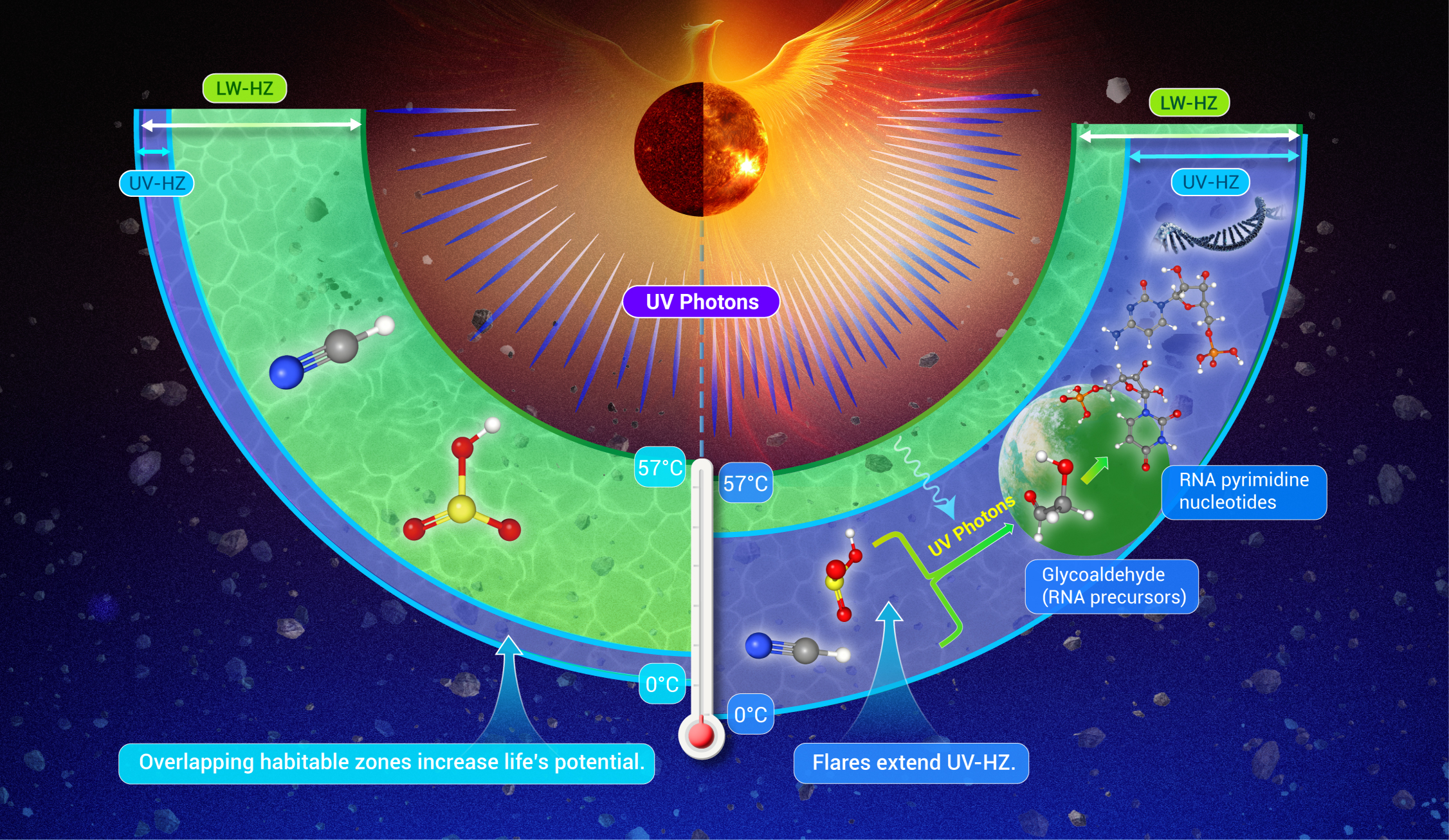}
\end{figure} 

\section{PUBLIC SUMMARY}

\newcommand{\coloredbox}[1]{\textcolor{#1}{\rule{0.2cm}{0.2cm}}}
\definecolor{darkwarmred}{HTML}{9C4847}

\begin{enumerate}[label=\protect\coloredbox{darkwarmred}]
\item An improved flare spectrum model evaluates the effect of stellar flares on planetary habitability.
\item Photochemical reaction rates varying with temperature constrain a physical ultraviolet habitable zone (UV-HZ).
\item Flares extend the HZ around low-mass stars to sustain both liquid water and prebiotic UV environments.
\item One rocky planet around \textit{Kepler} flaring stars is identified in the UV-HZ and is worth further observation.
\end{enumerate} 
 
    \clearpage
    
    \maketitle
    \makeAbstract
    
    \section{\colorbox{yellow}{INTRODUCTION}}
Detecting habitable planets has been a primary objective of the astronomical community for decades. Recent discoveries of numerous exoplanets have revealed various planetary architectures, with a higher prevalence of terrestrial worlds over giant planets. \supercite{2016RPPh...79c6901B} To assess the habitability of the terrestrial planets, it is essential to gain a deeper understanding of their intrinsic properties and the characteristics of their host stars. \supercite{1959PASP...71..421H} A critical aspect of this assessment is the potential for the rocky planets to sustain liquid water on their surfaces, which defines the liquid water habitable zone (LW-HZ). \supercite{2013ApJ...765..131K}

Searching for exoplanets in the habitable zone is crucial for identifying promising targets for follow-up observations and understanding their planetary properties.\supercite{2018ApJ...856..122K,2023AJ....165...34H} Previous studies have defined the conservative habitable zone's boundaries through simulations of water loss and the maximum greenhouse effect, while the optimistic habitable zone is based on empirical criteria from ``Recent Venus'' and ``Early Mars''.\supercite{2013ApJ...765..131K, 2014ApJ...787L..29K} However, these definitions primarily consider the star's steady incident flux and overlook the ultraviolet (UV) radiation, which is particularly significant for stars with frequent high-energy flares.

Excessive UV radiation can inhibit photosynthesis and damage biological systems, while moderate UV radiation supports essential biological processes. \supercite{1999Icar..141..399C,2006Icar..183..491B} According to the `Principle of Mediocrity', Buccino et al.\supercite{2006Icar..183..491B} proposed that planets in the UV radiation habitable zone (UV-HZ) should receive a UV flux between half and twice that of early Archean Earth. Since pyrimidine RNA precursors are essential building blocks for life, understanding the pathway to form pyrimidines is crucial.\supercite{C8CC01499J,2018SciA....4.3302R} This pathway involves seven steps with similar photochemical rate constants, starting from a mixture of HCN and SO$^{2-}_{3}$ or HS$^{-}$.  Rimmer et al.\supercite{2018SciA....4.3302R} estimated the amount of UV light for photochemistry by experimentally measuring the rate constants for UV-driven photochemical reactions (light chemistry, required for prebiotic synthesis) and bimolecular reactions occurring in the absence of UV light (dark chemistry). They selected the region where UV radiation (>45 erg cm$^{-2}$ s$^{-1}$) can provide at least 50\% yield per step, resulting in an overall yield (>0.1\%) sufficient to sustain stable prebiotic chemical reactions. According to the Arrhenius Equation,\supercite{1900C&T....20..389A} chemical reaction rates are highly sensitive to temperature. Rimmer et al.\supercite{2018SciA....4.3302R} also provide the NUV (200-280 nm) flux required for abiogenesis at different temperatures, referred to as the abiogenesis flux ($f_{\rm Abio}$).

As stellar temperature ($T_{\rm eff}$) decreases with decreasing stellar mass, UV radiation from low-mass stars is weaker. Thus, quiet low-mass stars may not provide sufficient UV energy for synthesizing biological mechanisms within the liquid water habitable zone.\supercite{2010Ap&SS.325...25G,2023MNRAS.522.1411S} 

Since many low-mass stars are more magnetically active, these flares, caused by magnetic reconnection on the stellar surface, release bolometric radiation\supercite{1989SoPh..121..299P,2010ARA&A..48..241B,2016ASSL..427..373S}. The flares also emit enhanced UV radiation compared to their quiescent stages,\supercite{2020AJ....159...60G,2022AJ....164..213G,2023MNRAS.519.3564J} which may support biogenetic processes.\supercite{C8CC01499J} However, frequent superflares can deplete the ozone layer, allowing much more harmful UV doses to reach the planetary surface and damage proteins and lipids.\supercite{2017ApJ...843...31Y,2019AsBio..19...64T} This raises a critical question: Is the UV radiation moderate due to stellar flares, to sustain the habitable zone around low-mass stars? Understanding habitability, especially for planets around low-mass flaring stars, requires more information about the UV radiation and flare frequency of the host star.

The flare frequency distribution (FFD) is a key indicator of stellar magnetic activity and crucial for assessing planetary habitability due to flares.\supercite{2020AJ....159...60G,2021MNRAS.504.3246J} However, instrument precision and observational cadence can lead to incomplete detection of low-energy flares.\supercite{2019ApJS..241...29Y,2022AJ....164..213G} Additionally, the lack of simultaneous multi-wavelength observations introduces significant inaccuracies in modeling flare spectral energy distributions (SEDs).\supercite{2023ApJ...944....5B,2023MNRAS.519.3564J} These limitations hinder accurate FFD determination for different stars. To estimate total flare energy and obtain flare SEDs from optical light curves, both radiative-hydrodynamic simulations and empirical models are used.\supercite{2022FrASS...934458K,2018ApJ...867...70L}

In this study, we focus on flaring stars, which exhibit frequent flare events that boost UV and bolometric radiation on orbiting planets. We reassess the UV-HZ around these stars using an empirical and corrected cumulative FFD (CFFD) for different stellar masses and also construct an SED model for stellar flares in the NUV band. Unlike previous studies,\supercite{2023MNRAS.522.1411S,2024ApJ...966...69L} we provide a more comprehensive approach by considering both the minimum requirements of UV radiation that vary with planetary surface temperatures and improving the estimation of UV-HZ boundaries based on the UV radiation required for RNA precursor synthesis. Additionally, we apply our methods to known {\it Kepler} planets around flaring stars to evaluate their habitability. This paper aims to identify which types of flaring stars can sustain temperate regions where liquid water and moderate UV radiation can enable the synthesis of genetic molecules or organisms.

 
\section{\colorbox{yellow}{MATERIALS AND METHODS}} \label{sect:method}
\subsection{Flare SED model for different flaring stars}
The flare spectral energy distributions (SEDs) are generated via both thermal and non-thermal mechanisms, depending on various environments, e.g., the local magnetic structure, local plasma conditions, and the rate of energy deposition.\supercite{2017ApJ...850..204W,2020ApJ...891...88W, 2021ApJ...906...72O} These SEDs are expected to explain the observed emissions and provide clues to physical processes in solar and stellar flares. We collect the characteristics of flare SEDs as follows:
\begin{enumerate}
\item Continuum radiation in the optical and NUV wavebands during flares is thought to be the response of the atmospheric heating, magnetic energy release, and electron acceleration at coronal altitudes.\supercite{2023ApJ...943L..23K} The hot-blackbody assumption (a thermal photospheric spectrum with T $\approx$9,000 K) is able to replicate some spectral regions of flare spectra (e.g., the white-light band), while ignoring emission lines.\supercite{2011A&A...530A..84K} The NUV continuum is significantly higher than the hot-blackbody assumption, likely caused by hydrogen recombination.\supercite{2023MNRAS.519.3564J}
\item Hydrogen Balmer lines are enhanced and broadened during flares, particularly in the NUV band. Below the Balmer jump wavelength (at $\sim$0.365 μm), flare fluxes can increase by several times compared to predictions from hot-blackbody models. However, only a few solar flares exhibit obvious Balmer jump.\supercite{2016ApJ...816...88K}
\item Emission lines contribution, especially in the NUV band, is non-negligible.\supercite{2019ApJ...871..167K,2023ApJ...944....5B} Fortunately, this contribution can be included by  integrating the hot-blackbody spectrum with an appropriate pseudo-continuum temperature.\supercite{2023MNRAS.519.3564J}
\end{enumerate}

Several flare SED models based on the 9,000 K hot-blackbody spectrum, combined with the FUV and NUV  emission lines taken by Hubble Space Telescope (HST), better match actual observations than the hot blackbody model (e.g., the MUSCLES model\supercite{2018ApJ...867...70L}). Based on the research of TESS white light flares for M dwarfs, and combined with the GALEX UV photometric data, Jackman et al.\supercite{2023MNRAS.519.3564J} extrapolated stellar FFD at the UV-band to test different flare SED models. They found that the GALEX NUV (0.177-0.283 μm) energies on M0-M2 stars predicted by the 9,000 K hot-blackbody model are underestimated by up to a factor of 2.3±0.6. More accurate SED of flares requires multiple waveband observations for each event, including low-energy flares.\supercite{2016ApJ...820...89F,2022ApJ...929..169R}

To improve the flare SED model on various star types, we employed a power-law continuum superimposed with a hot-blackbody spectrum from 0.100 to 0.365 μm in the UV band and a hot-blackbody model in the white-light band, considering the Balmer jump. Our flare SED model is described by Equation~(\ref{equation:flare_sed}), 
\newcounter{TempEqCnt} 
\setcounter{TempEqCnt}{\value{equation}} 
\setcounter{equation}{0} 

\begin{equation}
\small
F_{\lambda}(T_{\rm flare})= \left\{
\begin{aligned}
& F_{\lambda,\ \rm HBB}(T_{\rm flare}),\ \text{for}\ \lambda>0.365\ \rm{\mu m}, \\[1em]
& F_{\lambda,\ \rm HBB}(T_{\rm flare})+F_{\rm{0.365\ \mu m,\ HBB}}(T_{\rm flare})\times f_{\rm b}\times\left(\frac{\lambda}{\rm{0.365\ \mu m}}\right)^{\gamma},\text{for}\ 0.1 \ \rm{\mu m} \leq \lambda \leq \rm{0.365\ \mu m},
\end{aligned}
\right.
\label{equation:flare_sed}
\end{equation}
where $F_{\lambda,\ \rm HBB}(T_{\rm flare})= B_{\lambda}(T_{\rm flare}) + S_{\lambda}$, i.e. the combination of the hot-blackbody spectrum $B_{\lambda}(T_{\rm flare})$, and the stellar spectra at quiet state $S(\lambda)$. $T_{\rm flare}$ is the temperature of the stellar flare active region. The factor $f_{\rm b}$ is the enhancement ratio at 0.365 $\mu m$, and $\gamma$ is the power-law index related to the wavelength dependency.

We set $T_{\rm flare}$ to 9,000 K, leaving two parameters for the flare SED model: the enhancement of the continuum at wavelengths shorter than the Balmer jump ($f_{\rm b}$), and the power-law index ($\gamma$) describing the UV continuum trend.
This model corrects the underestimation of UV radiation using the 9,000 K hot-blackbody assumption.\supercite{2023ApJ...944....5B,2023MNRAS.519.3564J}  

Stellar and solar flares are manifestations of magnetic activity and share a similar physical process.\supercite{2015ApJ...809...79O} However, on stars of different masses, with varying convective layer thicknesses, flares exhibit distinct characteristics.\supercite{2017ApJ...837..125K, 2023ApJ...943L..23K} 
We fitted $f_{\rm b}$ and $\gamma$ with observations and spectrum model of M and G stellar flares, and use linear interpolation to model the SED for different types of stars (see Table~\ref{tab:sedpars} in the Supplemental Information).

\begin{figure}[htbp]
\centering
\includegraphics[width=1.0\linewidth]{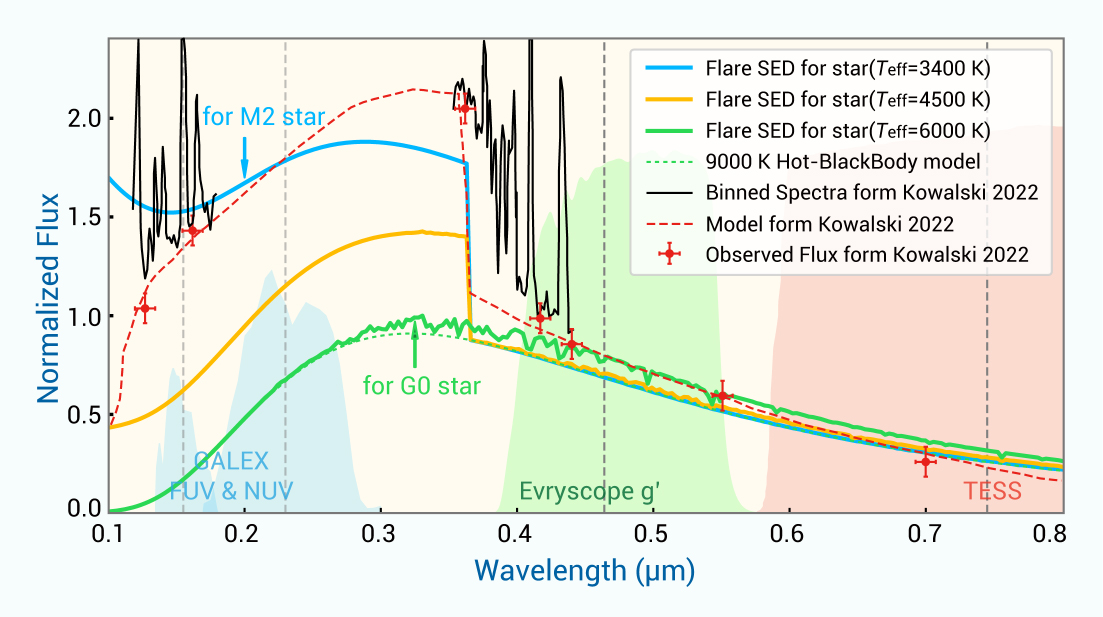}
\caption{
\textbf{Flare SEDs on different spectral types.} 
The blue, orange, and green solid lines represent flare SED models for stars with different effective temperatures of 3,400 K, 4,500 K, and 6,000 K, respectively. Two arrows highlight the flare SEDs for M2 ($T_{\rm eff}=$3,400 K) and G0 ($T_{\rm eff}=$6,000 K) type stars, respectively. For stars with $T_{\rm eff}<$ 6,000 K, flares can enhance the NUV band compared to the $T_{\rm flare}$=9,000 K hot-blackbody model (the green dotted line). The black solid line represents the ``great flare'' spectra of AD Leo, with points with error bars marking its wavelength-binned observed continuum fluxes from Kowalski.\supercite{2022FrASS...934458K} The brown dashed line is the best fit using a two-component radiative-hydrodynamic model from the same study. Colored backgrounds show the relative band-passes of different filters, with gray vertical dashed lines marking the mean wavelengths for GALEX FUV band (0.155 $\mu m$), GALEX NUV band (0.230 $\mu m$), Evryscope g$^{\prime}$ band (0.464 $\mu m$), and TESS band (0.745 $\mu m$). 
\label{fig:SED1}}
\end{figure}

Figure \ref{fig:SED1} shows several flare SED models. Previous observations\supercite{2016ApJ...816...88K, 2023MNRAS.519.3564J} indicate that for M stars, $f_{\rm b}$ is higher than for solar flares, while $\gamma$ is lower. This suggests that UV-optical radiation contrast during flares is more pronounced on M stars compared to G and K stars. 
In our model,  as $T_{\rm eff}$ ranges from 3,400 K to 6,000 K, $f_{\rm b}(T_{\rm eff})$ varies from 1.013 to 0.011, and $\gamma(T_{\rm eff})$ from -0.500 to 1.032 (see
Figure~\ref{fig:SED} in Supplemental Information). Note that this work investigates changes in the habitable zone considering flares for specific spectral types. Therefore, $f_{\rm b}$ and $\gamma$ represent the typical or average flare SED for a stellar type, not an individual star. 
The energy range for detected solar and stellar flares spans from 10$^{28}$ to 10$^{38}$ erg.\supercite{2011LRSP....8....6S,2019ApJS..241...29Y,2020AJ....159...60G,2020ApJ...890...46T}
Details on our flare SED model validation and the increase in stellar luminosity due to flares are in Section ``\hyperref[sect:flaresed]{Average luminosity of flaring stars}" in the Supplemental Information. 
The averaged stellar NUV radiation is a necessary condition for the accumulation of photochemical reaction products, which are essential for forming the macromolecular prebiotic inventory (see ``\hyperref[sect:HZ_diffMs]{Habitable zones around different flaring stars}" subsection for details).

\subsection{Calculation of Habitable Zone around Flaring Stars}\label{subsect:methods_HZofFlaring}
\subsubsection{\textit{Liquid Water Habitable Zone}}

For quiet stars, the LW-HZ boundaries are determined using the methods of Kopparapu et al.\supercite{2013ApJ...765..131K, 2014ApJ...787L..29K} The inner boundary corresponds to ``recent Venus'' at \SI{57}{\celsius},\supercite{2016NatCo...710627P} and the outer boundary to ``early Mars'' at \SI{0}{\celsius}. For flaring stars, stellar flares increase the flux reaching the planet. Although magnetic active regions are randomly distributed across the stellar surface, only flares directed towards the planet affect its habitability.  We calculated the increase in stellar luminosity caused by flares, as detailed in the Supplemental Information. Considering existing flare spectra and the SEDs for M to G stars, the integrated wavelength range is set from 0.1 to 3.0 $\mu m$.

By adjusting the LW-HZ boundaries for quiet stars considering the increased stellar bolometric luminosity, we determined the LW-HZ boundaries for flaring stars. This adjustment is uniformly applied to both the inner and outer limits,
\begin{equation}
d_{\rm flare,\ LW} = d_{\rm quiet,\ LW} \times (1+\frac{L_{\rm flare}}{L_{*}})^{0.5} ,\label{equation:d_flaring}
\end{equation}
where $d_{\rm flare,\ LW}$ is the distance from the host star to the boundary of LW-HZ when considering stellar flares, and $d_{\rm quiet,\ LW}$ is the distance from the host star to the boundary of LW-HZ without considering stellar flares.
$L_{\rm flare}$ is the time-averaged bolometric luminosity of the star with stellar flares (as explained with Equation~\ref{equation:Lflare} in the Supplemental Information), and $L_{*}$ is the bolometric luminosity in the quiescent state.

\subsubsection{\textit{Ultraviolet Radiation Habitable Zone} }\label{subsect:UVHZ}
The Ultraviolet Radiation Habitable Zone (UV-HZ) mainly focuses on biological processes. On the Archean Earth about 3.8 billion years ago, NUV radiation provided essential energy sources for synthesizing biochemical compounds. Previous studies defined the UV-HZ considering the UV flux required for abiogenesis only at \SI{0}{\celsius}. However, according to the Arrhenius Equation\supercite{1900C&T....20..389A,2018SciA....4.3302R}, chemical reaction rates are highly sensitive to temperature. On planets like Earth, with an average temperature of \SI{15}{\celsius}, reaction rates significantly increase. Consequently, more UV radiation is needed to sustain RNA synthesis.

To obtain the average NUV flux required for abiogenesis ($f_{\rm Abio}$), we fitted the relationship between $f_{\rm Abio}$ and the planet surface temperature ($T_{\rm surf}$) using the Arrhenius Equation, according to Figure 3A in Rimmer et al.\supercite{2018SciA....4.3302R} Similarly, we chose the 50\% curve from the figure to ensure sufficient flux for robust prebiotic chemistry. The fitting results are as follows:
\begin{equation}\label{equation:fAZ1}
f_{\rm Abio}=f_0\times e^{(T_{\rm surf}-273\ \rm K)/10\ K}\ \rm {photons\ cm^{-2}\ s^{-1}},
\end{equation}	
where $f_0=4.08\times 10^{12}$. 

\begin{figure}[htbp]
\centering
\includegraphics[width=0.7\linewidth]{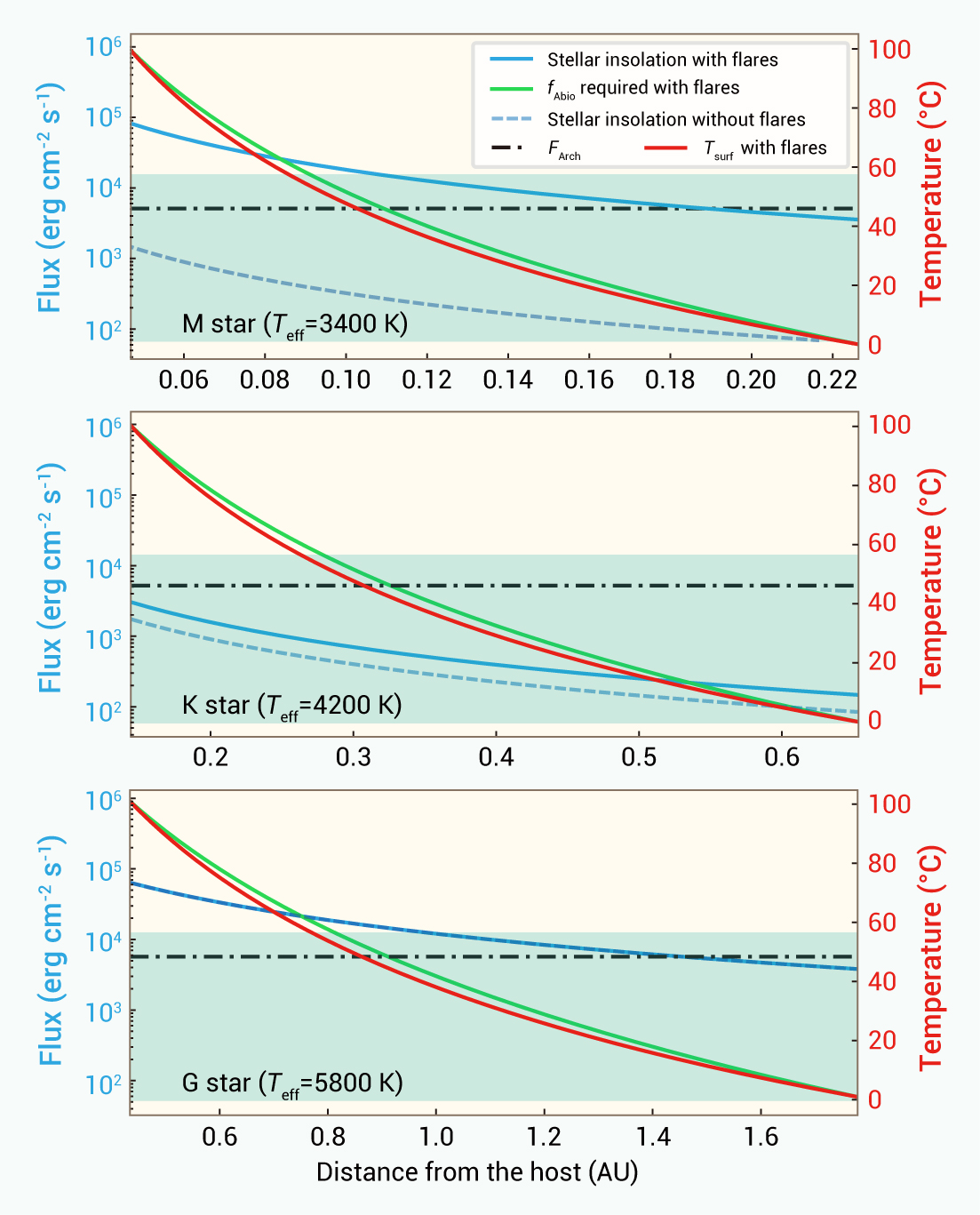}
\caption{\textbf{The required NUV flux for prebiotic chemistry in LW-HZ around typical M, K, and G stars.} The green solid line represents the $f_{\rm Abio}$ requirements at different locations, while the blue solid and dashed lines represent the stellar NUV insolation with and without flares, respectively. The inner boundary of the UV-HZ is the intersection of the green and blue solid lines. The NUV flux of Archean Earth ($F_{\rm Arch}$) is also shown using a dot-dashed line. The red line shows the variation of modeled surface temperature ($T_{\rm surf}$) on the planet, and the green background marks the region with temperatures between \SI{0}{\celsius} (273 K) and \SI{57}{\celsius} (330 K).} 
\label{fig:star_flux}\end{figure}

The surface equilibrium temperature $T_{\rm surf}$ is proportional to $d^{-0.5}$, if we ignore the atmospheric greenhouse effect of a planet. However, when considering the planet's atmosphere, the temperature decrease with distance is less significant. To model the correlation between $T_{\rm surf}$ and $d$, we adopted an exponential function:
\begin{equation} \label{equation:fAZ2}
T_{\rm surf}=T_{0} (\frac{d}{1\ \rm {AU}})^{\chi}.
\end{equation}
To determine $T_0$ and $\chi$ in Equation \eqref{equation:fAZ2}, we use the inner and outer boundaries of the LW-HZ defined by Kopparapu et al.\supercite{2013ApJ...765..131K,2014ApJ...787L..29K} The inner boundary is set at the Moist Greenhouse limit, where the surface temperature is set as 330 K,\supercite{2016NatCo...710627P}, while the outer boundary is at the freezing point of water ($T_{\rm surf}=273$ K). By comparing these boundary temperatures (i.e. $T_{\rm surf}$) with their respective distances (i.e. $d$), we can derive the value of $\chi$ using Equation \eqref{equation:fAZ2} for different stars. In the case of a solar-type star, $T_{0}=310.12$ K and $\chi=-0.22$. Consequently, by substituting Equation \eqref{equation:fAZ2} into \eqref{equation:fAZ1}, and adopting the first order approximation, we have $f_{\rm Abio}\propto{d^{-7}}$ within the LW-HZ. Figure~\ref{fig:star_flux} shows the $f_{\rm Abio}$ and modeled surface temperatures at different locations for typical M, K, and G stars. 

To obtain the quiescent NUV flux of different stars, stellar photospheric models (e.g., PHOENIX\supercite{2017ApJ...843..110R,2020MNRAS.494L..69A}) are usually adopted, which only include stellar photospheric radiation. However, radiation contributions from the chromosphere and transition region cannot be neglected. Especially for cooler stars (K to M stars), the photospheric flux between 200 and 280 nm is relatively weak in comparison to chromospheric emission.\supercite{2016ApJ...824..102L} 
To account for contributions from other mechanisms, we corrected the NUV flux obtained from the PHOENIX model by dividing by different factors. These factors are the fraction of modeled radiation to observed radiation in the GALEX NUV band (0.177-0.283 $\mu m$). We calculated the median factors for different stars according to the statistic sample in Wang et al.\supercite{2024ApJ...976...43W} (see Figure~\ref{fig:quiet_NUVflux}). After correction, the total NUV radiation of different stars in the quiescent stage can be obtained, including contributions from the photosphere, chromosphere, transition region, and corona.
 
To estimate the contribution of NUV due to stellar flares, we adopted the flare SED model and the empirical CFFD based on a sample of young stars ranging from G to M types \supercite{2022AJ....164..213G}. We calculated the total energy of flaring stars over a long-term period and added the time-averaged increase in NUV flux to the quiescent stellar flux (see ``\hyperref[subsect:method_enhancement]{Enhancement of stellar luminosity caused by flares}" in Supplemental Information). Note the NUV enhancement due to stellar flares is time-dependent. The typical timescale of flare events is several hours. According to previous studies, the reaction timescales of certain prebiotic precursor molecule, like 2-aminooxazole, range from minutes to hours.\supercite{2020MNRAS.494L..69A} However, for planetary atmosphere or surface, the subsequent chemistry of the products is often coupled with the global dynamics of materials, which typically operate on timescales much longer than flare durations. For instance, the global impact of stellar flares on the atmospheric ozone column depth can last for months to years, according to Segura et al.\supercite{2010AsBio..10..751S} Thus, adopting the time-averaged increase due to stellar flares can represent the secular and global influence on photochemistry.

After obtaining the averaged NUV flux of flaring stars, we used the lower limit of $f_{\rm Abio}$ to estimate the UV-HZ boundary. By comparing the average stellar insolation and $f_{\rm Abio}$ (see Equation~\ref{equation:fAZ1}) above the planet's atmosphere at different distances and surface temperatures, we determined the inner boundary of UV-HZ with and without flares (see Figure~\ref{fig:AZ_flux_host} in Supplemental Information).
As the distance from the star decreases, the abiogenesis flux rapidly increases ($f_{\rm Abio}$$\propto$ $d^{-7}$ for solar-type stars), while stellar radiation increases more gradually ($\propto$ $d^{-2}$). Therefore, there is a physical inner boundary of UV-HZ to guarantee the minimum requirements of $f_{\rm Abio}$. Beyond the inner boundary, stellar radiation consistently meets the abiogenesis flux required for life (see Figure~\ref{fig:star_flux}, Figure~\ref{fig:AZ_flux_diff} and \ref{fig:AZ_flux_host} in Supplemental Information). Considering the need for liquid water to support life, the outer boundary of UV-HZ is therefore set to align with the outer boundary of LW-HZ.

\subsection*{The limits of CFFD for ozone depletion}\label{subsect:ozone}
Since the inner boundary of UV-HZ is defined by the lower limit of UV flux, there should also be an upper limit of the UV flux, as excessive UV radiation can dissociate prebiotic molecules. According to Abrevaya et al.,\supercite{2020MNRAS.494L..69A} even if the received UVC (200-280 nm) flux is $\sim$15 times higher than the current Earth's flux outside the atmosphere (i.e. 92 W m$^{-2}$), a small fraction ($\sim 10^{-4}-10^{-5}$) of microorganisms can still survive. However, this UVC flux refers to radiation outside the atmosphere. Since atmospheric components, such as ozone, have varying abilities to shield UV radiation, the surface UV flux should be lower than the estimated flux outside the atmosphere. Furthermore,  UV-photounstable molecules can extend their lifetimes through a self-shielding mechanism (e.g., 2-aminooxazole) or the
presence of other UV-absorbing molecules (e.g., purine
ribonucleosides).\supercite{2021ESC.....5..239T} Additionally, turbid water containing UV-absorbing species (e.g., Fe(CN)$^{4-}_{6}$) can strongly shield UV radiation.\supercite{2022AsBio..22..242R} Thus, some molecules dissolved or accumulated in water may be preserved. We therefore did not really use the upper limit on UV flux via prebiotic chemistry.

However, for mature planetary systems, such as terrestrial planets around \textit{Kepler} flaring stars, ozone in Earth-like atmospheres becomes crucial for shielding UV radiation and protecting potential life. Thus, we constrained the CFFD of superflares based on previous studies to avoid significant ozone depletion. Frequent superflares can severely deplete the ozone layer via energetic particles and UV radiation. An Earth-like ozone layer around M dwarf stars like GJ1243 would be greatly depleted by flares of $10^{34}$ erg at a frequency of 0.4 day$^{-1}$.\supercite{2020AJ....159...60G,2019AsBio..19...64T} 
Therefore, we scaled an upper limit on CFFD to account for ozone depletion. I.e., if the CFFD of a given flaring star exceeds $10^{34}\times(\frac{d}{0.16\ \rm {AU}})^2$ erg flares, the planet is considered inhabitable. $d$ represents the distance between the planet and host stars. Fortunately, all the empirical CFFDs adopted for different stars are below these upper limits, even when considering the uncertainties in the fitted correlations.









\generateFigSubpanels{figureSuggestions}{figureSuggestions.pdf}{1}
    {Figure title which should not end in a period} 
    {{
        {Succinct caption for Panel A. Note that figure captions will not split across pages.}, 
        {Caption for panel B. Add additional subpanel captions as needed depending on the figure.} 
    }}

\section{\colorbox{yellow}{RESULTS}}
\subsection{Habitable zones around different flaring stars} \label{sect:HZ_diffMs}
Life on Earth originated 3.8 Gyr ago,\supercite{Mojzsis1996} when the Sun was approximately 0.8 Gyr old. Therefore, a timescale of $\sim$1.0 Gyr is required for life to originate.\supercite{2017Natur.543...60D,2021SciA....7.3963C}  Chromospheric activity and coronal emission of G, K, and M stars gradually keep a high level before 1 Gyr and decrease due to the reduced dynamo production of magnetic fields as the star spins down. \supercite{2014AJ....148...64S} It's hard to know when the prebiotic molecules can be produced for Archean Earth, but the M stars with longer lifetime in the main-sequence stage can have adequate time to accumulate the prebiotic molecules when the UV radiation becomes moderate. 
To estimate the NUV radiation of young stars during life origination, we selected 1.0 Gyr as the typical age for calculating stellar insolation. 

\bigskip
For the general case, we consider flaring stars ranging from 0.3 to 1.1 solar masses ($M_\odot$), spanning spectral types M2 to G0. We employed the stellar evolution model at 1.0 Gyr age from Baraffe et al.\supercite{2015A&A...577A..42B} to calculate the effective temperature ($T_{\rm eff}$) and radius ($R_{\rm *}$) of these stars. Using flaring samples from Gao et al.,\supercite{2022AJ....164..213G} we established an empirical relationship between the CFFD and $T_{\rm eff}$, and calculated the HZ boundaries considering the averaged luminosity of flaring stars. More calculation details can be found in ``\hyperref[sect:method]{MATERIALS AND METHODS}" Section. 

\begin{figure}[htp]
\centering
\includegraphics[width=1.0\linewidth]{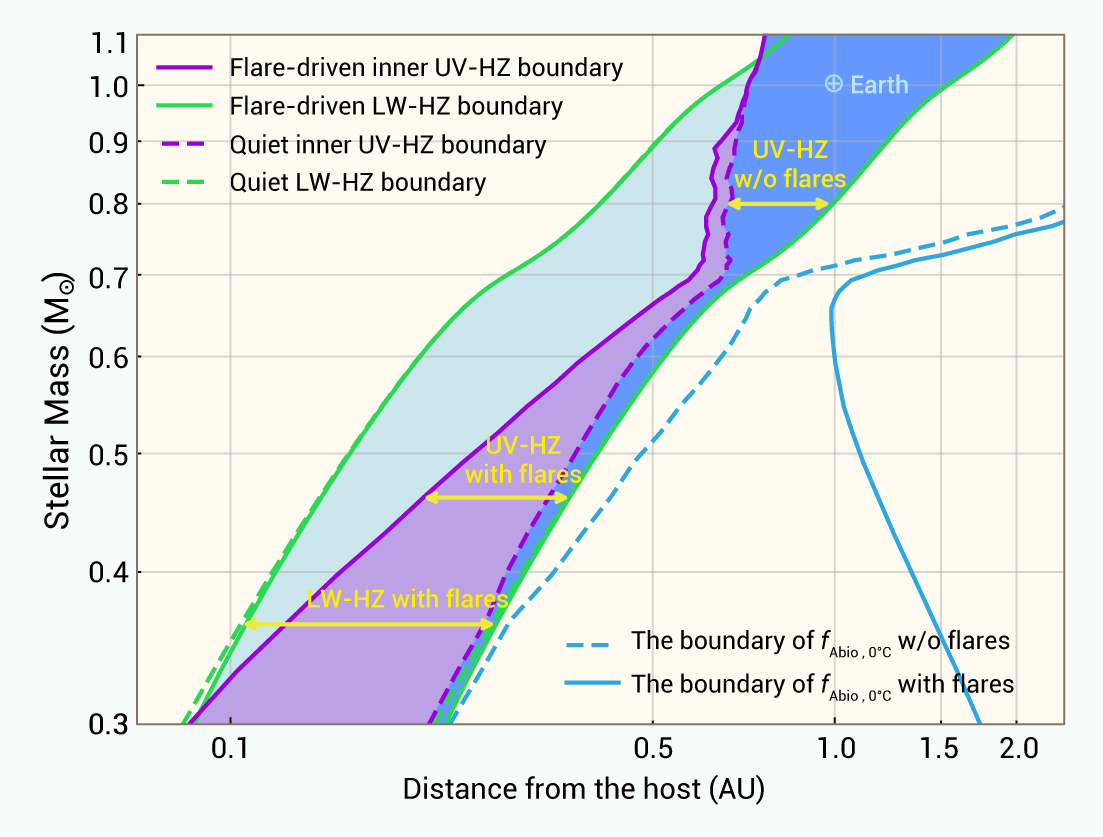}
\caption{\textbf{The habitable zones (HZs) around flaring stars at the age of 1.0 Gyr.} 
The magenta solid line indicates the flare-driven inner UV-HZ boundary, while the magenta dashed line shows the quiet inner UV-HZ boundary. 
The lime solid line indicates the flare-driven LW-HZ boundary, while the lime dashed line shows the quiet LW-HZ boundary. 
The boundary of the flux required for abiogenesis at \SI{0}{\celsius} without (w/o)  and with flares is plotted as blue dashed and solid lines, respectively. Considering that prebiotic chemistry also requires a liquid water environment, we adopt UV-HZ with the same outer boundary as LW-HZ.
The position of Earth is also marked with a circled cross in the figure.
\label{fig:HZofDiffMass}}
\end{figure} 

As shown in Figure \ref{fig:HZofDiffMass}, the UV-HZ always exists around early K and G-type stars ($M_{*}$ = 0.76--1.10 M$_{\odot}$, $T_{\rm eff}$ = 4,700--6,000 K), regardless of flares. 
{Considering the quiescent stellar NUV radiation, including contributions from the photosphere, chromosphere, transition region, and corona, the UV-HZ is narrow (e.g., less than 0.007 AU for a quiet star below $\sim$0.36 M$_{\odot}$), and the planet can hardly be located inside. If only considering the NUV flux contributed by the photosphere, the UV-HZ disappears for quiet stars below $\sim$0.76 M$_{\odot}$. 
However, if the star has flares, the increased UV flux can extend the UV-HZ. For G-type flaring stars, the UV-HZ extends slightly because the enhancement of UV flux due to flares is only a small fraction of quiescent radiation. Due to a higher flare frequency, the UV-HZ extends more noticeably for lower-mass stars. Additionally, K-type stars around 0.68 M$_{\odot}$ exhibit a narrow UV-HZ. Notably, the flare SED significantly enhances NUV luminosity rather than the total luminosity; thus, the LW-HZ experiences negligible change (\textless 0.006 AU) across all stellar types.

To assess the impact of flare SED models on UV-HZ changes, we examined two specific flare SEDs: the G0-type flare SED, which includes only the traditional 9,000 K hot-blackbody assumption, and the M2-type flare SED, which accounts for the Balmer jump of flares on M-type stars and is thus more accurate. As shown in Figure~\ref{fig:traverse1}, although the UV-HZ range using the M2-type flare SED is significantly larger, both SEDs can extend the UV-HZ obviously, and lead to a wide overlapping range with LW-HZ. 

\subsection{Habitable zones around \textit{Kepler} flaring hosts} \label{subsect:HZofHosts}
Yang and Liu \supercite{2019ApJS..241...29Y} presented a catalog of {\it{Kepler}} flaring stars, including five with planetary candidates and four with confirmed exoplanets listed as habitable planetary hosts in the NASA Exoplanet Archive [\url{https://exoplanetarchive.ipac.caltech.edu/}]. Their habitability is defined by an equilibrium temperature of 180 K to 310 K or received insolation between 0.25 and 2.2 times that of Earth.

Using the stellar and planetary parameters (see Table~\ref{tab:samples} in Supplemental Information), we calculated the LW-HZ boundaries based on Equation 3 from Kopparapu et al.\supercite{2013ApJ...765..131K} 
We performed flare detection on the \textit{Kepler} lightcurves of these hosts, and fitted the CFFD parameters for each star (see Table~\ref{tab:alpofHosts} in the Supplemental Information).
We measured the NUV radiation flux of five hosts using GALEX NUV data (see Table~\ref{tab:samples}). For four hosts without GALEX NUV or Swift UVOT observations, we adopted their NUV radiation flux as the statistical median values from Wang et al.\supercite{2024ApJ...976...43W} (see Figure~\ref{fig:quiet_NUVflux}).
For the Sun, we adopted 1.4 times the modern solar spectrum from Willmer\supercite{2018ApJS..236...47W} to account for the NUV radiation evolution of G-type stars with an age of 1.0 Gyr.\supercite{2025ApJS..281...13L}
We then derived the LW-HZ and UV-HZ boundaries, considering both the presence and absence of stellar flares. Figure \ref{fig:AZflux} shows the changes in stellar insolation and the changes in the habitable zone boundary before and after considering flares around Kepler-438. During the quiescent stage, the UV-HZ is as narrow as 0.04 AU. After accounting for the enhancement of flares, the width of the UV-HZ increases to 0.16 AU, resulting in a fourfold increase in the likelihood of accommodating an additional terrestrial planet in the HZ.

\begin{figure}[h]
\centering
\includegraphics[width=1.0\linewidth]{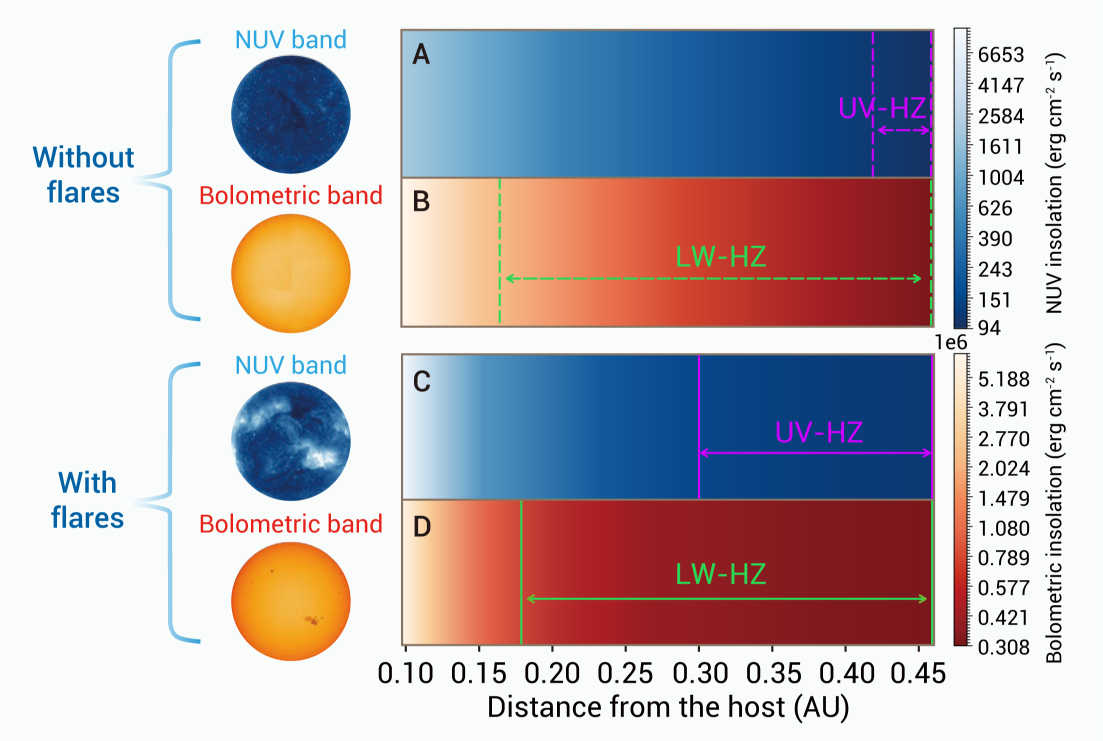} 
\caption{
\textbf{Stellar insolation around Kepler-438 without and with flares.} Panel A shows the NUV band insolation (200--280 nm) without flares, while Panel B presents the bolometric insolation under the same quiet conditions. Panels C and D show these insolation types with flares. Lime and magenta vertical lines mark the LW-HZ and UV-HZ boundaries without/with flares, respectively.}
\label{fig:AZflux}
\end{figure}

\begin{figure}[htp]
\centering
\includegraphics[width=1.0\linewidth]{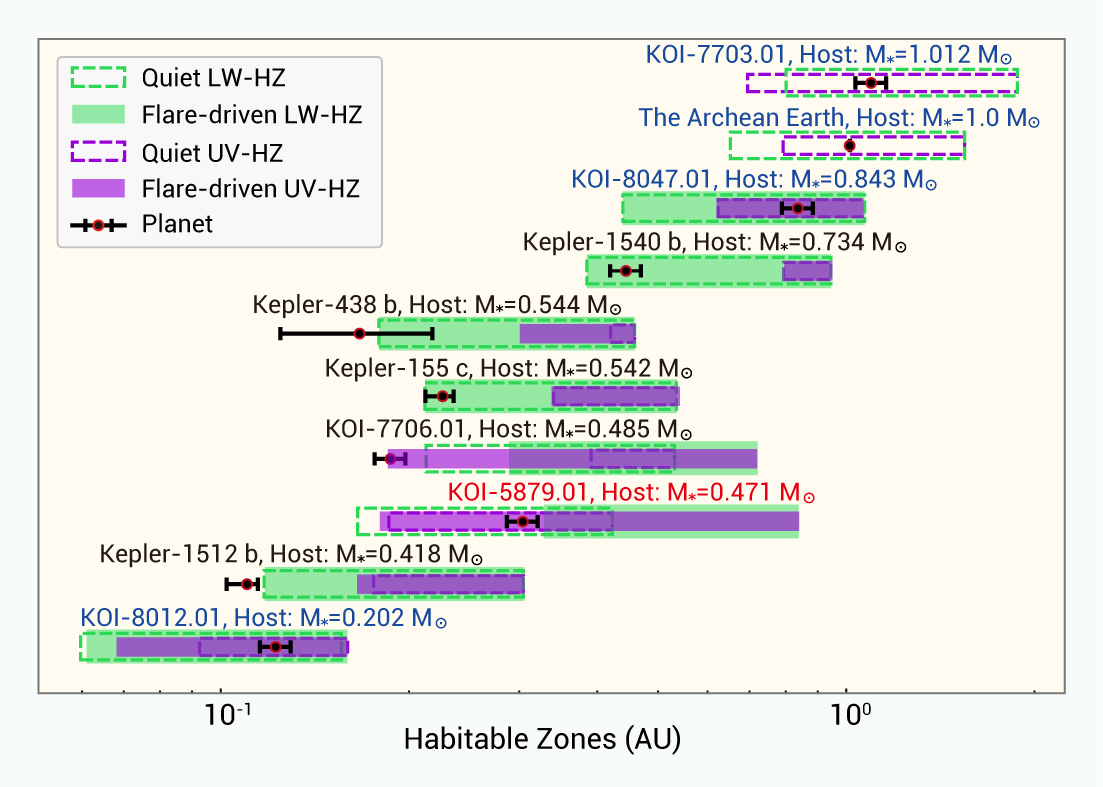} 
\caption{
\textbf{LW-HZs and UV-HZs sorted by stellar mass.}  Lime dashed boxes represent Quiet LW-HZs. Lime backgrounds represent Flare-driven LW-HZs. Magenta dashed boxes represent Quiet UV-HZs. Magenta backgrounds represent Flare-driven UV-HZs. Black dots represent planets or candidates located at their semi-major axes with errors (see Table~\ref{tab:HZofHosts}). Due to unavailable CFFD power-law indices for KOI-7703 and the Archean Sun, their flare-driven HZs are not considered.
\label{fig:HZofHosts}}
\end{figure} 

\begin{table}
  \caption{\normalsize Boundaries of the habitability zones for candidate and confirmed exoplanets orbiting \textit{Kepler} flaring stars.}
  \label{tab:HZofHosts}
  \begin{center}
  \resizebox{\textwidth}{!}{%
  \begin{tabular}{c|ccccc|ccc}
  \hline \hline
 The planet&   $a$$^{\textcolor{blue}{a}}$ & $d_{\rm in,\ FLW}$ & $d_{\rm out,\ FLW}$ & $d_{\rm in,\ FUV}$ &  located in$^{\textcolor{blue}{b}}$ & 
 TSM$^{\textcolor{blue}{c}}$   \\ 
 &   
(AU) & (AU)  & (AU) & (AU) & &   \\ \hline 
Kepler-1540 b  & 0.4426$\pm$0.0251 & 0.383 & 0.941  & 0.788 &  LW-HZ & 2.271   \\
KOI-7703.01$^{\textcolor{blue}{d}}$  & 1.0911$\pm$0.0618 & 0.798 & 1.869 & 0.693   &  LW-HZ, UV-HZ& 0.956 \\
KOI-8047.01  &  0.8332$\pm$0.0472 & 0.438 & 1.065 & 0.621 &  LW-HZ, UV-HZ & 0.576   \\
Kepler-155 c &  0.2254$^{+0.0092}_{-0.0139}$ & 0.211 & 0.533  &  0.333  & LW-HZ & 0.553  \\
KOI-5879.01$^{\textcolor{blue}{e}}$ & 0.3027$\pm$0.0172 & 0.327 & 0.835 & 0.179 &  -  & 0.332  \\
Kepler-1512 b &  0.1097$^{+0.0045}_{-0.0079}$ & 0.117 & 0.304 & 0.164 & - & 0.059  \\
Kepler-438 b  &  0.1660$^{+0.0510}_{-0.0420}$ & 0.179 & 0.459 & 0.300 &  LW-HZ? $^{\textcolor{blue}{f}}$ & 0.059   \\
KOI-7706.01$^{\textcolor{blue}{g}}$  &  0.1860$\pm$0.011  & 0.288 & 0.719 & 0.185 & -  & 0.028   \\
KOI-8012.01  &  0.1219$\pm$0.0069 & 0.062 & 0.162 & 0.069 &  UV-HZ, LW-HZ & 0.006   \\ \hline
The Archean Earth$^{\textcolor{blue}{d}}$ &  1.0 & 0.650 & 1.539 & 0.790 &  UV-HZ, LW-HZ  &  -   \\ \hline \hline
\end{tabular} }
\end{center}
\begin{singlespacing}
\setstretch{1.0}
Note: $d_{\rm in,\ FLW}$ and $d_{\rm out,\ FLW}$ denote the inner and outer boundaries of the flare-driven LW-HZ, respectively; $d_{\rm in,\ FUV}$ represents the inner boundaries of the flare-driven UV-HZ. Note the outer boundaries of the flare-driven UV-HZ are set the same with $d_{\rm out,\ FLW}$.\\
$^{\textcolor{blue}{a}}$ For KOI-8012.01 and KOI-7706.01, which lack semi-major axis uncertainty in the source reference, we applied a typical 17\% uncertainty for stellar mass from Thompson et al.\supercite{2018ApJS..235...38T} to estimate the uncertainty via error propagation.\\
$^{\textcolor{blue}{b}}$ ``UV-HZ'' and ``LW-HZ'' represent the ultraviolet habitable zone and the liquid water habitable zone driven by flares, respectively.\\
$^{\textcolor{blue}{c}}$ The Transmission Spectroscopy Metric (TSM) is calculated following Kempton et al.\supercite{2018PASP..130k4401K} to assess the viability of follow-up exo-atmospheric observations.\\
$^{\textcolor{blue}{d}}$ KOI-7703 and the Archean Sun lack sufficient flare events to fit the CFFD. Therefore, the boundaries for scenarios without stellar flares are presented. \\ 
$^{\textcolor{blue}{e}}$ For KOI-5879, a more appropriate integration range for flare energies was chosen from $10^{30}$ to $10^{38}$ erg (see the Notice b of Table~\ref{tab:alpofHosts} for details). \\
$^{\textcolor{blue}{f}}$ Measurement uncertainties in the semi-major axis prevent a clear conclusion of whether this planet resides within the LW-HZ (see Figure~\ref{fig:HZofHosts}).  \\
$^{\textcolor{blue}{g}}$ Flares on KOI-7706 may potentially deplete ozone layers of KOI-7706.01 (see Figure~\ref{fig:AZ_all}).\\
(This table is available in its entirety in machine-readable form.)
\end{singlespacing}
\end{table}

Figure \ref{fig:HZofHosts} illustrates the HZ boundaries of the planet hosts. KOI-7706.01 and Kepler-1512 b are no longer in the LW-HZ with updated stellar parameters (see Table \ref{tab:samples}). Although Kepler-1540 b, Kepler-438 b, and Kepler-155 c are in the LW-HZ, they are not in the UV-HZ. The three planets, i.e., KOI-7703.01, KOI-8047.01, and KOI-8012.01, are all located in the overlapping region of the two HZs (regardless of whether flaring is considered or not). Coincidentally, these three stars are G, K, and M stars, respectively.  
Note that KOI-5879.01 was originally in the LW-HZ. After considering the flares, the NUV radiation can satisfy the abiogenesis flux, but it is no longer in the LW-HZ.

As described in Subsection ``\hyperref[subsect:ozone]{The limits of CFFD for ozone depletion}", we also examined the limits of CFFD for ozone depletion, based on the fitted CFFD of these 9 planet hosts via their \textit{Kepler} light curves.  The frequency of superflares with certain energy must not exceed 0.4 day$^{-1}$. Otherwise, such superflares would lead to ozone depletion on their planets.\supercite{2019AsBio..19...64T}
Figure \ref{fig:AZ_1} shows an example of the observed and fitted CFFD of KOI-8012, considering the constraints from the ozone depletion. For KOI-7706, its flare frequency could potentially cause ozone depletion on its planet, because the CFFD with energy of 1.35$\times$10$^{34}$ erg)  is very close to the limit of 0.4 day$^{-1}$ (see the details in Figure~\ref{fig:AZ_all}).

\begin{figure}[h]
\centering
\includegraphics[width=0.7\linewidth]{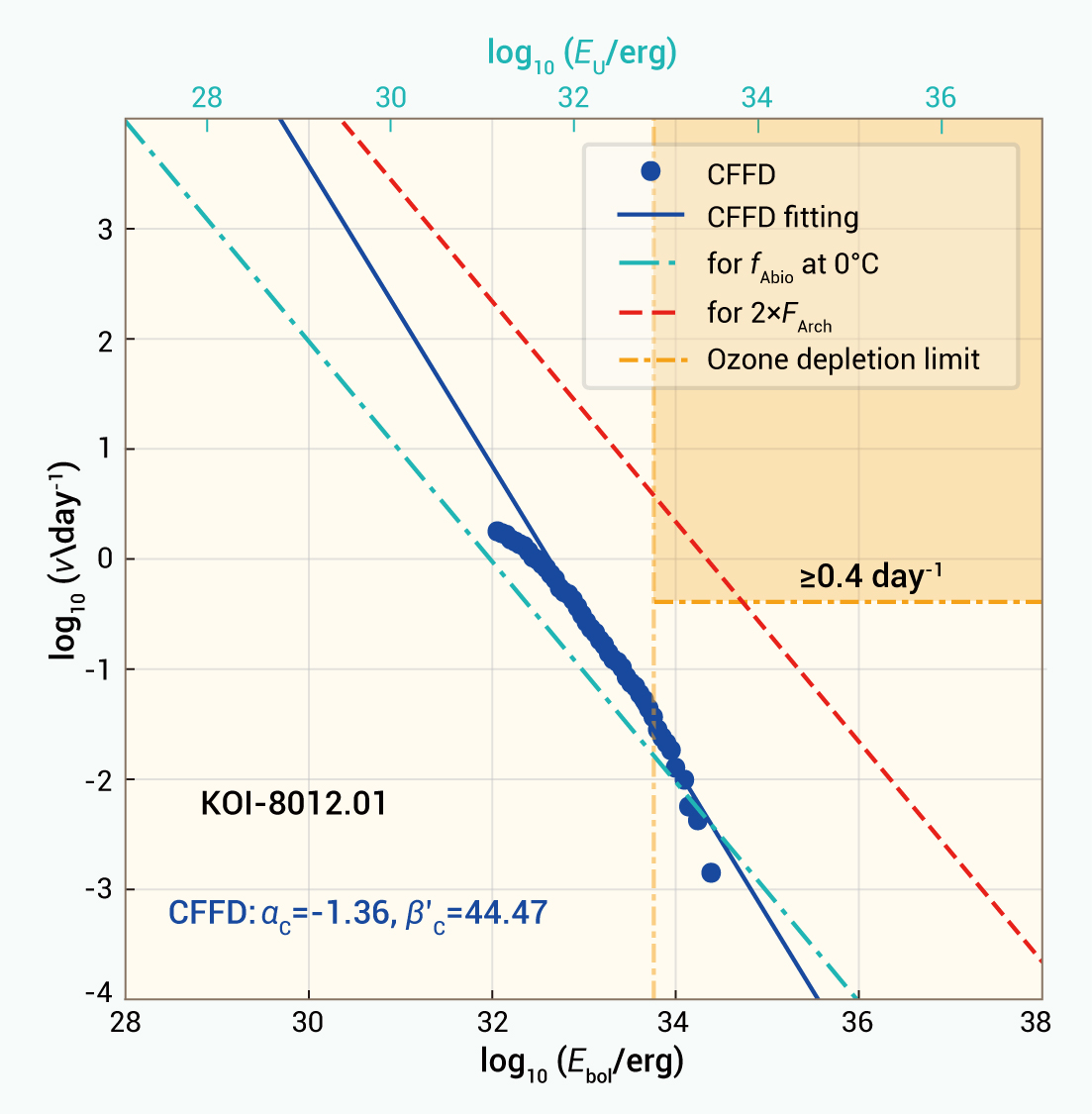}
\caption{
\textbf{The Cumulative FFD (CFFD) of KOI-8012.} The blue dot represents the CFFD directly from observations. The blue line illustrates the fitted CFFD, with two parameters ($\alpha_{\rm C}$ and $\beta'_{\rm C}$, in blue text) marked in the lower left corner. The orange area denotes the ozone depletion forbidden region, where flare rates are $\geq$ 0.4 day$^{-1}$ with flare bolometric energy larger than 0.58$\times$10$^{34}$ erg. 
The cyan dot-dashed line represents the lower limit of flaring required by the abiogenesis flux at \SI{0}{\celsius}. The red dashed line represents the frequency for twice the NUV flux received by the Archean Earth.\supercite{2006Icar..183..491B}
Note that the cyan dot-dashed and red dashed lines are only drawn in this figure for comparison with previous studies, and are not used in the definition of the UV-HZ. }
\label{fig:AZ_1}
\end{figure} 

\bigskip
In summary, the habitability of the nine planets around \textit{Kepler} flaring hosts is as follows:
\begin{itemize}
    \item One rocky planet (KOI-8012.01) and two super-earths (KOI-8047.01 and KOI-7703.01) are in both the UV-HZ and the LW-HZ. The CFFDs of their host stars do not exceed the limit of ozone depletion. Thus, they provide radiation environments more suitable for abiogenesis.
    \item Three planets (Kepler-1540 b, Kepler-438 b, and Kepler-155 c) are located in the LW-HZ. However, they are not within the UV-HZ defined by abiogenesis requirements under modeled surface temperatures. For Kepler-438 b and Kepler-155 c, if the temperature in some regions of their surface can be maintained at \SI{0}{\celsius}, the received UV radiation can satisfy the requirements for abiogenesis (see Figure~\ref{fig:AZ_all} for more explanation).
    \item KOI-7706.01 and Kepler-1512 b are not located in the LW-HZ or the UV-HZ. Moreover, the CFFD of KOI-7706 nearly reaches the threshold for ozone depletion, making these two planets less likely to support prebiotic chemistry. 
    \item KOI-5879.01 was originally within the LW-HZ and the UV-HZ if its host star were quiescent. However, flares from the host star cause an outward expansion of the LW-HZ, and it is therefore no longer within the HZ. 
\end{itemize}
    \section{\colorbox{yellow}{DISCUSSION}} 
Although many exoplanets have been studied statistically, assessing the habitability of individual planets in the habitable zone is still challenging from both astrobiological and observational perspectives. Evaluating habitable zones around stars in various aspects helps us better understand exoplanet habitability.\supercite{2023AJ....165...34H,2023ApJ...948L..26H} By re-evaluating the habitable zones and creating a comprehensive catalog of planets within them, we can infer that terrestrial planets in both liquid water and UV radiation habitable zones are more likely to support life.

In this paper, we highlight that the requirements of UV radiation for abiogenesis vary rapidly with temperature. According to the analysis of planets around flaring stars, we use a modeled temperature to estimate the inner boundary for UV-HZ. Since the surface temperature is correlated with different aspects, e.g., atmosphere effect, surface albedo, and thermal redistribution on day and night side.\supercite{2021NatGe..14..832R} Thus, Kepler-1540 b, Kepler-438 b, and Kepler-155 c deserve further investigation by large telescopes to better constrain their temperatures. Even KOI-7703.01, KOI-8047.01, and KOI-8012.01, which are among the most promising targets, need to be characterized in terms of their atmospheric properties. We also calculate the TSM of these planets, as shown in Table \ref{tab:HZofHosts}.

An averaged UV flux is adopted due to flare events in this paper. However, the influence of time-dependent enhancements of UV flux on life or chemistry needs to be compared with the continuum UV enhancements. Through the simulation of phenomena such as solar activity, Ranjan et al.\supercite{2016AsBio..16...68R} found that secular time-dependent variations in the UV band have little influence on prebiotic chemistry, but they also suggested caution when extrapolating from high-fluence regimes to natural conditions.
By modeling flares via pulsed UV radiation, microorganisms can tolerate high fluences of UV radiation in quantities, and at even shorter wavelengths than not experienced by microorganisms on present-day Earth.\supercite{2025SoSyR..59...72A} Thus, our assumption of an averaged UV flux is reasonable. Of course, more quantitative studies on the influence of time-dependent UV radiation on prebiotic chemistry would be preferred to refine the average assumption.

To accurately model or determine the UV radiation of stars, we need more NUV observations of various stars, especially young stars. In the quiescent stage, a survey including a large sample of stars is necessary to characterize the NUV radiation correlated with different stellar parameters, e.g., rotation period, age, metallicity, and effective temperature. Secular monitoring like \textit{Kepler} and \textit{TESS} would also be beneficial for characterizing the CFFD of flaring stars. Additionally, time-series observations or spectra during stellar flare events are crucial for understanding the physical processes and refining the SED model of flares.

\bigskip
Astrobiology experiments to explore the influence of UV flux on prebiotic chemistry are also crucial to  determine the limits of UV radiation. UV radiation has both beneficial and detrimental effects on the accumulation of biological macromolecules such as RNA precursors. Extreme UV radiation can destroy probiotic chemistry. However, the formation of life is a process of selective evolution over long periods. Certain mechanisms protect these biomacromolecules, ensuring their survival at specific rates and probabilities, either in particular locations or within specific environments. Studying the shielding mechanisms \supercite{2022AsBio..22..242R,2023CmChe...6..259S}  in different environments can help refine the UV requirements for abiogenesis. Furthermore, testing the amount of cumulative products over a longer timescale under a wide range of UV radiation levels is critical for habitability studies. The quantitative assessment of the effects of UV radiation on the origin of life remains an open issue, requiring more data and interdisciplinary collaborations.

After the origin of life, many other aspects can also influence the sustenance of life. For example, the synthesis of genetic molecules is influenced by various factors. X-rays, coronal mass ejections (CMEs)\supercite{2016ApJ...826..195K,2007AsBio...7..167K,2024A&A...688A.138P}, and stellar energetic particles (SEPs)\supercite{2019ApJ...881..114Y,2022SciA....8I9743H} can all cause water loss and affect planetary habitability via chemistry. Optimistically, when life exists on a planet, it can resist high UV radiation through mechanisms such as clumping and biofilm formation.\supercite{2020MNRAS.494L..69A} Therefore, further laboratory research is needed to study how microbes, such as Deinococcus radiodurans\supercite{2019AsBio..19...64T} and tardigrade, perform in extreme UV environments to determine the upper limit of UV radiation for sustaining life. 

\section{\colorbox{yellow}{CONCLUSION}} 
Traditional habitable zones for liquid water are widely accepted for detecting habitable planets, while the UV radiation is considered by a lot of works to show its importance for life origins via prebiotic chemistry. Adopting the NUV (200-280 nm) requirements from previous experimental requirements of
abiogenesis, and considering the effects on chemical reaction rate at different temperatures, we constrain the inner boundary of UV-HZ naturally. After correcting the quiescent stellar NUV radiation for young stars ($\sim$1 Gyr), based on the observation from GALEX, the UV-HZ can overlap with the LW-HZ in a narrow region for low mass stars. However, planets around these low-mass stars can hardly be located in the overlapped regions. To calculate the UV enhancement due to flares, we construct an SED model for flares and adopt an empirical model of CFFD correlated with different stars. We find that flaring stars can tolerate both LW-HZ and UV-HZ in the same region, much wider than the quiescent low mass stars. Therefore, the flaring low-mass stars can sustain the habitability of planets in the overlapped HZ. Specifically, we investigate nine Kepler flaring stars and refine their LW-HZ and UV-HZ, according to the observed CFFD. One rocky planet (KOI-8012.01) and two super-Earths (KOI-8047.01 and KOI-7703.01) are located within both habitable zones, avoiding the erosion of ozone in the atmosphere. The other three planets, i.e., Kepler-1540 b, Kepler-438 b, and Kepler-155 c, deserve further investigation for the surface temperature to assess their habitability. Our work provides a positive perspective on searching for habitable planets and potential life around flaring low-mass stars.

\section{\colorbox{yellow}{RESOURCE AVAILABILITY}}
\subsection{Lead contact}
Dong-Yang Gao: https://orcid.org/0000-0001-6643-2138; Hui-Gen Liu: https://orcid.org/0000-0001-5162-1753; Ming Yang: https://orcid.org/0000-0002-6926-2872; Ji-Lin Zhou: https://orcid.org/0000-0003-1680-2940.
\subsection{Data and Code Availability}
All data needed to evaluate the conclusions in the paper are present in the paper and/or the
Supplemental Information. Additional data and code related to this paper are available from the lead contact upon request.

\section{ACKNOWLEDGMENTS}
These results are based on observations obtained with \textit{Kepler}. We are grateful to the \texttt{Lightkurve}\supercite{2018ascl.soft12013L}} for the valuable assistance in gathering our \textit{Kepler} data. 
We thank Song Wang, Xue Li, and Jia-Hui Wang from the National Astronomical Observatories, Chinese Academy of Sciences, for their helpful discussions on stellar NUV radiation and stellar ages.
We thank National Key R\&D Program of China (No. 2024YFA1611801), National Natural Science Foundation of China (No. 12150009), China Manned Space Project (No. CMS-CSST-2025-A16), the Instrument Education Funds of Shandong University (No. yr20240205), the Civil Aerospace Technology Research Project (No. D010102), Fundamental and Interdisciplinary Disciplines Breakthrough Plan of the Ministry of Education of China (JYB2025XDXM105), and the Shanghai Municipal Education Commission's Artificial Intelligence Innovation Program for Empowering Discipline Development and Reforming Research Paradigms.

\section{AUTHOR CONTRIBUTIONS}
H.-G.L. proposed and designed this study. D.-Y.G., H.-G.L., and M.Y. investigated the observation data and pertinent literature. D.-Y.G. conducted the calculations and analysis. J.-L. Z. provides key suggestions on this work. All authors discussed the results and reviewed the manuscript.

\section{DECLARATION OF INTERESTS} %
The authors declare no competing interests.

\section{SUPPLEMENTAL INFORMATION}
Supplemental Text, Figures S1 to S11, Tables S1 to S4, Supplemental References S1-S32. \\ 

    \begin{singlespace}
        \printbibliography[title={REFERENCES}, heading=bibnumbered]
    \end{singlespace}
    \clearpage
    
    \thispagestyle{empty}
\begin{center} 
{\LARGE Supplemental Information for\newline
Flare-driven habitability: Expanding life's potential around low-mass stars} \newline
\end{center}
Dong-Yang~Gao$^{1,2,6}$,
Hui-Gen~Liu$^{1,4,6,\ast}$,
Ming~Yang$^{3,5,6,\ast}$, and
Ji-Lin~Zhou$^{1,4}$\\ 
\small$^6$These authors contributed equally to this work.\\
\small$^\ast$Corresponding author. Email: \href{mailto:huigen@nju.edu.cn}{huigen@nju.edu.cn} (H.-G.L.); \href{mailto:myang@tongji.edu.cn}{myang@tongji.edu.cn} (M.Y.)

\subsubsection*{This PDF file includes:}
Supplemental Text\\
Figures S1 to S11\\
Tables S1 to S4\\
Supplemental References S1-S32


    \clearpage
    \begin{refsection}
    \renewcommand{\thefigure}{S\arabic{figure}}
    \renewcommand{\thetable}{S\arabic{table}}
    \renewcommand{\theequation}{S\arabic{equation}}
    \renewcommand{\thepage}{S\arabic{page}}

    \setcounter{figure}{0}
    \setcounter{table}{0}
    \setcounter{equation}{0}
    \setcounter{page}{1}

    \DeclareFieldFormat{labelnumber}{S#1}  



\section*{\colorbox{yellow}{Supplemental Text}}
\subsection*{Average luminosity of flaring stars}\label{sect:flaresed}
\subsubsection*{\textit{Flare spectral energy distribution} }
In the SED model, flare radiation in the UV band (0.100 to 0.365 $\mu m$) follows a power-law continuum superimposed with a hot-blackbody spectrum, while the white-light band follows a 9,000 K hot-blackbody model. A Balmer jump separates the two regions. This model is based on previous studies,\supercite{2017ApJ...837..125K,2023ApJ...944....5B,2023MNRAS.519.3564J} and has integrated emission-line contributions into the continuum. The detailed flare spectral energy distribution model of the flare is shown in Equation (\ref{equation:flare_sed}) in the main paper.

\begin{sidewaysfigure}
\centering
\includegraphics[width=1.0\linewidth]{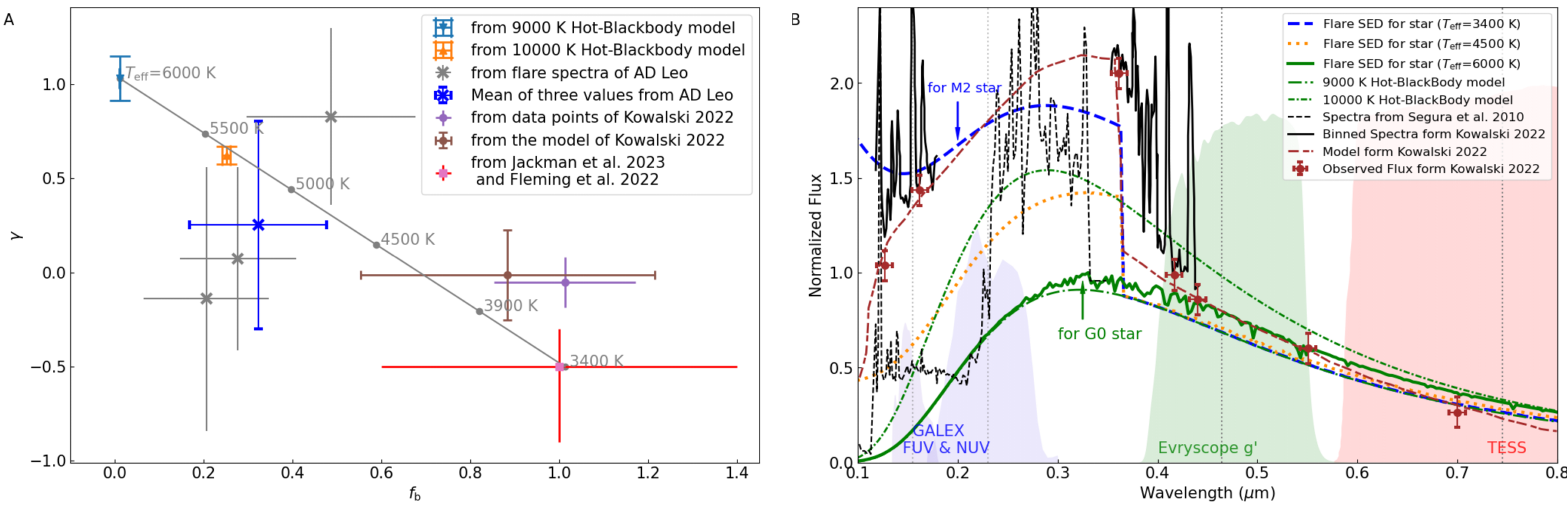} 
\caption{\textbf{(A) Parameters of flare SED.} The parameters are derived from flare spectra and models in Table \ref{tab:sedpars}. The gray line represents the linearly scaled model for the flare SED with varying stellar effective temperature ($T_{\rm eff}$). \textbf{(B) Flare SED for different stars.} 
The solid, dotted, and dashed lines in various colors represent flare SED models for different stars with $T_{\rm eff}=$6,000 K, 4,500 K, and 3,400 K, respectively. The black solid line represents the adjusted ``great flare'' spectra of AD Leo, with plus symbols marking the data from wavelength-binned observed continuum fluxes by Kowalski.\supercite{2022FrASS...934458K} The red long-short-dashed line is the best fit using a two-component radiative-hydrodynamic model from the same study.
The gray dashed line is the ``great flare'' spectra from Segura et al.\supercite{2010AsBio..10..751S}
The green long-dash-dotted and short-dash-dotted lines show the 9,000 and 10,000 K hot-blackbody model, respectively. 
Colored backgrounds show the relative band-passes of different filters, with gray vertical dotted lines marking the mean wavelengths for GALEX FUV band (0.155 $\mu m$), GALEX NUV band (0.230 $\mu m$), Evryscope g$^{\prime}$ band (0.464 $\mu m$), and TESS band (0.745 $\mu m$). 
\label{fig:SED}}
\end{sidewaysfigure}

We derived the parameters $f_{\rm b}$ and $\gamma$ of the flare SED model based on the spectral observations of flares on AD Leo and the simulated flare SED of the G-type star. The fitted results are shown in Table~\ref{tab:sedpars} and illustrated in Figure \ref{fig:SED}A. Adopting a linear correlation assumption, as $T_{\rm eff}$ ranges from 3,400 K to 6,000 K, $f_{\rm b}$ decreases from 1.013 to 0.011, while $\gamma$ varies from -0.500 to 1.032. Figure \ref{fig:SED}B displays several flare SED models. Note that this study primarily investigates the effects of flares on the habitable zone across different spectral types, where $f_{\rm b}$ and $\gamma$ are indicative of characteristics of certain spectral types rather than individual stars.

\begin{sidewaystable}
    \caption{\normalsize The specrum and fitted values of $f_{\rm b}$ and $\gamma$ in the flare SED model. \label{tab:sedpars}}
    \begin{center}
    \begin{tabular}{c|cl|c|c} \hline \hline
Spectral Type &  spectrum & & $f_{\rm b}$  &  $\gamma$   \\ 
\hline
\multirow{6}{*}{M-type star} & \multirow{3}{*}{ AD Leo: flare spectrum near H$\gamma$ peak$^{\textcolor{blue}{a}}$ } &  800 s &  0.486$\pm$0.190  & 0.829$\pm$0.470\\
& &  915 s (Peak)& 0.277$\pm$0.131 & 0.072$\pm$0.487 \\
& &  1,038 s & 0.206$\pm$0.141 & -0.140$\pm$0.700 \\ \cmidrule{2-5} 
& \multirow{2}{*}{AD Leo: flare spectrum at U band peak$^{\textcolor{blue}{b}}$ } & Observation & 1.013$\pm$0.159 & -0.052$\pm$0.133  \\ 
& & Model & 0.884$\pm$0.331 & -0.014$\pm$0.240 \\ \cmidrule{2-5}
& UV flare photometry$^{\textcolor{blue}{c}}$ &  & 1.0$\pm$0.4 &  -0.5$_{-0.4}^{+0.2}$ \\ \hline 
\multirow{2}{*}{G-type star} & \multirow{2}{*}{ Hot-BlackBody Model } & 9,000 K  & 0.011$\pm$0.001   & 1.032$\pm$0.118 \\ 
 & & 10,000 K  & 0.251$\pm$0.009  &  0.623$\pm$0.048 \\ \hline \hline
\end{tabular}  \\ \vspace{1mm}
\end{center}
$^{\textcolor{blue}{a}}$ Segura et al.\supercite{2010AsBio..10..751S} provide synthetic flare spectra near the peak of the H$\gamma$ line around 915 seconds. \\
$^{\textcolor{blue}{b}}$ Kowalski\supercite{2022FrASS...934458K} provides a synthetic flare spectrum and continuum observations at the U band peak around 542 seconds. \\
$^{\textcolor{blue}{c}}$ Fleming et al.\supercite{2022ApJ...928....8F} provide 21 flare events observed on GJ 65 in both GALEX FUV ($\lambda_{\rm eff}=0.1539\ \mu m, \Delta \lambda = 0.1344 -0.1786\ \mu m$) and NUV ($\lambda_{\rm eff}=0.2316\ \mu m, \Delta \lambda = 0.1771 -0.2831\ \mu m$) bands. Fleming et al.\supercite{2022ApJ...928....8F} provide the clues of $\gamma$, which is used to determine $f_{\rm b}$ by incorporating the Energy Correction Factor (ECF) of 2.7$\pm$0.6 from Jackman et al.\supercite{2023MNRAS.519.3564J} \\
\end{sidewaystable}

\begin{figure}[h]
\centering
\includegraphics[width=1.0\linewidth]{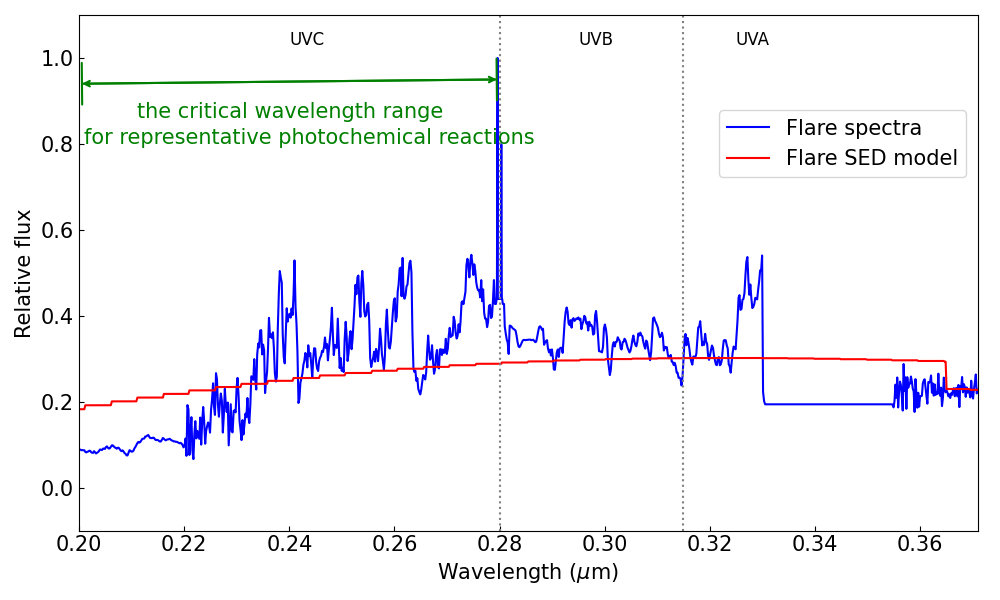}
\caption{
\textbf{Comparison of observed flare spectrum with the SED model.}
The blue line represents the multi-band composite spectrum synthesized by Segura et al.\supercite{2010AsBio..10..751S} at 915 s. The red line represents the SED model described in the main paper. Grey vertical dotted lines demarcate the boundaries between UVA and UVB (0.315 $\mu$m), UVB and UVC (0.28 $\mu$m). The green horizontal arrow highlights the NUV wavelength range (0.20--0.28 $\mu$m), which is vital for photochemical reactions.\supercite{2018SciA....4.3302R} In this range, the flux ratio of the flare spectra to the SED model is approximately 1.07. 
\label{fig:BAS_S}}
\end{figure}

To validate the flare SED model, we compared the calculated NUV flux with the observation results. The observed spectra are chosen as the peak flare spectra on AD Leo at 915 s from Segura et al.\supercite{2010AsBio..10..751S} Figure~\ref{fig:BAS_S} illustrates the comparison result. The total energy of the SED model is only 7\% less than the observed spectral energy in the UVC band (0.2 - 0.28 $\mu m$). Furthermore, we estimated the total energy correction factor for M stars, i.e., the flare energy ratio between the SED model and the 9,000K hot-blackbody spectrum. Our results are consistent with the estimated values of different previous works in various UV bands, as evidenced in Table \ref{tab:SEDcampares}. Therefore, the continuum flare SED model in this paper is suitable. Since there is lack of observed SED in the NUV band for K stars, it is difficult to validate the SED model for K stars. However, the flare enhancement in the NUV band usually contributes less for the G and K stars; therefore, we linearly interpolate the SED model for the K star.

\begin{table}
\caption{\normalsize Comparison of the flare SED model with previous models and observations for M stars. \label{tab:SEDcampares}}
\begin{center}
\begin{tabular}{c|l|c} \hline \hline
Parameters  &  Previous Work & This Work \\ 
\hline
\multirow{3}{*}{$ECF^{\textcolor{blue}{a}}$ in NUV} & 2.7$\pm$0.6\supercite{2023MNRAS.519.3564J}  &  \multirow{3}{*}{2.87}  \\
& 3.86\supercite{2024MNRAS.532.4436B} &  \\
& 2--3\supercite{2019ApJ...871..167K} & \\ \hline
\multirow{2}{*}{$ECF$ in FUV} & 6.5$\pm$0.7\supercite{2023MNRAS.519.3564J}  &  \multirow{2}{*}{9.38}\\
& 12.6\supercite{2024MNRAS.532.4436B} &    \\
\hline
\multirow{2}{*}{$F_{\rm FUV}$/$F_{\rm NUV}$}$^{\textcolor{blue}{b}}$ & 0.85$\pm$0.52\supercite{2022ApJ...928....8F} & \multirow{2}{*}{0.85} \\ 
 & 0.5--2\supercite{2024MNRAS.532.4436B} & \\ \hline
\multirow{2}{*}{$F_{0.2\ \mu m }$/$F_{0.28\ \mu m}$ $^{\textcolor{blue}{c}}$ }  & 1.3\supercite{2022FrASS...934458K} & \multirow{2}{*}{1.2} \\
 &  $\sim$0.9\supercite{2010AsBio..10..751S} & \\ \hline
$F_{\rm 0.464\ \mu m,\ g^{\prime}}$/$F_{\rm 0.745\ \mu m,\ TESS}$  $^{\textcolor{blue}{d}}$ & 2.78$\pm$0.1 \supercite{2020ApJ...902..115H} & 2.65 \\ \hline \hline
\end{tabular} 
\end{center}
$^{\textcolor{blue}{a}}$ Energy Correction Factor (ECF) under the 9,000 K hot-blackbody assumption. \\
$^{\textcolor{blue}{b}}$ Flux ratio derived by GALEX FUV and NUV bands, including 21 flare events from GJ 65 \supercite{2022ApJ...928....8F} and 182 events from primarily M-type stars.\supercite{2024MNRAS.532.4436B} \\
$^{\textcolor{blue}{c}}$ Flux ratio derived from the SED models of Kowalski\supercite{2022FrASS...934458K} and Segura et al.\supercite{2010AsBio..10..751S} for the ``great flare'' of AD Leo. \\
$^{\textcolor{blue}{d}}$ Flux ratio derived from multiple superflares observed by Evryscope and TESS, detailed in Howard et al.\supercite{2020ApJ...902..115H}
\end{table}

\clearpage

\subsubsection*{\textit{Quiescent NUV radiation correction based on observations}}
\begin{figure}[h]
\centering
\includegraphics[width=1.0\linewidth]{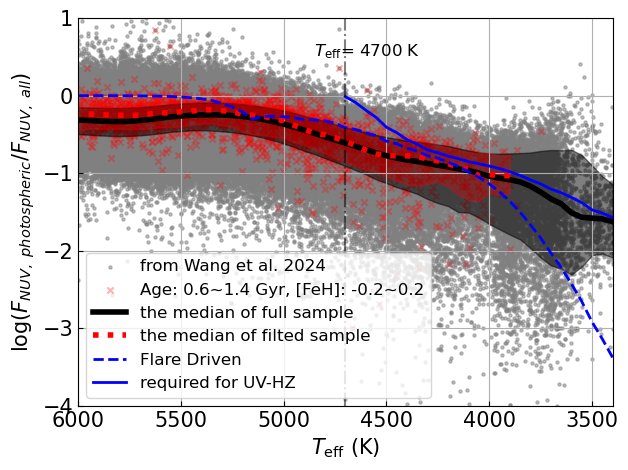}
\caption{
\textbf{Ratio of photospheric emission to total emission in GALEX NUV band (0.177--0.283 $\mu m$).}  
The gray dots are from Wang et al.\supercite{2024ApJ...976...43W} The black line indicates the binned medians. The grey area represents the 1-$\sigma$ dispersion. To examine stellar radiation with an age of 1.0 Gyr, we filtered the stars with ages from 0.6 to 1.4 Gyr and [FeH] from -0.2 to +0.2. The selected stars are shown by pink crosses. The red dotted line indicates the binned medians of filtered samples, which are consistent with the median of the full sample. The dashed blue line indicates the ratio of the NUV radiation flux of the stellar photosphere (from the PHOENIX model) to the radiation flux driven by the flare (from this work). The blue solid line indicates the ratio of photospheric NUV radiation to total NUV radiation when the outer boundary of LW-HZ just meets the NUV radiation required by the biological mechanism (i.e., abiogenesis flux). Stars above the blue solid lines can not sustain a UV-HZ. Stars hotter than 4,700 K (separated by the black vertical dot-dashed line) have UV-HZ in the quiescent state.
\label{fig:quiet_NUVflux}}
\end{figure}

To accurately estimate the NUV emission of stars in the quiescent state, we used the ratio of photospheric emission to total observed emission via GALEX NUV band \supercite{2024ApJ...976...43W} (see Table~\ref{tab:samples}).
Also, to match the general case used in this paper (i.e., the stellar age is 1.0 Gyr, and the metallicity [Fe/H] is 0), we used the age information provided by Wang J. H.\supercite{2025ApJS..280...13W} and the [Fe/H] parameter provided by Wang S.\supercite{2024ApJ...976...43W}.
We found that the median values of the full sample and the filtered sample (Age: 0.6--1.4 Gyr, [Fe/H]$\sim$-0.2--0.2) were close (see Figure~\ref{fig:quiet_NUVflux}). Due to the lack of stars with temperatures below 4,000 K in the filtered sample, we adopted the statistical median values of the full sample in the NUV flux correction.

\subsubsection*{\textit{Enhancement of stellar luminosity caused by flares}} \label{subsect:method_enhancement}
The cumulative flare frequency distribution (CFFD), denoted by $\nu$, can be described as
\begin{equation}
\nu = k_{\rm c} E^{\alpha_{\rm c}}
\label{equation:cffd}
\end{equation}
or 
\begin{equation}
{\rm log}_{10}(\nu) = \beta_{\rm c} + \alpha_{\rm c} \times {\rm log}_{10}(E), \label{equation:cffd}
\end{equation}
where $\beta_{\rm c}$ = log$_{10}(k_{\rm c})$.
The occurrence frequency of flares with energies between $E$ and $E + dE$ is given by 
\begin{equation}
d\nu(E) = k_{\rm c}\alpha_{\rm c} E^{(\alpha_{\rm c}-1)} dE.
\label{equation:cffd}
\end{equation}
If the unit of $k_{\rm c}$ is set as $day^{-1}$, then the total flare energy per day is 
\begin{align}
 E_{\rm total} &= \int_{E_{\rm low}}^{E_{\rm high}} Ed\nu(E)  dE=\int_{E_{\rm low}}^{E_{\rm high}} k_{\rm c}\alpha_{\rm c} E^{\alpha_{\rm c}}dE &  \notag \\ \\
           &= 
\begin{cases} 
\alpha_{\rm c}10^{\beta_{\rm c}}\times(E_{\rm high}^{(\alpha_{\rm c}+1)}-E_{\rm low}^{(\alpha_{\rm c}+1)})/(\alpha_{\rm c}+1),
 & \mbox{for} \ \alpha_{\rm c} \neq -1 \\ \\
\alpha_{\rm c}10^{\beta_{\rm c}}\times ln(E_{\rm high}/E_{\rm low}) ,& \mbox{for} \ \alpha_{\rm c} = -1
\end{cases}
\label{equation:FlareEnergy}
\end{align}
where $E_{\rm low}$ and $E_{\rm high}$ are the lower and upper energy limits considered in the calculation, respectively.
The energy range for detected solar and stellar flares spans from $10^{28}$ to $10^{38}$ erg.\supercite{2011LRSP....8....6S,2019ApJS..241...29Y,2020AJ....159...60G,2020ApJ...890...46T}

The averaged increase in stellar luminosity due to flares is:
\begin{equation}
L_{\rm flare}  = E_{\rm total} /({\rm {1\ day}}) = E_{\rm total} /({\rm {86400\ s}}).
\label{equation:Lflare}
\end{equation}

Using the $\alpha_{\rm c}$ and $\beta_{\rm c}$ values from our previous study,\supercite{2022AJ....164..213G} we obtained the distribution relationship of $\alpha_{\rm c}$ and $\beta_{\rm c}$ with $T_{\rm eff}$ for stars ranging from M2 to G1 as follows: \\
\begin{equation}\label{equation:alpvsteff}
\alpha_{\rm c}= \left\{ \begin{array}{rcl}
0.44(\pm0.01)\times\frac{T_{\rm eff}}{1000\ \rm  K}-2.84(\pm0.05), & \mbox{for}
& 3400\ \rm K\leq\ \mathit{T}_{\rm eff} < 5150\ \rm K \\ \\
-0.53(\pm0.03)\times\frac{T_{\rm eff}}{1000\ \rm  K}+2.18(\pm0.18), & \mbox{for} & 5150\ \rm K\leq\ \mathit{T}_{\rm eff}\leq 5900\ \rm K
\end{array}\right.
\end{equation}
\begin{equation}\label{equation:betavsteff}
\beta_{\rm c}= -32.0\times\alpha_{\rm c}-1.09\times10^{-7}(\pm1.10\times10^{-4} )\times\frac{T_{\rm eff}}{1000\ \rm K}+1.12(\pm0.54).
\end{equation}

In our previous study, which relied on the assumption of a 9,000 K hot-blackbody model,\supercite{2022AJ....164..213G} the contribution of UV radiation was underestimated. Adopting the validated SED model in the main paper, we can calculate a correction factor for the modeled flux in the NUV band, i.e.,
\begin{equation}
R_{\rm corrected}= \frac{\int_{\rm bol} F_{\lambda}(T_{\rm flare})d\lambda}{\int_{\rm bol} B_{\lambda}(T_{\rm flare})d\lambda},
 \end{equation}
where $T_{\rm flare}$ represents the flare temperature and is set to be 9,000 K, $F_{\rm \lambda}$ is the flare spectrum acccording to the flare SED model (Equation~\ref{equation:flare_sed} in the main paper), and $B_{\lambda}(T_{\rm flare})$ is the Planck function evaluated at the flare temperature. For instance, for the G-type star KOI-7703, $R_{\rm corrected}$ = 1.14; while for the M-type star Kepler-1512, $R_{\rm corrected}$ = 1.87.

After the correction of the flare energy, the slope of CFFD is unchanged ( i.e., $\alpha_{\rm c}$ remains constant), while $\beta_{\rm c}$ increases. The updated $\beta^{'}_{\rm c}$ after correction can be described by the fitted parameter before correction: 
\begin{equation} \label{equation:beta_correct}
\beta^{'}_{\rm c}=-{\rm log}_{10}(R_{\rm corrected})\times\alpha_{\rm c} +  \beta_{\rm c}.
\end{equation}

\begin{figure}[h]
\centering
\includegraphics[width=0.8\linewidth]{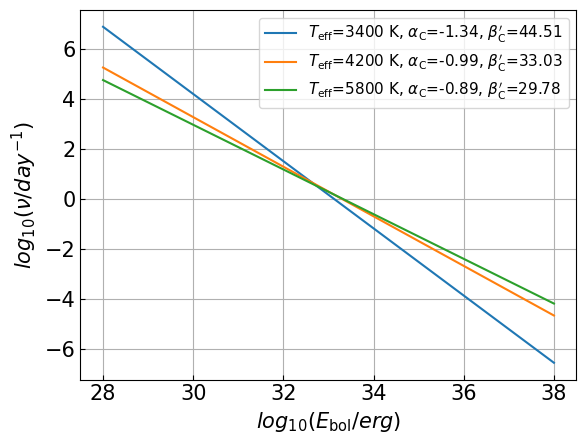}
\caption{\textbf{The CFFDs of three types of stars with temperatures of 3,400 K, 4,200 K, and 5,800 K, respectively.} The power-law index $\alpha_{\rm C}$ and $\beta^{'}_{\rm C}$ are obtained from Equation \ref{equation:alpvsteff} - \ref{equation:beta_correct}. \label{fig:alpha_show}}
\end{figure}

The empirical CFFDs of typical G, K, and M stars after correction are shown in Figure \ref{fig:alpha_show}. For most M-type flaring stars, we find $\alpha_{\rm C} \textless$ -1, indicating that low-energy flares contribute the majority of the total flare energy. On the contrary, for stars hotter than K5, we find $\alpha_{\rm C}$ $\textgreater$ -1, indicating that high-energy flares are dominant.\supercite{2014ApJ...797..121H,2018ApJ...858...55P,2021MNRAS.504.3246J, 2022AJ....164..213G}

Due to the CFFD of different stars, the averaged enhancement of bolometric luminosity of a star due to flares ($L_{\rm flare}$) can be obtained directly by Equation (\ref{equation:Lflare}). To estimate the averaged enhancement of UV luminosity, we need to combine the SED model to estimate the area of active regions during flares, $A_{\rm flare}$. According to the following equation:
\begin{equation}
\frac{L_{\rm flare}}{L_{*}} = \frac{A_{\rm flare}\int_{\rm bol}F_{\rm flare}(\lambda)d\lambda}{\pi R_{*}^2 \int_{\rm bol}S_{*}(\lambda)d\lambda},
\end{equation}
where $F_{\rm flare}(\lambda)$ is the flare SED as defined in Equation (\ref{equation:flare_sed}) in the main paper, and $S_{*}(\lambda)$ is the stellar spectra in quiescent stage after correction. Since we already know the values of both $L_{\rm flare}$ and the luminosity of star in quiescent stage $L{*}$, we can calculate the dimensionless ratio:
\begin{equation}
\frac{A_{\rm flare}}{\pi R_{*}^2 } = \frac{\frac{L_{\rm flare}}{L_{*}}}{\frac{\int_{\rm bol}F_{\rm flare}(\lambda)d\lambda}{\int_{\rm bol}S_{*}(\lambda)d\lambda}}.
\end{equation}
The bolometric integral of the spectra is integrated from 0.1 to 3 $\mu m$. Then we can estimate the average enhancement in the NUV band as follows by integration from 0.2 to 0.28 $\mu m$:
\begin{equation}
f_{\rm flare,\ NUV} = \frac{A_{\rm flare}}{\pi R_{*}^2}\frac{\int_{\rm NUV}F_{\rm flare}(\lambda)d\lambda}{\int_{\rm NUV}S_{*}(\lambda)d\lambda}.
\label{equation:tuidao_flareUVHZ}
\end{equation}

\subsection*{UV Habitable Zone of Flaring stars}\label{sect:HZofFS}
\subsubsection*{\textit{Habitable zones around different types of flaring stars}}\label{sect:appl_differentMass}
\begin{figure}[h]
\centering
\includegraphics[width=1.0\linewidth]{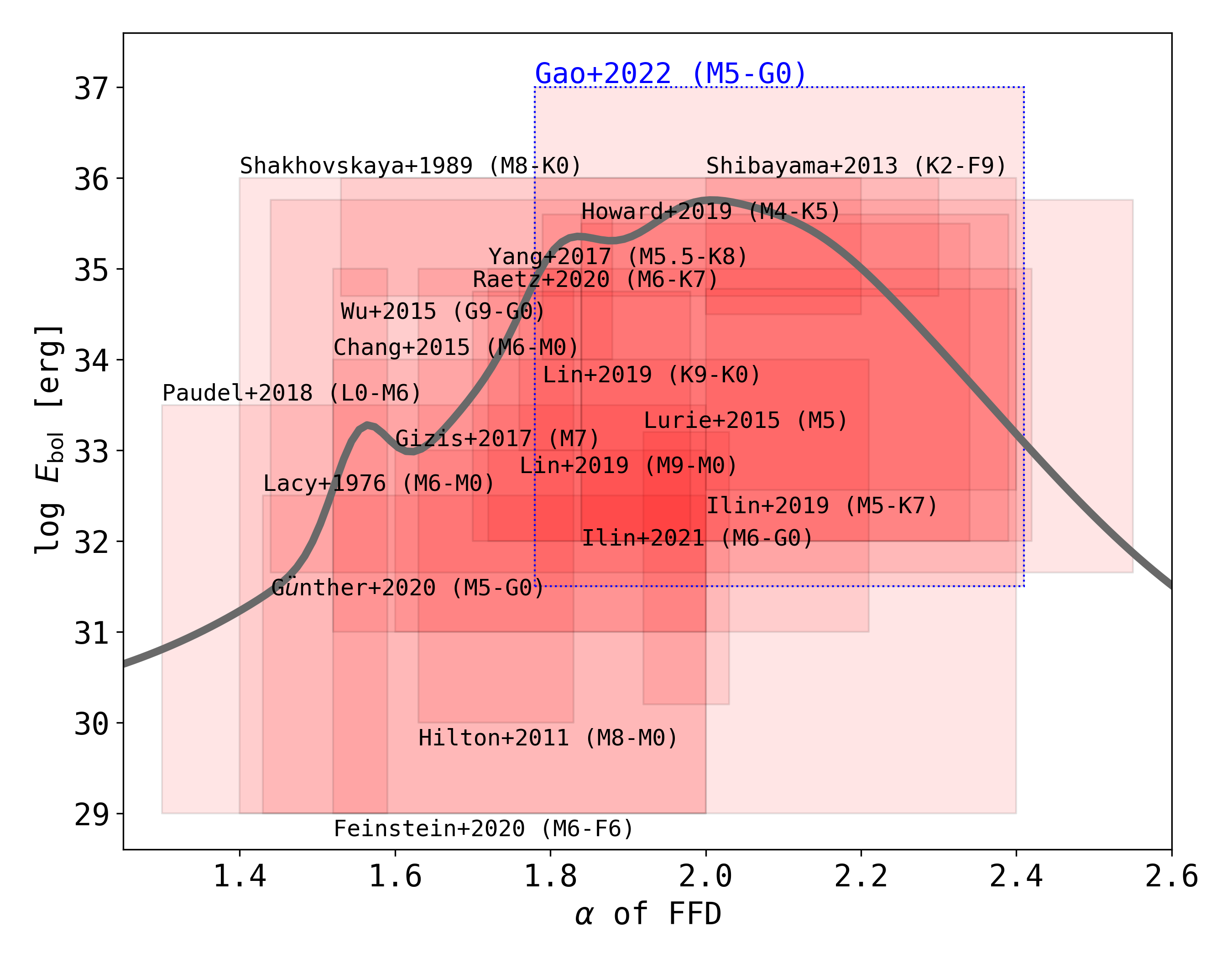}
\caption{\textbf{Overview of FFD fitting.} Red rectangles indicate the flare energy range on the vertical axis and the $\alpha$ range on the horizontal axis, with references positioned in the upper or lower left corner of each rectangle.
The gray line illustrates the superposition of these results, with each result depicted by a Gaussian fit.\supercite{2021A&A...645A..42I} The rectangle marked with a blue dashed line shows the FFD parameters used in this work, sourced from Gao et al.\supercite{2022AJ....164..213G} Note that the value illustrated here is the positive power-law index $\alpha$ of the FFD, distinct from the negative power-law index of the CFFD ($\alpha_{\rm c}$).
In the case of complete flare detection, $\alpha_{\rm c}\sim -\alpha+1$. 
\label{fig:FFDcompare}}
\end{figure}

We adopt our previous study\supercite{2022AJ....164..213G} to describe the CFFD for different flaring stars with age $\sim$1 Gyr. Based on 121 faring star samples, including G, K, and M stars, we obtained consistent $\alpha$ with other research as shown in Figure~\ref{fig:FFDcompare}. An empirical relationship between the CFFD and stellar effective temperature ($T_{\rm eff}$) is achieved, as shown in Equation~\ref{equation:alpvsteff} and \ref{equation:betavsteff}. The rotating period of the stars are between 1 and 5 day, which induced an age of around 0.1 to 3 Gyr via empirical gyrochronology isochrones.\supercite{2008ApJ...687.1264M,2019ApJS..241...29Y} Furthermore, 57 (47\%) of the 121 samples also have ages from FLAME (Final Luminosity Age Mass Estimator\supercite{2023A&A...674A..28F})  with 95\% of them being older than 1 Gyr. Therefore, the correlations are adopted to represent the CFFD of stars around 1.0 Gyr old.

\begin{figure}[h]
\centering
\includegraphics[width=1.0\linewidth]{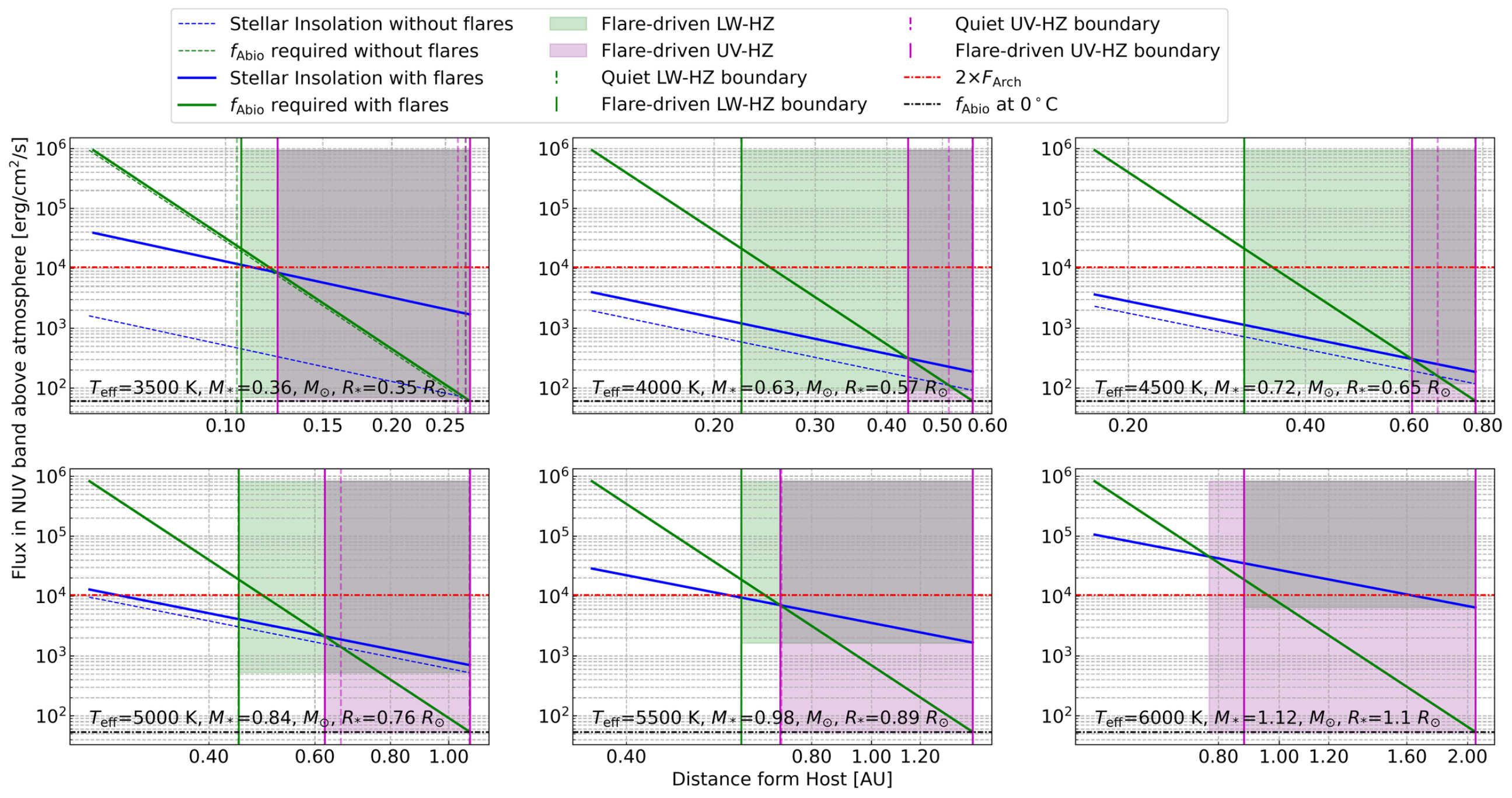}
\caption{\textbf{The flare-driven LW-HZ (green area) and UV-HZ (purple area) for flaring stars across various $T_{\rm eff}$.}
Different colors and line styles are explained at the top.
The dashed and solid  blue lines represent the stellar insolation without flares and with flares, respectively.
The dashed and solid green lines represent the required abiogenesis flux without flares and with flares, respectively.
The dashed and solid green vertical lines represent the boundary of the LW-HZ without flares and with flares, respectively.
The dashed and solid magenta vertical lines represent the boundary of the UV-HZ without flares and with flares, respectively.
The dot-dashed red horizontal line represents twice the NUV flux received by the Archean Earth ($2\times F_{\rm Arch}$).\supercite{2006Icar..183..491B}
The dot-dashed black horizontal line represents the minimum NUV flux necessary for abiogenesis ($f_{\rm Abio}$) at \SI{0}{\celsius}.\supercite{2018SciA....4.3302R}
Note that the dot-dashed red and black horizontal lines are only drawn in the figure for comparison with previous work.
Our definition of UV-HZ does not use the values corresponding to these two lines.}
\label{fig:AZ_flux_diff}
\end{figure}

To estimate the stellar parameters around 1 Gyr, we use the stellar evolution model described by Baraffe et al.\supercite{2015A&A...577A..42B} The stellar masses are set between 1.1 and 0.3 solar masses to represent different G, M, and K stars, and we calculate their effective temperatures ($T_{\rm eff}$) and radii ($R_{\rm *}$) at 1 Gyr. The photospheric NUV flux is generated by the Phoenix model according to the stellar parameters ([Fe/H]=0). Then we do correction for quiescent NUV radiation by dividing the median value of the full sample in Figure \ref{fig:quiet_NUVflux}. These parameters were subsequently utilized to calculate the inner boundary of the UV-HZ with flares for the general case. Considering the planetary atmosphere transmittance, ozone and oxygen are absent in the atmosphere of Archean Earth. Thus, we adopt the atmosphere transmittance as 0.84 for K and G stars (with effective temperatures $T_{\rm eff} \geq$ 5,000 K and masses $M_{*} \geq$ 0.84 M$_{\odot}$), and 0.74 for cooler stars ($T_{\rm eff}<$ 5,000 K).\supercite{2006Icar..183..491B} Figure~\ref{fig:AZ_flux_diff} illustrates LW-HZ (green area) and UV-HZ (purple area) for flaring stars across various $T_{\rm eff}$. For a better view, Figure~\ref{fig:HZofDiffMass} in the main paper illustrates the boundaries of the habitable zones both with and without flares for different stellar masses.

\subsubsection*{\textit{Habitable Zones around \textit{Kepler} flaring hosts}}\label{sect:appl_hosts}
\begin{figure}[h]
\centering
\includegraphics[width=1.0\linewidth]{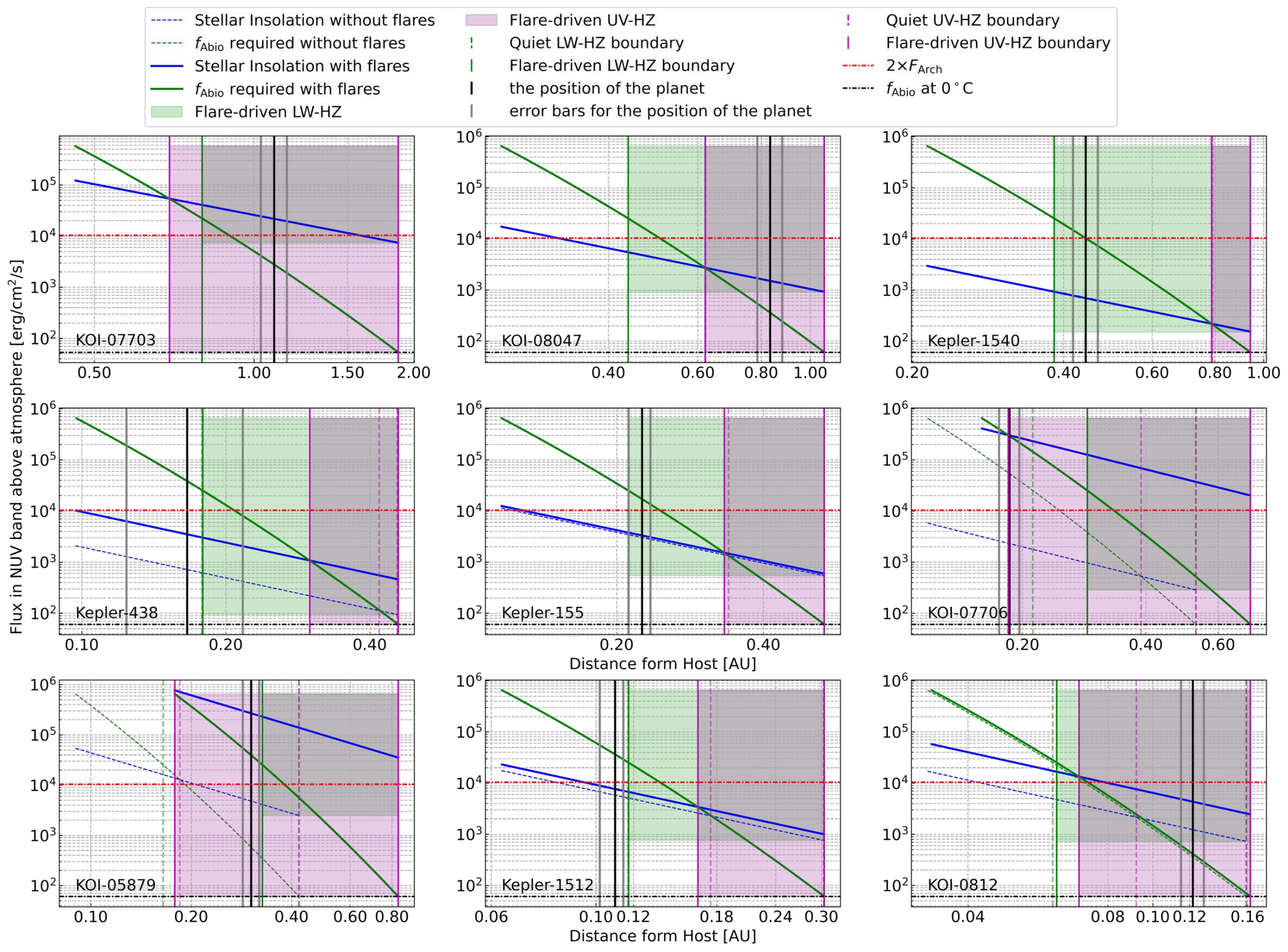}
\caption{\textbf{The flare-driven LW-HZ (green area) and UV-HZ (purple area) for flaring Kepler planetary systems analyzed in this paper.} The black solid vertical lines represent the positions of planets/candidates and their error bars (gray solid vertical lines).
Other colors and line styles are the same as Figure~\ref{fig:AZ_flux_diff}.
\label{fig:AZ_flux_host}}
\end{figure}

For \textit{Kepler} flaring stars, we fit the CFFD parameters $\alpha_{\rm C}$ and $\beta^{'}_{\rm C}$ for all nine stars, as shown in Table \ref{tab:alpofHosts}. The updated stellar and planetary parameters are presented in Table \ref{tab:samples}. We obtained their boundaries of the LW-HZs and the UV-HZs as shown in Table \ref{tab:HZofHosts} in the main paper and Figure~\ref{fig:AZ_flux_host}. Figure~\ref{fig:HZofHosts} in the main paper also illustrates these boundaries for a better view.

\begin{table}
\caption{\normalsize CFFD parameters of Kepler flaring stars}
\label{tab:alpofHosts}
\begin{center}
\begin{tabular}{cccc}
\hline \hline
KIC ID & Planet Name & $\alpha_{\rm C}$ & $\beta'_{\rm C}$ \\ \hline 
 4742188 & KOI-7703.01$^{\textcolor{blue}{a}}$  & - & -   \\
 11294822 & KOI-8047.01 & -1.08$\pm${0.34} & 35.18$\pm${11.28}  \\
 5938970 & Kepler-1540 b  & -1.10$\pm${0.03} & 35.49$\pm${1.12}  \\
 6497146 & Kepler-438 b & -1.70$\pm${0.25} & 54.42$\pm${8.16}   \\
 12302530 & Kepler-155 c  & -0.42$\pm${0.14} & 12.51$\pm${4.65}  \\
 4762283 & KOI-7706.01 & -1.58$\pm${0.07} & 53.00$\pm${2.58}   \\
11198419 & KOI-5879.01$^{\textcolor{blue}{b}}$  & -2.35$\pm{0.07}$ & 77.94$\pm${2.44}\\
8424002 & Kepler-1512 b & -1.64$\pm${0.01} & 52.12$\pm${0.41}  \\
10452252 & KOI-8012.01 & -1.36$\pm${0.04} & 44.47$\pm${1.34}   \\  \hline \hline
\end{tabular} 
\end{center} 
$^{\textcolor{blue}{a}}$ For KIC 4742188 (KOI-7703), only one flare event was detected, so the CFFD parameters could not be fitted.\\
$^{\textcolor{blue}{b}}$ To calculate the total flare energy, we established an integration range from $10^{28}$ to $10^{38}$ erg for all stars, except KOI-5879. For KOI-5879, due to its minimum $\alpha_{\rm C}$ value of -2.35, the luminosity from flares ($L_{\rm flare}$) approximates the star's own luminosity ($L_{*}$) when setting the lower limit at $10^{29.85}$ erg. We therefore chose $10^{30}$ erg for this star as a more appropriate lower limit for flare energy, resulting in a ratio of $L_{\rm flare}$/$L_{*}$ equal to 0.62.\\
(This table is available in its entirety in machine-readable form.)
\end{table}

\begin{sidewaystable}
\caption{\normalsize Stellar and planetary parameters of the nine Kepler stars.}
\label{tab:samples}
\begin{center} \small
\begin{tabular}{c|ccccccc|cc|cc} \hline \hline
The planets/candidates &  $T_{\rm eff}$ & $R_{*}$& $[Fe/H]$ &log$g$ & $M_{*}$ & Distance$^{\textcolor{blue}{a}}$& $f_{\rm NUV,\ obs}$/$f_{\rm NUV,\ ph}$ $^{\textcolor{blue}{b}}$ & $R_{\rm p}$ &  Ref. $^{\textcolor{blue}{c}}$& $T_{\rm eq}$$^{\textcolor{blue}{d}}$  \\ 
  & (K) & (R$_{\odot}$) & (dex) & (dex) &(M$_{\odot}$) &(pc) &  & (R$_{\oplus}$) &  & (K) \\ \hline
KOI-8012.01  & 3374$^{+112}_{-82}$ & 0.218$^{+0.09}_{-0.06}$ & -0.36$^{+0.30}_{-0.25}$ & 5.066$^{+0.103}_{-0.126}$ & 0.202$^{+0.111}_{-0.06}$ & 101.197 $^{+0.479}_{-0.475}$ & 452& 0.42$^{+0.17}_{-0.12}$ &  \cite{2018ApJS..235...38T} & 216    \\
Kepler-1512 b  & 3484$^{+91}_{-91}$ & 0.393$^{+0.048}_{-0.072}$ & 0.15$\pm$0.11 &4.865$^{+0.077}_{-0.045}$ & 0.418$^{+0.051}_{-0.090}$ & 315.276$\pm$23.903 & 402 &0.80$\pm$0.13&  \cite{2017AJ....154..264T} & 318     \\
KOI-5879.01 & 3827$^{+56}_{-75}$ & 0.462$^{+0.05}_{-0.07}$ & -0.40$^{+0.1}_{-0.3}$ &4.782$^{+0.06}_{-0.07}$ &0.471$^{+0.06}_{-0.09}$ & 414.134$\pm$5.229 & 365 &1.60$^{+0.17}_{-0.24}$  & \cite{2015ApJS..217...31M} &  228 \\
KOI-7706.01  &  4281$^{+115}_{-140}$ & 0.480$^{+0.034}_{-0.064}$& -1.18$^{+0.30}_{-0.35}$ &4.761$^{+0.077}_{-0.033}$ & 0.485$^{+0.046}_{-0.3}$ & 436.123$^{+5.203}_{-5.084}$ & 7.36 & 1.19$^{+0.08}_{-0.16}$ & \cite{2018ApJS..235...38T} &332   \\
Kepler-155 c & 4057$^{+251}_{-64}$ & 0.529$^{+0.045}_{-0.135}$& -0.06$^{+0.090}_{-0.085}$ &4.721$^{+0.147}_{-0.027}$ & 0.542$^{+0.053}_{-0.125}$ & 293.499$\pm$2.279 & 61.4 & 1.97$^{+0.19}_{-0.48}$  & \cite{2017AJ....154..264T} & 300 \\
Kepler-438 b & 3748$^{+112}_{-112}$ & 0.520$^{+0.038}_{-0.061}$ & 0.16$\pm$0.14 &4.740$^{+0.059}_{-0.029}$ & 0.544$^{+0.041}_{-0.061}$ & 179.882$^{+3.241}_{-3.128}$ &  33.9 &1.12$^{+0.16}_{-0.17}$ & \cite{2017AJ....154..264T} &  320  \\
Kepler-1540 b  & 4641$^{+74}_{-83}$ & 0.747$^{+0.02}_{-0.04}$& 0.16$\pm$0.15 &4.557$^{+0.049}_{-0.014}$ & 0.734$^{+0.044}_{-0.028}$ & 244.960$^{+0.903}_{-0.896}$ & 3.9 &3.14$^{+0.08}_{-0.17}$ & \cite{2018ApJS..235...38T} &  291  \\
KOI-8047.01 & 4829$^{+144}_{-129}$ & 0.794$^{+0.046}_{-0.051}$ & 0.48$^{+0.05}_{-0.3}$ &4.564$^{+0.032}_{-0.048}$ & 0.843$^{+0.036}_{-0.056}$ & 587.233$^{+6.626}_{-6.480}$ & 9.4 &1.98$^{+0.11}_{-0.13}$  & \cite{2018ApJS..235...38T} &  227  \\
KOI-7703.01& 6031$^{+180}_{-198}$ & 0.992$^{+0.311}_{-0.111}$ & -0.14$\pm$0.3 &4.45$^{+0.07}_{-0.21}$ & 1.012$^{+0.144}_{-0.131}$ & 1906.89$^{+135.25}_{-118.76}$ & 1.95 & 1.96$^{+0.61}_{-0.22}$  & \cite{2018ApJS..235...38T} &  277  \\ \hline
The Archean Earth  & 5607 & 0.9113 & 0.00 & 4.44 &  1.00 & 1.00 & 1.41 & 1.0 & - & 258  \\ \hline \hline
\end{tabular} 
\end{center} 
$^{\textcolor{blue}{a}}$ The distance values of Kepler-438 and KOI-8047 are derived using parallaxes from GAIA DR3,\supercite{2022yCat.1355....0G} while the distance values for other stars come from NASA Exoplanet Archive[\textcolor{cyan}{\url{https://exoplanetarchive.ipac.caltech.edu/}}]. \\
$^{\textcolor{blue}{b}}$ This column lists the ratios of total NUV emission to photospheric NUV emission. The ratios of five hosts (KOI-8012, Kepler-1512, KOI-5879, Kepler-155, and KOI-8047) are measured using GALEX NUV data. For four hosts (KOI-7706, Kepler-438, Kepler-1540, and KOI-7703) without GALEX NUV and Swift UVOT observations, we estimated their ratios using the statistical values of Wang et al.\supercite{2024ApJ...976...43W} (binned median curves in Figure~\ref{fig:quiet_NUVflux}). For the Sun, we used the solar spectrum of Willmer\supercite{2018ApJS..236...47W} and the age evolution of NUV radiation of G-type stars to obtain the ratio of the Archean Sun with an age of 1.0 Gyr.\supercite{2025ApJS..281...13L} \\
$^{\textcolor{blue}{c}}$ This column lists the references for planetary parameters.\\
$^{\textcolor{blue}{d}}$ The planet's equilibrium temperature ($T_{\rm eq}$) is calculated assuming zero albedo and complete redistribution of heat between the day and night sides.\supercite{2018PASP..130k4401K}\\
(This table is available in its entirety in machine-readable form.)
\end{sidewaystable}

\subsection*{The requirements of observed CFFD for abiogenesis} \label{sect:AZ}
For flaring stars, flares in certain bands are directly observed; thus, we can obtain the flare energy in such a band, rather than the total energy. To estimate the abiogenesis limit of CFFD in the NUV band (200-280 nm), we need to convert the flare energy to the NUV band via the SED model. Here, we demonstrate how to derive the CFFD needed for abiogenesis at a given planetary surface temperature. We adopt the U-band (320-400 nm), for instance, because there are previous observations for flares on M stars, e.g., AD Leo.\supercite{2014ApJ...797..121H} The observed flare energy flux in U-band $F_{\rm U}$ can be converted to the energy flux in NUV band, i.e., the correction factor is
\begin{equation}
r_{\rm U} = \frac{F_{\rm U}}{\int_{\rm {200\ nm}}^{\rm {280\ nm}}F_{\rm flare}(\lambda)d\lambda}.
\label{equation:k_U'}
\end{equation}
For G type star like KOI-7703，$r_{\rm U}=1.94$， while for M type star like KOI-8012, $r_{\rm U}=1.19$. Considering the NUV flux due to flares dominates for M stars, we can ignore the flux for the quiet star. For a surface temperature of $T_{\rm surf}$, the required energy for abiogenesis in NUV band is $F_{\rm 0}\times e^{(T_{\rm surf}-273\ \rm K)/10\ \rm K}$, where $F_{\rm 0}=8 \times 10^{27}\ \rm {erg\ s^{-1}}$ is the required flux according to Equation \ref{equation:fAZ1}. Thus, for a given $T_{\rm surf}$ of a planet located $a_{\rm p}$ from the star, we can obtain a CFFD limit in U-band for flaring stars as follows,
\begin{equation}
v \geq \frac{8 \times 10^{27}\ \rm {erg\ s^{-1}}}{E_{\rm U}} (\frac{a_{\rm p}}{1\ \rm {AU}})^2\times e^{(T_{\rm surf}-273\ {\rm K})/10\ {\rm K}} \times \frac{r_{\rm U}}{\eta}, 
\end{equation}
where $\eta$ is the planetary atmosphere transmittance. Note for G and K stars, when the quiescent NUV flux is not ignorable, it should be deducted from $F_{\rm 0}$. Especially, if the quiescent NUV flux is larger than $F_{\rm 0}$, the limit is satisfied, and there is no limit for the flare frequency. Setting the surface temperature $T_{\rm surf}$= \SI{0}{\celsius} (273 K) and $\eta=1$, we can obtain a minimum CFFD limit,
\begin{equation}
v \geq 6.9\ {\rm day^{-1}}\times(\frac{10^{32}\ {\rm erg} }{E_{\rm U}})(\frac{a_{\rm p}}{1\ {\rm AU}})^2\times r_{\rm U}.
\label{equation:az}
\end{equation}
Note, if the observed CFFD in the U-band intersects with the CFFD limit, there should be enough NUV energy to drive the prebiotic chemistry for abiogenesis.

\begin{figure}[htb]
\centering
\includegraphics[width=1.0\linewidth]{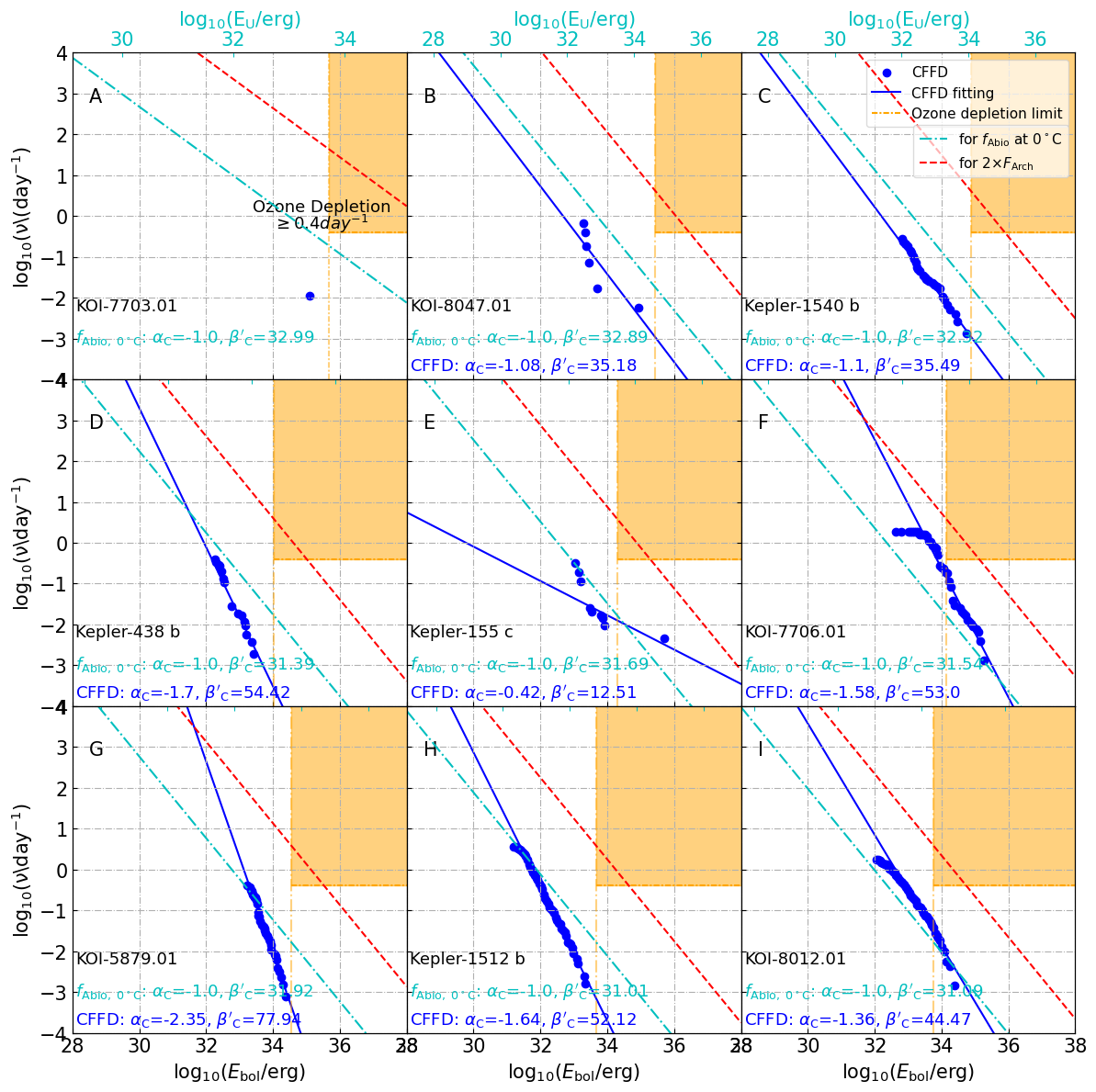}
\caption{\textbf{CFFDs of exoplanet host stars.}
The blue dots represent the CFFD directly from observations. The blue line illustrates the fitted CFFD, with fitted parameters ($\alpha_{\rm C}$ and $\beta'_{\rm C}$, in blue text) marked in the lower left corner of the figure. 
The orange area denotes the ozone depletion region, where flare frequency $\geq$0.4 day$^{-1}$ for flare bolometric energy larger than $10^{34}\times(\frac{d}{0.16\ \rm {AU}})^2$ erg. The cyan dot-dashed line represents the minimum limit of CFFD required by the abiogenesis flux according to Equation \ref{equation:az}. The red dashed line represents the frequency for twice the NUV flux received by the Archean Earth. For KOI-7703, only one flare was detected during Quarter 7 (89.4 days), with an energy of 1.28$\times$10$^{35}$ erg, resulting in a flare frequency of approximately 0.01 day$^{-1}$.
Note that the cyan dot-dashed and red dashed lines are only drawn in the figure for comparison with previous work. Our definition of UV-HZ does not use the values corresponding to these two lines. Since the surface temperature of a planet can vary with different atmospheric effects, the cyan line provides a minimum requirement for abiogenesis at \SI{0}{\celsius}. If the observed CFFD in the U-band intersects with the cyan line, like Kepler-438 b and Kepler-155 c, there should be enough NUV energy to drive the prebiotic chemistry on the planet. Note, the double x-axis show the bolometric flare energy $E_{\rm bol}$ and the flare energy in U-band $E_{\rm U}$ via the SED model in Equation \ref{equation:flare_sed}. 
\label{fig:AZ_all}}
\end{figure}

We plot the limit of CFFD in Figure \ref{fig:AZ_1}  and Figure~\ref{fig:AZ_all}, to demonstrate the minimum requirements if the planets have a surface temperature of \SI{0}{\celsius}. Also, we plot the CFFD to achieve the flux twice that of the Archean Earth, i.e., 2 $\times$ 5,200 erg cm$^{-2}$ s$^{-1}$ for reference \supercite{2006Icar..183..491B}.

\clearpage
\subsection*{Uncertainty of the  calculations}\label{subsect:uncertainty}
\subsubsection*{\textit{Impact of Flare SEDs}}
\begin{figure}[h]
\centering
\includegraphics[width=1.0\linewidth]{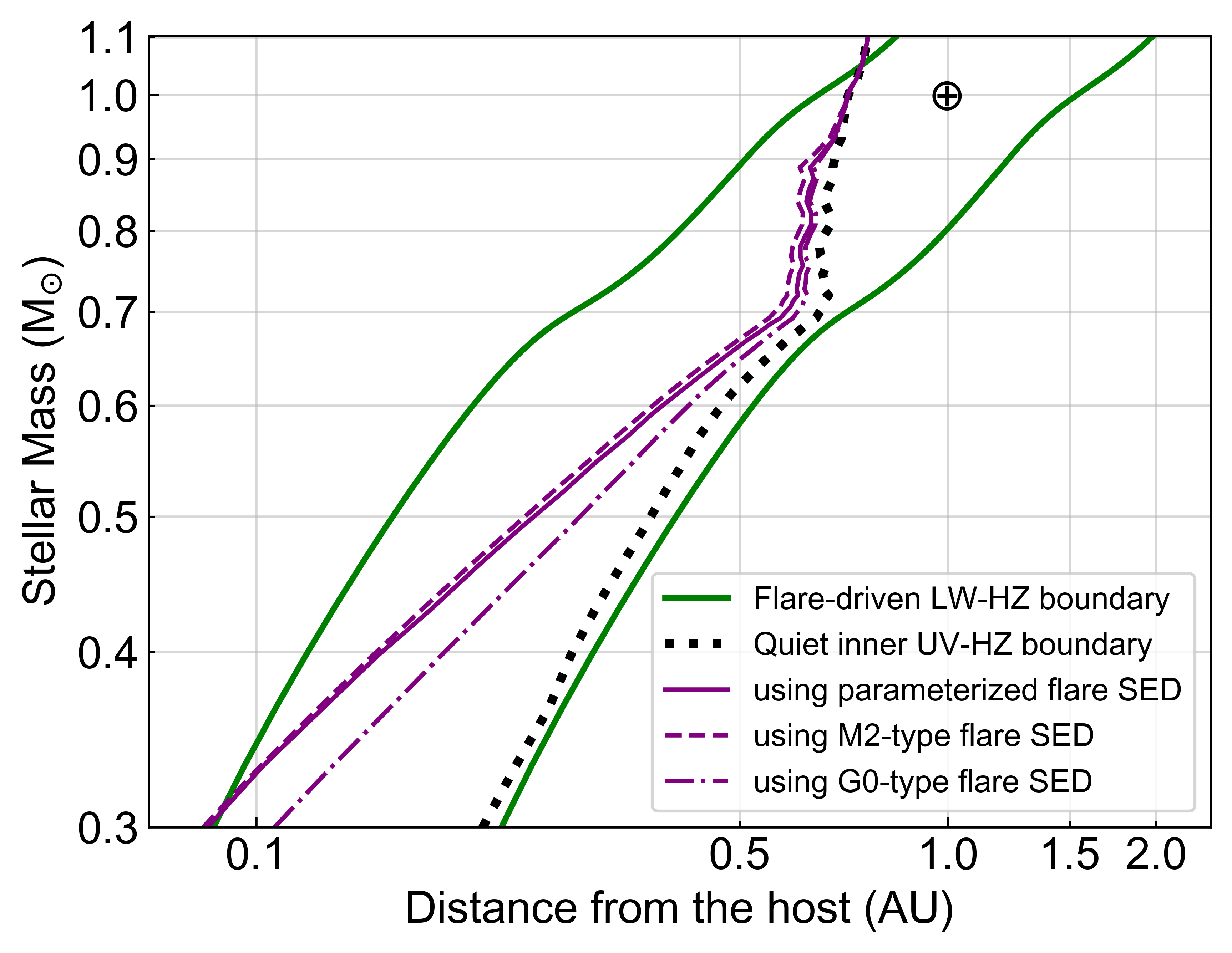} 
\caption{\textbf{Change of UV-HZ and LW-HZ with different flare SEDs.}
The green solid lines show the inner and outer boundary of the LW-HZ with flares. 
The dark dotted line indicates the quiet inner boundary of UV-HZ. 
The purple solid line indicates the flare-driven inner boundary as determined by the parameterized flare SED model (refer to Equation \ref{equation:flare_sed} in the main paper).
The purple dashed and dot-dashed lines illustrate the inner boundary using the M2 ($T_{\rm eff}$=3,400  K) and G2 ($T_{\rm eff}$=6,000 K) flare SED model, respectively.
The changes in the SED influence the UV-HZ more obviously for low-mass stars. The chosen G2 SED with no Balmer jump leads to fewer NUV contributions and then has a further inner UV-HZ.}
\label{fig:traverse1}
\end{figure}

We test two different flare SEDs, i.e, G0-type ($T_{\rm eff}$=6,000 K) and M2-type ($T_{\rm eff}$=3,400 K). No matter what the stellar type is, the variation in flare SED parameters alters the UV-HZ, as depicted in Figure~\ref{fig:traverse1}. No matter which SEDs are adopted, the UV-HZ extended obviously, while the M2 SED leads to a closer inner boundary for all types of stars, because of the more obvious Balmer Jump in the SED model. Further spectrum observations of the flare SED are favored to determine the UV-HZ.

\subsubsection*{\textit{Impact of CFFDs}} \label{subsect:uncertainty_ffds}
\begin{figure}[h]
\centering
\includegraphics[width=0.8\linewidth]{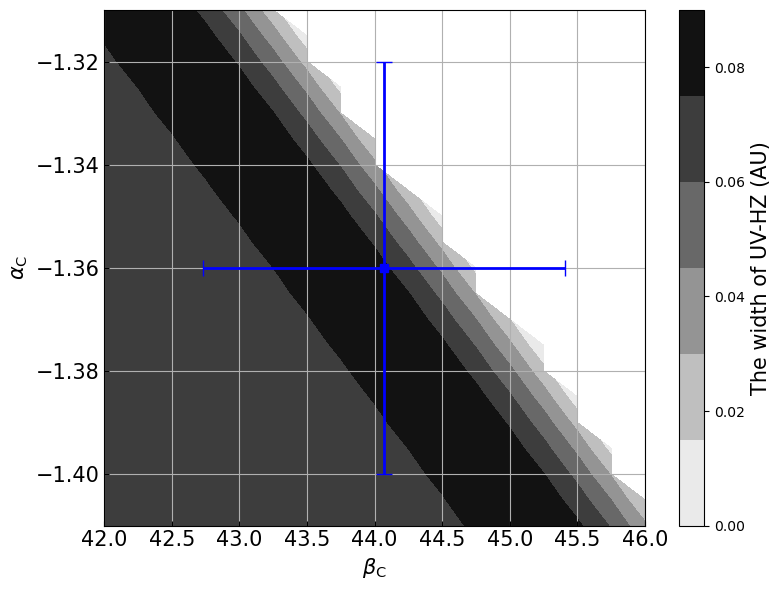}
\caption{\textbf{Variation of UV-HZ width with different CFFD parameters for KOI-8012.}
Blue cross indicate the $\alpha_{\rm C}$ and $\beta_{\rm C}$ values for KOI-8012, obtained using the method of Gao et al.\supercite{2022AJ....164..213G} Here, we artificially cut the ozone depletion limit. When the $\beta_{\rm C}$ increases by $\sim$0.7, or $\alpha_{\rm C}$ increases by 0.2, the width of UV-HZ decreases from $\sim$0.09 AU to 0 AU, and disappear. However, for most cases inside the 1-sigma uncertainty, the UV-HZ can sustain with a width larger than 0.05 AU.
\label{fig:traverse2}}
\end{figure}

The CFFD of a star usually has large uncertainties. We test the influence of CFFD on the UV-HZ, using KOI-8012 as an example. The uncertainties of the two CFFD parameters are shown as the blue cross in Figure~\ref{fig:traverse2}. In most cases, the UV-HZ can exist, with a width of $\sim 0.05$ AU, while for large $\alpha_{\rm C}$ and $\beta_{\rm C}$, the UV-HZ vanishes because of the limit of ozone depletion. Thus, the uncertainties of CFFD can affect the width of UV-HZ, but the influence of the uncertainty would be limited, unless the flare energy reaches the limits of ozone depletion.

The values and uncertainties of the CFFD parameters $\alpha_{\rm C}$ and $\beta^{'}_{\rm C}$ are listed in Table \ref{tab:alpofHosts} for nine \textit{Kepler} flaring hosts. To determine the UV-HZ width in an accurate way, both flare SEDs and the CFFDs need to be measured or modeled more precisely.

\begin{figure}[h]
\centering
\includegraphics[width=1.0\linewidth]{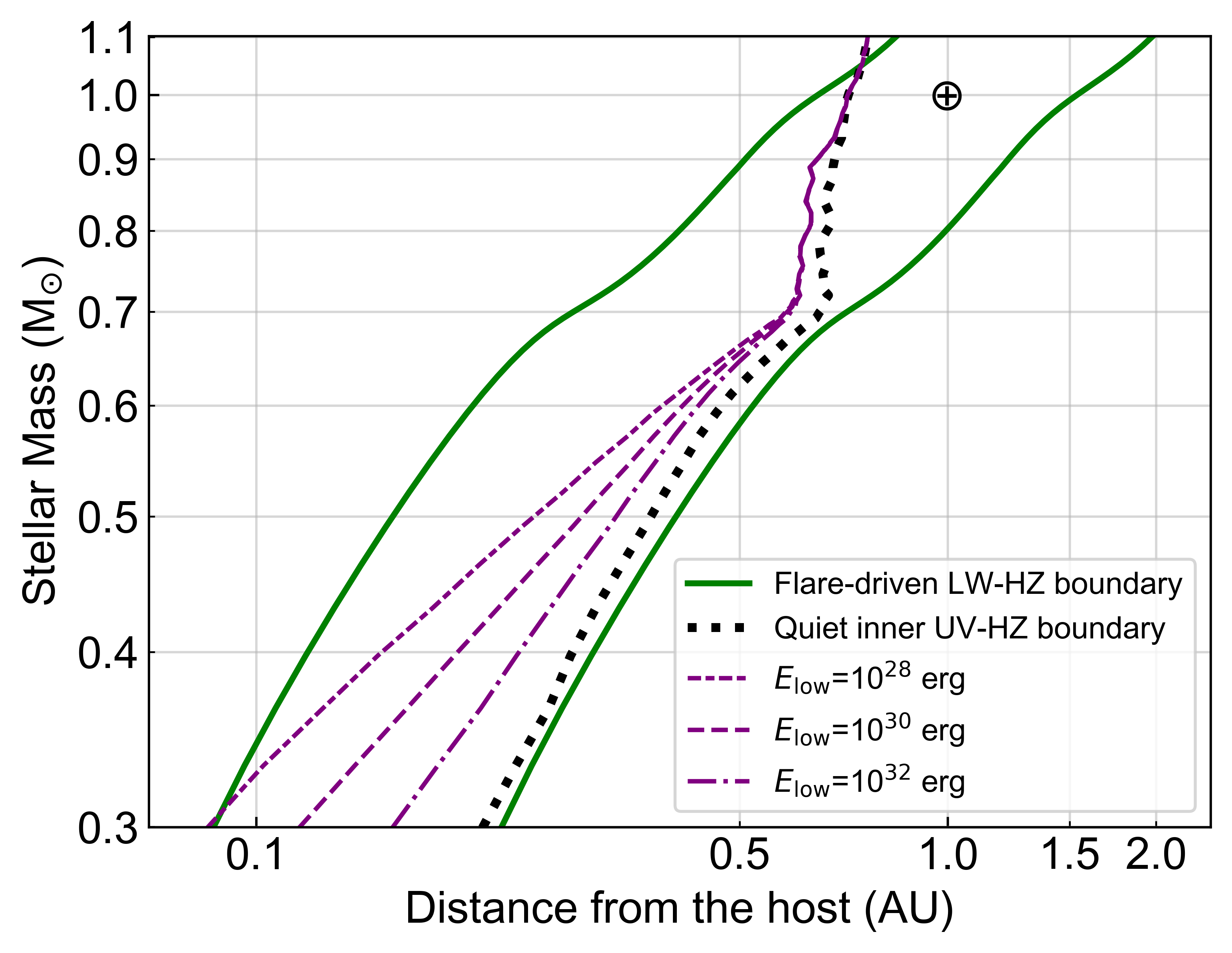}
\caption{\textbf{Variation of the UV-HZ with the lower energy threshold of flare detection.} 
The flare-driven inner boundary of UV-HZ is shown for flare energies down to 10$^{28}$ erg (short-dashed), 10$^{30}$ erg (long-dashed), and 10$^{32}$ erg (dot-dashed).
Other colors and line styles are the same as Figure \ref{fig:traverse1}. 
\label{fig:traverse3}}
\end{figure}

Additionally, when we estimate the increases of NUV flux due to flares via Equation \ref{equation:FlareEnergy}, the upper and lower energy limits we choose will affect the total integrated energy due to flares. Since most of $\alpha_{\rm C}<-1$, the flares with low energy will dominantly contribute to the total energy. Thus, we test different choices for the lower energy limit. Figure \ref{fig:traverse3} shows how the UV-HZ varies when a low-energy flare changes from $10^{28}$ to $10^{32}$ erg. The results of G stars are not sensitive to the choices, while for M stars, the choices become important. Even with a higher limit as high as $10^{32}$ erg, the UV-HZ expanded by an average of $\sim$0.05 AU. As the detection of low-energy flares becomes more complete, the UV-HZ expands to the inner region.

    \clearpage
    \begin{singlespace}
        \printbibliography[title={Supplemental References}, heading=bibnumbered]

@string{apj = "The Astrophysical Journal"}

@string{mnras = "Monthly Notices of the Royal Astronomical Society"}

@string{ aj = "The Astronomical Journal"}

@string{icarus = "Icarus"}

@string{apjl = "The Astrophysical Journal Letters"}

@string{apjs = "The Astrophysical Journal Supplement Series"}

@string{pasp = "Publications of the Astronomical Society of the Pacific"}

@string{aap = "Astronomy \& Astrophysics"}

@string{araa = "Annual Review of Astronomy and Astrophysics"}

@string{solphys = "Solar Physics"}

@string{science = "Science"}

@string{ana = "Analytical Chemistry"}

@string{life = "Life"}

@Misc{methods,
  note = {Materials and methods are available as supplementary material},
}

@ARTICLE{2021NatGe..14..832R,
       author = {{Riihel{\"a}}, Aku and {Bright}, Ryan M. and {Anttila}, Kati},
        title = "{Recent strengthening of snow and ice albedo feedback driven by Antarctic sea-ice loss}",
      journal = {Nature Geoscience},
         year = 2021,
        month = oct,
       volume = {14},
       number = {11},
        pages = {832-836},
          doi = {10.1038/s41561-021-00841-x},
       adsurl = {https://ui.adsabs.harvard.edu/abs/2021NatGe..14..832R}
}

@ARTICLE{2016NatCo...710627P,
       author = {{Popp}, Max and {Schmidt}, Hauke and {Marotzke}, Jochem},
        title = "{Transition to a Moist Greenhouse with CO$_{2}$ and solar forcing}",
      journal = {Nature Communications},
         year = 2016,
        month = feb,
       volume = {7},
          eid = {10627},
        pages = {10627},
          doi = {10.1038/ncomms10627},
archivePrefix = {arXiv},
       eprint = {1610.02283},
 primaryClass = {astro-ph.EP},
       adsurl = {https://ui.adsabs.harvard.edu/abs/2016NatCo...710627P}
}

@ARTICLE{2016RPPh...79c6901B,
       author = {{Borucki}, William J.},
        title = "{KEPLER Mission: development and overview}",
      journal = {Reports on Progress in Physics},
         year = 2016,
        month = mar,
       volume = {79},
       number = {3},
          eid = {036901},
        pages = {036901},
          doi = {10.1088/0034-4885/79/3/036901},
       adsurl = {https://ui.adsabs.harvard.edu/abs/2016RPPh...79c6901B}
}

@ARTICLE{2023AJ....165...34H,
       author = {{Hill}, Michelle L. and {Bott}, Kimberly and {Dalba}, Paul A. and {Fetherolf}, Tara and {Kane}, Stephen R. and {Kopparapu}, Ravi and {Li}, Zhexing and {Ostberg}, Colby},
        title = "{A Catalog of Habitable Zone Exoplanets}",
      journal = aj,
         year = 2023,
        month = feb,
       volume = {165},
       number = {2},
          eid = {34},
        pages = {34},
          doi = {10.3847/1538-3881/aca1c0},
       adsurl = {https://ui.adsabs.harvard.edu/abs/2023AJ....165...34H}
}

@ARTICLE{2013ApJ...765..131K,
       author = {{Kopparapu}, Ravi Kumar and {Ramirez}, Ramses and {Kasting}, James F. and {Eymet}, Vincent and {Robinson}, Tyler D. and {Mahadevan}, Suvrath and {Terrien}, Ryan C. and {Domagal-Goldman}, Shawn and {Meadows}, Victoria and {Deshpande}, Rohit},
        title = "{Habitable Zones around Main-sequence Stars: New Estimates}",
      journal = apj,
         year = 2013,
        month = mar,
       volume = {765},
       number = {2},
          eid = {131},
        pages = {131},
          doi = {10.1088/0004-637X/765/2/131},
archivePrefix = {arXiv},
       eprint = {1301.6674},
 primaryClass = {astro-ph.EP},
       adsurl = {https://ui.adsabs.harvard.edu/abs/2013ApJ...765..131K}
}

@ARTICLE{2018ApJ...856..122K,
       author = {{Kopparapu}, Ravi Kumar and {H{\'e}brard}, Eric and {Belikov}, Rus and {Batalha}, Natalie M. and {Mulders}, Gijs D. and {Stark}, Chris and {Teal}, Dillon and {Domagal-Goldman}, Shawn and {Mandell}, Avi},
        title = "{Exoplanet Classification and Yield Estimates for Direct Imaging Missions}",
      journal = apj,
         year = 2018,
        month = apr,
       volume = {856},
       number = {2},
          eid = {122},
        pages = {122},
          doi = {10.3847/1538-4357/aab205},
archivePrefix = {arXiv},
       eprint = {1802.09602},
 primaryClass = {astro-ph.EP},
       adsurl = {https://ui.adsabs.harvard.edu/abs/2018ApJ...856..122K}
}

@ARTICLE{2023MNRAS.522.1411S,
       author = {{Spinelli}, R. and {Borsa}, F. and {Ghirlanda}, G. and {Ghisellini}, G. and {Haardt}, F.},
        title = "{The ultraviolet habitable zone of exoplanets}",
      journal = mnras,
         year = 2023,
        month = jun,
       volume = {522},
       number = {1},
        pages = {1411-1418},
          doi = {10.1093/mnras/stad928},
archivePrefix = {arXiv},
       eprint = {2303.16229},
 primaryClass = {astro-ph.EP},
       adsurl = {https://ui.adsabs.harvard.edu/abs/2023MNRAS.522.1411S}
}

@ARTICLE{2017ApJ...843..110R,
       author = {{Ranjan}, Sukrit and {Wordsworth}, Robin and {Sasselov}, Dimitar D.},
        title = "{The Surface UV Environment on Planets Orbiting M Dwarfs: Implications for Prebiotic Chemistry and the Need for Experimental Follow-up}",
      journal = apj,
         year = 2017,
        month = jul,
       volume = {843},
       number = {2},
          eid = {110},
        pages = {110},
          doi = {10.3847/1538-4357/aa773e},
archivePrefix = {arXiv},
       eprint = {1705.02350},
 primaryClass = {astro-ph.EP},
       adsurl = {https://ui.adsabs.harvard.edu/abs/2017ApJ...843..110R}
}

@ARTICLE{1959PASP...71..421H,
       author = {{Huang}, Su-Shu},
        title = "{The Problem of Life in the Universe and the Mode of Star Formation}",
      journal = pasp,
         year = 1959,
        month = oct,
       volume = {71},
       number = {422},
        pages = {421},
          doi = {10.1086/127417},
       adsurl = {https://ui.adsabs.harvard.edu/abs/1959PASP...71..421H}
}

@ARTICLE{2022AJ....164..213G,
       author = {{Gao}, Dong-Yang and {Liu}, Hui-Gen and {Yang}, Ming and {Zhou}, Ji-Lin},
        title = "{Correcting Stellar Flare Frequency Distributions Detected by TESS and Kepler}",
      journal = aj,
         year = 2022,
        month = nov,
       volume = {164},
       number = {5},
          eid = {213},
        pages = {213},
          doi = {10.3847/1538-3881/ac937e},
archivePrefix = {arXiv},
       eprint = {2301.07552},
 primaryClass = {astro-ph.SR},
       adsurl = {https://ui.adsabs.harvard.edu/abs/2022AJ....164..213G}
}

@Article{C8CC01499J,
author ="Xu, Jianfeng and Ritson, Dougal J. and Ranjan, Sukrit and Todd, Zoe R. and Sasselov, Dimitar D. and Sutherland, John D.",
title  ="Photochemical reductive homologation of hydrogen cyanide using sulfite and ferrocyanide",
journal  ="Chem. Commun.",
year  ="2018",
volume  ="54",
issue  ="44",
pages  ="5566-5569",
publisher  ="The Royal Society of Chemistry",
doi  ="10.1039/C8CC01499J",
url  ="http://dx.doi.org/10.1039/C8CC01499J"}

@ARTICLE{2017ApJ...850..204W,
       author = {{Watanabe}, Kyoko and {Kitagawa}, Jun and {Masuda}, Satoshi},
        title = "{Characteristics that Produce White-light Enhancements in Solar Flares Observed by Hinode/SOT}",
      journal = apj,
         year = 2017,
        month = dec,
       volume = {850},
       number = {2},
          eid = {204},
        pages = {204},
          doi = {10.3847/1538-4357/aa9659},
archivePrefix = {arXiv},
       eprint = {1710.09531},
 primaryClass = {astro-ph.SR},
       adsurl = {https://ui.adsabs.harvard.edu/abs/2017ApJ...850..204W}
}

@ARTICLE{2020ApJ...891...88W,
       author = {{Watanabe}, Kyoko and {Imada}, Shinsuke},
        title = "{White-light Emission and Chromospheric Response by an X1.8-class Flare on 2012 October 23}",
      journal = apj,
         year = 2020,
        month = mar,
       volume = {891},
       number = {1},
          eid = {88},
        pages = {88},
          doi = {10.3847/1538-4357/ab711b},
       adsurl = {https://ui.adsabs.harvard.edu/abs/2020ApJ...891...88W}
}

@ARTICLE{2014ApJ...787L..29K,
       author = {{Kopparapu}, Ravi Kumar and {Ramirez}, Ramses M. and {SchottelKotte}, James and {Kasting}, James F. and {Domagal-Goldman}, Shawn and {Eymet}, Vincent},
        title = "{Habitable Zones around Main-sequence Stars: Dependence on Planetary Mass}",
      journal = apjl,
         year = 2014,
        month = jun,
       volume = {787},
       number = {2},
          eid = {L29},
        pages = {L29},
          doi = {10.1088/2041-8205/787/2/L29},
archivePrefix = {arXiv},
       eprint = {1404.5292},
 primaryClass = {astro-ph.EP},
       adsurl = {https://ui.adsabs.harvard.edu/abs/2014ApJ...787L..29K}
}

@ARTICLE{2022FrASS...934458K,
       author = {{Kowalski}, Adam F.},
        title = "{Near-Ultraviolet Continuum Modeling of the 1985 April 12 Great Flare of AD Leo}",
      journal = {Frontiers in Astronomy and Space Sciences},
         year = 2022,
        month = nov,
       volume = {9},
          eid = {351},
        pages = {351},
          doi = {10.3389/fspas.2022.1034458},
archivePrefix = {arXiv},
       eprint = {2210.16980},
 primaryClass = {astro-ph.SR},
       adsurl = {https://ui.adsabs.harvard.edu/abs/2022FrASS...934458K}
}

@ARTICLE{2023ApJ...943L..23K,
       author = {{Kowalski}, Adam F.},
        title = "{Bridging High-density Electron Beam Coronal Transport and Deep Chromospheric Heating in Stellar Flares}",
      journal = apjl,
         year = 2023,
        month = feb,
       volume = {943},
       number = {2},
          eid = {L23},
        pages = {L23},
          doi = {10.3847/2041-8213/acb144},
archivePrefix = {arXiv},
       eprint = {2301.03477},
 primaryClass = {astro-ph.SR},
       adsurl = {https://ui.adsabs.harvard.edu/abs/2023ApJ...943L..23K}
}

@ARTICLE{2016ApJ...816...88K,
       author = {{Kleint}, Lucia and {Heinzel}, Petr and {Judge}, Phil and {Krucker}, S{\"a}m},
        title = "{Continuum Enhancements in the Ultraviolet, the Visible and the Infrared during the X1 Flare on 2014 March 29}",
      journal = apj,
         year = 2016,
        month = jan,
       volume = {816},
       number = {2},
          eid = {88},
        pages = {88},
          doi = {10.3847/0004-637X/816/2/88},
archivePrefix = {arXiv},
       eprint = {1511.04161},
 primaryClass = {astro-ph.SR},
       adsurl = {https://ui.adsabs.harvard.edu/abs/2016ApJ...816...88K}
}

@ARTICLE{2023ApJ...944....5B,
       author = {{Brasseur}, C.~E. and {Osten}, Rachel A. and {Tristan}, Isaiah I. and {Kowalski}, Adam F.},
        title = "{Constraints on Stellar Flare Energy Ratios in the NUV and Optical from a Multiwavelength Study of GALEX and Kepler Flare Stars}",
      journal = apj,
         year = 2023,
        month = feb,
       volume = {944},
       number = {1},
          eid = {5},
        pages = {5},
          doi = {10.3847/1538-4357/acab59},
archivePrefix = {arXiv},
       eprint = {2212.08696},
 primaryClass = {astro-ph.SR},
       adsurl = {https://ui.adsabs.harvard.edu/abs/2023ApJ...944....5B}
}

@ARTICLE{2023MNRAS.519.3564J,
       author = {{Jackman}, James A.~G. and {Shkolnik}, Evgenya L. and {Million}, Chase and {Fleming}, Scott and {Richey-Yowell}, Tyler and {Loyd}, R.~O. Parke},
        title = "{Extending optical flare models to the UV: results from comparing of TESS and GALEX flare observations for M Dwarfs}",
      journal = mnras,
         year = 2023,
        month = mar,
       volume = {519},
       number = {3},
        pages = {3564-3583},
          doi = {10.1093/mnras/stac3135},
archivePrefix = {arXiv},
       eprint = {2210.15688},
 primaryClass = {astro-ph.SR},
       adsurl = {https://ui.adsabs.harvard.edu/abs/2023MNRAS.519.3564J}
}

@ARTICLE{2020ApJ...902..115H,
       author = {{Howard}, Ward S. and {Corbett}, Hank and {Law}, Nicholas M. and {Ratzloff}, Jeffrey K. and {Galliher}, Nathan and {Glazier}, Amy L. and {Gonzalez}, Ramses and {Vasquez Soto}, Alan and {Fors}, Octavi and {del Ser}, Daniel and {Haislip}, Joshua},
        title = "{EvryFlare. III. Temperature Evolution and Habitability Impacts of Dozens of Superflares Observed Simultaneously by Evryscope and TESS}",
      journal = apj,
         year = 2020,
        month = oct,
       volume = {902},
       number = {2},
          eid = {115},
        pages = {115},
          doi = {10.3847/1538-4357/abb5b4},
archivePrefix = {arXiv},
       eprint = {2010.00604},
 primaryClass = {astro-ph.SR},
       adsurl = {https://ui.adsabs.harvard.edu/abs/2020ApJ...902..115H}
}

@ARTICLE{1999Icar..141..399C,
       author = {{Cockell}, Charles S.},
        title = "{Carbon Biochemistry and the Ultraviolet Radiation Environments of F, G, and K Main Sequence Stars}",
      journal = icarus,
         year = 1999,
        month = oct,
       volume = {141},
       number = {2},
        pages = {399-407},
          doi = {10.1006/icar.1999.6167},
       adsurl = {https://ui.adsabs.harvard.edu/abs/1999Icar..141..399C}
}

@ARTICLE{2024ApJ...966...69L,
       author = {{Li}, Xue and {Wang}, Song and {Han}, Henggeng and {Yang}, Huiqin and {Zheng}, Chuanjie and {Huang}, Yang and {Liu}, Jifeng},
        title = "{Ultraviolet and Chromospheric Activity and Habitability of M Stars}",
      journal = apj,
         year = 2024,
        month = may,
       volume = {966},
       number = {1},
          eid = {69},
        pages = {69},
          doi = {10.3847/1538-4357/ad3038},
archivePrefix = {arXiv},
       eprint = {2402.17384},
 primaryClass = {astro-ph.SR},
       adsurl = {https://ui.adsabs.harvard.edu/abs/2024ApJ...966...69L}
}

@ARTICLE{2011A&A...530A..84K,
       author = {{Kretzschmar}, M.},
        title = "{The Sun as a star: observations of white-light flares}",
      journal = aap,
         year = 2011,
        month = jun,
       volume = {530},
          eid = {A84},
        pages = {A84},
          doi = {10.1051/0004-6361/201015930},
archivePrefix = {arXiv},
       eprint = {1103.3125},
 primaryClass = {astro-ph.SR},
       adsurl = {https://ui.adsabs.harvard.edu/abs/2011A&A...530A..84K}
}

@ARTICLE{2018ApJ...867...70L,
       author = {{Loyd}, R.~O. Parke and {Shkolnik}, Evgenya L. and {Schneider}, Adam C. and {Barman}, Travis S. and {Meadows}, Victoria S. and {Pagano}, Isabella and {Peacock}, Sarah},
        title = "{HAZMAT. IV. Flares and Superflares on Young M Stars in the Far Ultraviolet}",
      journal = apj,
         year = 2018,
        month = nov,
       volume = {867},
       number = {1},
          eid = {70},
        pages = {70},
          doi = {10.3847/1538-4357/aae2ae},
archivePrefix = {arXiv},
       eprint = {1810.03277},
 primaryClass = {astro-ph.SR},
       adsurl = {https://ui.adsabs.harvard.edu/abs/2018ApJ...867...70L}
}

@ARTICLE{2019ApJS..241...29Y,
       author = {{Yang}, Huiqin and {Liu}, Jifeng},
        title = "{The Flare Catalog and the Flare Activity in the Kepler Mission}",
      journal = apjs,
         year = 2019,
        month = apr,
       volume = {241},
       number = {2},
          eid = {29},
        pages = {29},
          doi = {10.3847/1538-4365/ab0d28},
archivePrefix = {arXiv},
       eprint = {1903.01056},
 primaryClass = {astro-ph.SR},
       adsurl = {https://ui.adsabs.harvard.edu/abs/2019ApJS..241...29Y}
}

@ARTICLE{1989SoPh..121..299P,
       author = {{Pettersen}, B.~R.},
        title = "{A Review of Stellar Flares and Their Characteristics}",
      journal = solphys,
         year = 1989,
        month = mar,
       volume = {121},
       number = {1-2},
        pages = {299-312},
          doi = {10.1007/BF00161702},
       adsurl = {https://ui.adsabs.harvard.edu/abs/1989SoPh..121..299P}
}

@ARTICLE{2010ARA&A..48..241B,
       author = {{Benz}, Arnold O. and {G{\"u}del}, Manuel},
        title = "{Physical Processes in Magnetically Driven Flares on the Sun, Stars, and Young Stellar Objects}",
      journal = araa,
         year = 2010,
        month = sep,
       volume = {48},
        pages = {241-287},
          doi = {10.1146/annurev-astro-082708-101757},
       adsurl = {https://ui.adsabs.harvard.edu/abs/2010ARA&A..48..241B}
}

@INPROCEEDINGS{2016ASSL..427..373S,
       author = {{Shibata}, K. and {Takasao}, S.},
        title = "{Fractal Reconnection in Solar and Stellar Environments}",
    booktitle = {Magnetic Reconnection: Concepts and Applications},
         year = 2016,
       editor = {{Gonzalez}, Walter and {Parker}, Eugene},
       series = {Astrophysics and Space Science Library},
       volume = {427},
        month = jan,
        pages = {373},
          doi = {10.1007/978-3-319-26432-5\_10},
archivePrefix = {arXiv},
       eprint = {1606.09401},
 primaryClass = {astro-ph.SR},
       adsurl = {https://ui.adsabs.harvard.edu/abs/2016ASSL..427..373S}
}

@ARTICLE{2006Icar..183..491B,
       author = {{Buccino}, Andrea P. and {Lemarchand}, Guillermo A. and {Mauas}, Pablo J.~D.},
        title = "{Ultraviolet radiation constraints around the circumstellar habitable zones}",
      journal = icarus,
         year = 2006,
        month = aug,
       volume = {183},
       number = {2},
        pages = {491-503},
          doi = {10.1016/j.icarus.2006.03.007},
archivePrefix = {arXiv},
       eprint = {astro-ph/0512291},
 primaryClass = {astro-ph},
       adsurl = {https://ui.adsabs.harvard.edu/abs/2006Icar..183..491B}
}

@ARTICLE{1900C&T....20..389A,
       author = {{Arrhenius}, S.},
        title = "{Les oscillations s{\'e}culaires de la temp{\'e}rature {\'a} la surface du glove terrestre}",
      journal = {Ciel et Terre},
         year = 1900,
        month = jan,
       volume = {20},
        pages = {389-396},
       adsurl = {https://ui.adsabs.harvard.edu/abs/1900C&T....20..389A}
}

@ARTICLE{2020MNRAS.494L..69A,
       author = {{Abrevaya}, X.~C. and {Leitzinger}, M. and {Oppezzo}, O.~J. and {Odert}, P. and {Patel}, M.~R. and {Luna}, G.~J.~M. and {Forte Giacobone}, A.~F. and {Hanslmeier}, A.},
        title = "{The UV surface habitability of Proxima b: first experiments revealing probable life survival to stellar flares}",
      journal = mnras,
         year = 2020,
        month = may,
       volume = {494},
       number = {1},
        pages = {L69-L74},
          doi = {10.1093/mnrasl/slaa037},
archivePrefix = {arXiv},
       eprint = {2003.00984},
 primaryClass = {astro-ph.SR},
       adsurl = {https://ui.adsabs.harvard.edu/abs/2020MNRAS.494L..69A}
}

@ARTICLE{2020AJ....159...60G,
       author = {{G{\"u}nther}, Maximilian N. and {Zhan}, Zhuchang and {Seager}, Sara and {Rimmer}, Paul B. and {Ranjan}, Sukrit and {Stassun}, Keivan G. and {Oelkers}, Ryan J. and {Daylan}, Tansu and {Newton}, Elisabeth and {Kristiansen}, Martti H. and {Olah}, Katalin and {Gillen}, Edward and {Rappaport}, Saul and {Ricker}, George R. and {Vanderspek}, Roland K. and {Latham}, David W. and {Winn}, Joshua N. and {Jenkins}, Jon M. and {Glidden}, Ana and {Fausnaugh}, Michael and {Levine}, Alan M. and {Dittmann}, Jason A. and {Quinn}, Samuel N. and {Krishnamurthy}, Akshata and {Ting}, Eric B.},
        title = "{Stellar Flares from the First TESS Data Release: Exploring a New Sample of M Dwarfs}",
      journal = aj,
         year = 2020,
        month = feb,
       volume = {159},
       number = {2},
          eid = {60},
        pages = {60},
          doi = {10.3847/1538-3881/ab5d3a},
archivePrefix = {arXiv},
       eprint = {1901.00443},
 primaryClass = {astro-ph.EP},
       adsurl = {https://ui.adsabs.harvard.edu/abs/2020AJ....159...60G}
}

@ARTICLE{2025ApJS..281...13L,
       author = {{Li}, Xue and {Wang}, Song and {Ma}, Jun and {Han}, Henggeng and {Huang}, Yang and {Liu}, Jifeng},
        title = "{Evolution of Stellar Activity and Habitable Zone. I. Ultraviolet Emission of Dwarfs in Open Clusters and Field Stars}",
      journal = apjs,
         year = 2025,
        month = nov,
       volume = {281},
       number = {1},
          eid = {13},
        pages = {13},
          doi = {10.3847/1538-4365/ae08a9},
archivePrefix = {arXiv},
       eprint = {2509.18559},
 primaryClass = {astro-ph.SR},
       adsurl = {https://ui.adsabs.harvard.edu/abs/2025ApJS..281...13L}
}

@ARTICLE{2018ApJ...858...55P,
       author = {{Paudel}, Rishi R. and {Gizis}, John E. and {Mullan}, D.~J. and {Schmidt}, Sarah J. and {Burgasser}, Adam J. and {Williams}, Peter K.~G. and {Berger}, Edo},
        title = "{K2 Ultracool Dwarfs Survey. III. White Light Flares Are Ubiquitous in M6-L0 Dwarfs}",
      journal = apj,
         year = 2018,
        month = may,
       volume = {858},
       number = {1},
          eid = {55},
        pages = {55},
          doi = {10.3847/1538-4357/aab8fe},
archivePrefix = {arXiv},
       eprint = {1803.07708},
 primaryClass = {astro-ph.SR},
       adsurl = {https://ui.adsabs.harvard.edu/abs/2018ApJ...858...55P}
}

@ARTICLE{2020ApJ...890...46T,
       author = {{Tu}, Zuo-Lin and {Yang}, Ming and {Zhang}, Z.~J. and {Wang}, F.~Y.},
        title = "{Superflares on Solar-type Stars from the First Year Observation of TESS}",
      journal = apj,
         year = 2020,
        month = feb,
       volume = {890},
       number = {1},
          eid = {46},
        pages = {46},
          doi = {10.3847/1538-4357/ab6606},
archivePrefix = {arXiv},
       eprint = {1912.11572},
 primaryClass = {astro-ph.SR},
       adsurl = {https://ui.adsabs.harvard.edu/abs/2020ApJ...890...46T}
}

@ARTICLE{2008ApJ...687.1264M,
       author = {{Mamajek}, Eric E. and {Hillenbrand}, Lynne A.},
        title = "{Improved Age Estimation for Solar-Type Dwarfs Using Activity-Rotation Diagnostics}",
      journal = apj,
         year = 2008,
        month = nov,
       volume = {687},
       number = {2},
        pages = {1264-1293},
          doi = {10.1086/591785},
archivePrefix = {arXiv},
       eprint = {0807.1686},
 primaryClass = {astro-ph},
       adsurl = {https://ui.adsabs.harvard.edu/abs/2008ApJ...687.1264M}
}

@ARTICLE{2021A&A...645A..42I,
       author = {{Ilin}, Ekaterina and {Schmidt}, Sarah J. and {Poppenh{\"a}ger}, Katja and {Davenport}, James R.~A. and {Kristiansen}, Martti H. and {Omohundro}, Mark},
        title = "{Flares in open clusters with K2. II. Pleiades, Hyades, Praesepe, Ruprecht 147, and M 67}",
      journal = aap,
         year = 2021,
        month = jan,
       volume = {645},
          eid = {A42},
        pages = {A42},
          doi = {10.1051/0004-6361/202039198},
archivePrefix = {arXiv},
       eprint = {2010.05576},
 primaryClass = {astro-ph.SR},
       adsurl = {https://ui.adsabs.harvard.edu/abs/2021A&A...645A..42I}
}

@ARTICLE{2014ApJ...797..121H,
       author = {{Hawley}, Suzanne L. and {Davenport}, James R.~A. and {Kowalski}, Adam F. and {Wisniewski}, John P. and {Hebb}, Leslie and {Deitrick}, Russell and {Hilton}, Eric J.},
        title = "{Kepler Flares. I. Active and Inactive M Dwarfs}",
      journal = apj,
         year = 2014,
        month = dec,
       volume = {797},
       number = {2},
          eid = {121},
        pages = {121},
          doi = {10.1088/0004-637X/797/2/121},
archivePrefix = {arXiv},
       eprint = {1410.7779},
 primaryClass = {astro-ph.SR},
       adsurl = {https://ui.adsabs.harvard.edu/abs/2014ApJ...797..121H}
}

@ARTICLE{2021MNRAS.504.3246J,
       author = {{Jackman}, James A.~G. and {Wheatley}, Peter J. and {Acton}, Jack S. and {Anderson}, David R. and {Bayliss}, Daniel and {Briegal}, Joshua T. and {Burleigh}, Matthew R. and {Casewell}, Sarah L. and {G{\"a}nsicke}, Boris T. and {Gill}, Samuel and {Gillen}, Edward and {Goad}, Michael R. and {G{\"u}nther}, Maximilian N. and {Henderson}, Beth A. and {Hodgkin}, Simon T. and {Jenkins}, James S. and {Pugh}, Chloe and {Queloz}, Didier and {Raynard}, Liam and {Tilbrook}, Rosanna H. and {Watson}, Christopher A. and {West}, Richard G.},
        title = "{Stellar flares detected with the Next Generation Transit Survey}",
      journal = mnras,
         year = 2021,
        month = jul,
       volume = {504},
       number = {3},
        pages = {3246-3264},
          doi = {10.1093/mnras/stab979},
archivePrefix = {arXiv},
       eprint = {2104.02648},
 primaryClass = {astro-ph.SR},
       adsurl = {https://ui.adsabs.harvard.edu/abs/2021MNRAS.504.3246J}
}

@ARTICLE{2018SciA....4.3302R,
       author = {{Rimmer}, Paul B. and {Xu}, Jianfeng and {Thompson}, Samantha J. and {Gillen}, Ed and {Sutherland}, John D. and {Queloz}, Didier},
        title = "{The origin of RNA precursors on exoplanets}",
      journal = {Science Advances},
         year = 2018,
        month = aug,
       volume = {4},
       number = {8},
        pages = {eaar3302},
          doi = {10.1126/sciadv.aar3302},
archivePrefix = {arXiv},
       eprint = {1808.02718},
 primaryClass = {astro-ph.EP},
       adsurl = {https://ui.adsabs.harvard.edu/abs/2018SciA....4.3302R}
}

@ARTICLE{2018ascl.soft12013L,
   author = {{Lightkurve Collaboration} and {Cardoso}, J.~V.~d.~M. and
             {Hedges}, C. and {Gully-Santiago}, M. and {Saunders}, N. and
             {Cody}, A.~M. and {Barclay}, T. and {Hall}, O. and
             {Sagear}, S. and {Turtelboom}, E. and {Zhang}, J. and
             {Tzanidakis}, A. and {Mighell}, K. and {Coughlin}, J. and
             {Bell}, K. and {Berta-Thompson}, Z. and {Williams}, P. and
             {Dotson}, J. and {Barentsen}, G.},
    title = "{Lightkurve: Kepler and TESS time series analysis in Python}",
journal = {Astrophysics Source Code Library, record ascl:1812.013},
     year = 2018,
    month = dec,
archivePrefix = "ascl",
   eprint = {1812.013},
   adsurl = {http://adsabs.harvard.edu/abs/2018ascl.soft12013L},
}

@ARTICLE{2024MNRAS.532.4436B,
       author = {{Berger}, Vera L. and {Hinkle}, Jason T. and {Tucker}, Michael A. and {Shappee}, Benjamin J. and {van Saders}, Jennifer L. and {Huber}, Daniel and {Reep}, Jeffrey W. and {Sun}, Xudong and {Yang}, Kai E.},
        title = "{Stellar flares are far-ultraviolet luminous}",
      journal = mnras,
         year = 2024,
        month = aug,
       volume = {532},
       number = {4},
        pages = {4436-4445},
          doi = {10.1093/mnras/stae1648},
archivePrefix = {arXiv},
       eprint = {2312.12511},
 primaryClass = {astro-ph.SR},
       adsurl = {https://ui.adsabs.harvard.edu/abs/2024MNRAS.532.4436B}
}

@ARTICLE{2021ApJ...906...72O,
       author = {{Okamoto}, Soshi and {Notsu}, Yuta and {Maehara}, Hiroyuki and {Namekata}, Kosuke and {Honda}, Satoshi and {Ikuta}, Kai and {Nogami}, Daisaku and {Shibata}, Kazunari},
        title = "{Statistical Properties of Superflares on Solar-type Stars: Results Using All of the Kepler Primary Mission Data}",
      journal = apj,
         year = 2021,
        month = jan,
       volume = {906},
       number = {2},
          eid = {72},
        pages = {72},
          doi = {10.3847/1538-4357/abc8f5},
archivePrefix = {arXiv},
       eprint = {2011.02117},
 primaryClass = {astro-ph.SR},
       adsurl = {https://ui.adsabs.harvard.edu/abs/2021ApJ...906...72O}
}

@ARTICLE{2019ApJ...871..167K,
       author = {{Kowalski}, Adam F. and {Wisniewski}, John P. and {Hawley}, Suzanne L. and {Osten}, Rachel A. and {Brown}, Alexander and {Fari{\~n}a}, Cecilia and {Valenti}, Jeff A. and {Brown}, Stephen and {Xilouris}, Manolis and {Schmidt}, Sarah J. and {Johns-Krull}, Christopher},
        title = "{The Near-ultraviolet Continuum Radiation in the Impulsive Phase of HF/GF-type dMe Flares. I. Data}",
      journal = apj,
         year = 2019,
        month = feb,
       volume = {871},
       number = {2},
          eid = {167},
        pages = {167},
          doi = {10.3847/1538-4357/aaf058},
archivePrefix = {arXiv},
       eprint = {1811.04021},
 primaryClass = {astro-ph.SR},
       adsurl = {https://ui.adsabs.harvard.edu/abs/2019ApJ...871..167K}
}

@ARTICLE{2016ApJ...820...89F,
       author = {{France}, Kevin and {Loyd}, R.~O. Parke and {Youngblood}, Allison and {Brown}, Alexander and {Schneider}, P. Christian and {Hawley}, Suzanne L. and {Froning}, Cynthia S. and {Linsky}, Jeffrey L. and {Roberge}, Aki and {Buccino}, Andrea P. and {Davenport}, James R.~A. and {Fontenla}, Juan M. and {Kaltenegger}, Lisa and {Kowalski}, Adam F. and {Mauas}, Pablo J.~D. and {Miguel}, Yamila and {Redfield}, Seth and {Rugheimer}, Sarah and {Tian}, Feng and {Vieytes}, Mariela C. and {Walkowicz}, Lucianne M. and {Weisenburger}, Kolby L.},
        title = "{The MUSCLES Treasury Survey. I. Motivation and Overview}",
      journal = apj,
         year = 2016,
        month = apr,
       volume = {820},
       number = {2},
          eid = {89},
        pages = {89},
          doi = {10.3847/0004-637X/820/2/89},
archivePrefix = {arXiv},
       eprint = {1602.09142},
 primaryClass = {astro-ph.SR},
       adsurl = {https://ui.adsabs.harvard.edu/abs/2016ApJ...820...89F}
}

@ARTICLE{2022ApJ...929..169R,
       author = {{Richey-Yowell}, Tyler and {Shkolnik}, Evgenya L. and {Loyd}, R.~O. Parke and {Jackman}, James A.~G. and {Schneider}, Adam C. and {Ag{\"u}eros}, Marcel A. and {Barman}, Travis and {Meadows}, Victoria S. and {Gibson}, Rose and {Douglas}, Stephanie T.},
        title = "{HAZMAT. VIII. A Spectroscopic Analysis of the Ultraviolet Evolution of K Stars: Additional Evidence for K Dwarf Rotational Stalling in the First Gigayear}",
      journal = apj,
         year = 2022,
        month = apr,
       volume = {929},
       number = {2},
          eid = {169},
        pages = {169},
          doi = {10.3847/1538-4357/ac5f48},
archivePrefix = {arXiv},
       eprint = {2203.15237},
 primaryClass = {astro-ph.SR},
       adsurl = {https://ui.adsabs.harvard.edu/abs/2022ApJ...929..169R}
}

@ARTICLE{2017ApJ...843...31Y,
       author = {{Youngblood}, Allison and {France}, Kevin and {Loyd}, R.~O. Parke and {Brown}, Alexander and {Mason}, James P. and {Schneider}, P. Christian and {Tilley}, Matt A. and {Berta-Thompson}, Zachory K. and {Buccino}, Andrea and {Froning}, Cynthia S. and {Hawley}, Suzanne L. and {Linsky}, Jeffrey and {Mauas}, Pablo J.~D. and {Redfield}, Seth and {Kowalski}, Adam and {Miguel}, Yamila and {Newton}, Elisabeth R. and {Rugheimer}, Sarah and {Segura}, Ant{\'\i}gona and {Roberge}, Aki and {Vieytes}, Mariela},
        title = "{The MUSCLES Treasury Survey. IV. Scaling Relations for Ultraviolet, Ca II K, and Energetic Particle Fluxes from M Dwarfs}",
      journal = apj,
         year = 2017,
        month = jul,
       volume = {843},
       number = {1},
          eid = {31},
        pages = {31},
          doi = {10.3847/1538-4357/aa76dd},
       adsurl = {https://ui.adsabs.harvard.edu/abs/2017ApJ...843...31Y}
}

@ARTICLE{2015ApJ...809...79O,
       author = {{Osten}, Rachel A. and {Wolk}, Scott J.},
        title = "{Connecting Flares and Transient Mass-loss Events in Magnetically Active Stars}",
      journal = apj,
         year = 2015,
        month = aug,
       volume = {809},
       number = {1},
          eid = {79},
        pages = {79},
          doi = {10.1088/0004-637X/809/1/79},
       adsurl = {https://ui.adsabs.harvard.edu/abs/2015ApJ...809...79O}
}

@ARTICLE{2017ApJ...837..125K,
       author = {{Kowalski}, Adam F. and {Allred}, Joel C. and {Uitenbroek}, Han and {Tremblay}, Pier-Emmanuel and {Brown}, Stephen and {Carlsson}, Mats and {Osten}, Rachel A. and {Wisniewski}, John P. and {Hawley}, Suzanne L.},
        title = "{Hydrogen Balmer Line Broadening in Solar and Stellar Flares}",
      journal = apj,
         year = 2017,
        month = mar,
       volume = {837},
       number = {2},
          eid = {125},
        pages = {125},
          doi = {10.3847/1538-4357/aa603e},
       adsurl = {https://ui.adsabs.harvard.edu/abs/2017ApJ...837..125K}
}

@ARTICLE{2010AsBio..10..751S,
       author = {{Segura}, Ant{\'\i}gona and {Walkowicz}, Lucianne M. and {Meadows}, Victoria and {Kasting}, James and {Hawley}, Suzanne},
        title = "{The Effect of a Strong Stellar Flare on the Atmospheric Chemistry of an Earth-like Planet Orbiting an M Dwarf}",
      journal = {Astrobiology},
         year = 2010,
        month = sep,
       volume = {10},
       number = {7},
        pages = {751-771},
          doi = {10.1089/ast.2009.0376},
       adsurl = {https://ui.adsabs.harvard.edu/abs/2010AsBio..10..751S}
}

@ARTICLE{2011LRSP....8....6S,
       author = {{Shibata}, Kazunari and {Magara}, Tetsuya},
        title = "{Solar Flares: Magnetohydrodynamic Processes}",
      journal = {Living Reviews in Solar Physics},
         year = 2011,
        month = dec,
       volume = {8},
       number = {1},
          eid = {6},
        pages = {6},
          doi = {10.12942/lrsp-2011-6},
       adsurl = {https://ui.adsabs.harvard.edu/abs/2011LRSP....8....6S}
}

@ARTICLE{2023A&A...674A..28F,
       author = {{Fouesneau}, M. and {Fr{\'e}mat}, Y. and {Andrae}, R. and {Korn}, A.~J. and {Soubiran}, C. and {Kordopatis}, G. and {Vallenari}, A. and {Heiter}, U. and {Creevey}, O.~L. and {Sarro}, L.~M. and {de Laverny}, P. and {Lanzafame}, A.~C. and {Lobel}, A. and {Sordo}, R. and {Rybizki}, J. and {Slezak}, I. and {{\'A}lvarez}, M.~A. and {Drimmel}, R. and {Garabato}, D. and {Delchambre}, L. and {Bailer-Jones}, C.~A.~L. and {Hatzidimitriou}, D. and {Lorca}, A. and {Le Fustec}, Y. and {Pailler}, F. and {Mary}, N. and {Robin}, C. and {Utrilla}, E. and {Abreu Aramburu}, A. and {Bakker}, J. and {Bellas-Velidis}, I. and {Bijaoui}, A. and {Blomme}, R. and {Bouret}, J. -C. and {Brouillet}, N. and {Brugaletta}, E. and {Burlacu}, A. and {Carballo}, R. and {Casamiquela}, L. and {Chaoul}, L. and {Chiavassa}, A. and {Contursi}, G. and {Cooper}, W.~J. and {Dafonte}, C. and {Demouchy}, C. and {Dharmawardena}, T.~E. and {Garc{\'\i}a-Lario}, P. and {Garc{\'\i}a-Torres}, M. and {Gomez}, A. and {Gonz{\'a}lez-Santamar{\'\i}a}, I. and {Jean-Antoine Piccolo}, A. and {Kontizas}, M. and {Lebreton}, Y. and {Licata}, E.~L. and {Lindstr{\o}m}, H.~E.~P. and {Livanou}, E. and {Magdaleno Romeo}, A. and {Manteiga}, M. and {Marocco}, F. and {Martayan}, C. and {Marshall}, D.~J. and {Nicolas}, C. and {Ordenovic}, C. and {Palicio}, P.~A. and {Pallas-Quintela}, L. and {Pichon}, B. and {Poggio}, E. and {Recio-Blanco}, A. and {Riclet}, F. and {Santove{\~n}a}, R. and {Schultheis}, M.~S. and {Segol}, M. and {Silvelo}, A. and {Smart}, R.~L. and {S{\"u}veges}, M. and {Th{\'e}venin}, F. and {Torralba Elipe}, G. and {Ulla}, A. and {van Dillen}, E. and {Zhao}, H. and {Zorec}, J.},
        title = "{Gaia Data Release 3. Apsis. II. Stellar parameters}",
      journal = aap,
         year = 2023,
        month = jun,
       volume = {674},
          eid = {A28},
        pages = {A28},
          doi = {10.1051/0004-6361/202243919},
archivePrefix = {arXiv},
       eprint = {2206.05992},
 primaryClass = {astro-ph.SR},
       adsurl = {https://ui.adsabs.harvard.edu/abs/2023A&A...674A..28F}
}

@ARTICLE{2007AsBio...7..167K,
       author = {{Khodachenko}, Maxim L. and {Ribas}, Ignasi and {Lammer}, Helmut and {Grie{\ss}meier}, Jean-Mathias and {Leitner}, Martin and {Selsis}, Franck and {Eiroa}, Carlos and {Hanslmeier}, Arnold and {Biernat}, Helfried K. and {Farrugia}, Charles J. and {Rucker}, Helmut O.},
        title = "{Coronal Mass Ejection (CME) Activity of Low Mass M Stars as An Important Factor for The Habitability of Terrestrial Exoplanets. I. CME Impact on Expected Magnetospheres of Earth-Like Exoplanets in Close-In Habitable Zones}",
      journal = {Astrobiology},
         year = 2007,
        month = feb,
       volume = {7},
       number = {1},
        pages = {167-184},
          doi = {10.1089/ast.2006.0127},
       adsurl = {https://ui.adsabs.harvard.edu/abs/2007AsBio...7..167K}
}

@ARTICLE{2010Ap&SS.325...25G,
       author = {{Guo}, Jianpo and {Zhang}, Fenghui and {Zhang}, Xianfei and {Han}, Zhanwen},
        title = "{Habitable zones and UV habitable zones around host stars}",
      journal = apss,
         year = 2010,
        month = jan,
       volume = {325},
       number = {1},
        pages = {25-30},
          doi = {10.1007/s10509-009-0173-9},
archivePrefix = {arXiv},
       eprint = {1003.1222},
 primaryClass = {astro-ph.SR},
       adsurl = {https://ui.adsabs.harvard.edu/abs/2010Ap&SS.325...25G}
}

@ARTICLE{2023ApJ...948L..26H,
       author = {{Hall}, C. and {Stancil}, P.~C. and {Terry}, J.~P. and {Ellison}, C.~K.},
        title = "{A New Definition of Exoplanet Habitability: Introducing the Photosynthetic Habitable Zone}",
      journal = apjl,
         year = 2023,
        month = may,
       volume = {948},
       number = {2},
          eid = {L26},
        pages = {L26},
          doi = {10.3847/2041-8213/acccfb},
       adsurl = {https://ui.adsabs.harvard.edu/abs/2023ApJ...948L..26H}
}

@ARTICLE{2015A&A...577A..42B,
       author = {{Baraffe}, Isabelle and {Homeier}, Derek and {Allard}, France and {Chabrier}, Gilles},
        title = "{New evolutionary models for pre-main sequence and main sequence low-mass stars down to the hydrogen-burning limit}",
      journal = aap,
         year = 2015,
        month = may,
       volume = {577},
          eid = {A42},
        pages = {A42},
          doi = {10.1051/0004-6361/201425481},
archivePrefix = {arXiv},
       eprint = {1503.04107},
 primaryClass = {astro-ph.SR},
       adsurl = {https://ui.adsabs.harvard.edu/abs/2015A&A...577A..42B}
}

@ARTICLE{2017Natur.543...60D,
       author = {{Dodd}, Matthew S. and {Papineau}, Dominic and {Grenne}, Tor and {Slack}, John F. and {Rittner}, Martin and {Pirajno}, Franco and {O'Neil}, Jonathan and {Little}, Crispin T.~S.},
        title = "{Evidence for early life in Earth{\textquoteright}s oldest hydrothermal vent precipitates}",
      journal = nat,
         year = 2017,
        month = mar,
       volume = {543},
       number = {7643},
        pages = {60-64},
          doi = {10.1038/nature21377},
       adsurl = {https://ui.adsabs.harvard.edu/abs/2017Natur.543...60D}
}

@ARTICLE{2021SciA....7.3963C,
       author = {{Cavalazzi}, Barbara and {Lemelle}, Laurence and {Simionovici}, Alexandre and {Cady}, Sherry L. and {Russell}, Michael J. and {Bailo}, Elena and {Canteri}, Roberto and {Enrico}, Emanuele and {Manceau}, Alain and {Maris}, Assimo and {Salom{\'e}}, Murielle and {Thomassot}, Emilie and {Bouden}, Nordine and {Tucoulou}, R{\'e}mi and {Hofmann}, Axel},
        title = "{Cellular remains in a  3.42-billion-year-old subseafloor hydrothermal environment}",
      journal = {Science Advances},
         year = 2021,
        month = jul,
       volume = {7},
       number = {29},
        pages = {eabf3963},
          doi = {10.1126/sciadv.abf3963},
       adsurl = {https://ui.adsabs.harvard.edu/abs/2021SciA....7.3963C}
}

@ARTICLE{2017AJ....154..264T,
       author = {{Torres}, Guillermo and {Kane}, Stephen R. and {Rowe}, Jason F. and {Batalha}, Natalie M. and {Henze}, Christopher E. and {Ciardi}, David R. and {Barclay}, Thomas and {Borucki}, William J. and {Buchhave}, Lars A. and {Crepp}, Justin R. and {Everett}, Mark E. and {Horch}, Elliott P. and {Howard}, Andrew W. and {Howell}, Steve B. and {Isaacson}, Howard T. and {Jenkins}, Jon M. and {Latham}, David W. and {Petigura}, Erik A. and {Quintana}, Elisa V.},
        title = "{Validation of Small Kepler Transiting Planet Candidates in or near the Habitable Zone}",
      journal = aj,
         year = 2017,
        month = dec,
       volume = {154},
       number = {6},
          eid = {264},
        pages = {264},
          doi = {10.3847/1538-3881/aa984b},
       adsurl = {https://ui.adsabs.harvard.edu/abs/2017AJ....154..264T}
}

@ARTICLE{2018PASP..130k4401K,
       author = {{Kempton}, Eliza M. -R. and {Bean}, Jacob L. and {Louie}, Dana R. and {Deming}, Drake and {Koll}, Daniel D.~B. and {Mansfield}, Megan and {Christiansen}, Jessie L. and {L{\'o}pez-Morales}, Mercedes and {Swain}, Mark R. and {Zellem}, Robert T. and {Ballard}, Sarah and {Barclay}, Thomas and {Barstow}, Joanna K. and {Batalha}, Natasha E. and {Beatty}, Thomas G. and {Berta-Thompson}, Zach and {Birkby}, Jayne and {Buchhave}, Lars A. and {Charbonneau}, David and {Cowan}, Nicolas B. and {Crossfield}, Ian and {de Val-Borro}, Miguel and {Doyon}, Ren{\'e} and {Dragomir}, Diana and {Gaidos}, Eric and {Heng}, Kevin and {Hu}, Renyu and {Kane}, Stephen R. and {Kreidberg}, Laura and {Mallonn}, Matthias and {Morley}, Caroline V. and {Narita}, Norio and {Nascimbeni}, Valerio and {Pall{\'e}}, Enric and {Quintana}, Elisa V. and {Rauscher}, Emily and {Seager}, Sara and {Shkolnik}, Evgenya L. and {Sing}, David K. and {Sozzetti}, Alessandro and {Stassun}, Keivan G. and {Valenti}, Jeff A. and {von Essen}, Carolina},
        title = "{A Framework for Prioritizing the TESS Planetary Candidates Most Amenable to Atmospheric Characterization}",
      journal = pasp,
         year = 2018,
        month = nov,
       volume = {130},
       number = {993},
        pages = {114401},
          doi = {10.1088/1538-3873/aadf6f},
archivePrefix = {arXiv},
       eprint = {1805.03671},
 primaryClass = {astro-ph.EP},
       adsurl = {https://ui.adsabs.harvard.edu/abs/2018PASP..130k4401K}
}

@ARTICLE{2022ApJ...928....8F,
       author = {{Fleming}, Scott W. and {Million}, Chase and {Osten}, Rachel A. and {Kolotkov}, Dmitrii Y. and {Brasseur}, C.~E.},
        title = "{New Time-resolved, Multi-band Flares in the GJ 65 System with gPhoton}",
      journal = apj,
         year = 2022,
        month = mar,
       volume = {928},
       number = {1},
          eid = {8},
        pages = {8},
          doi = {10.3847/1538-4357/ac5037},
archivePrefix = {arXiv},
       eprint = {2202.02861},
 primaryClass = {astro-ph.SR},
       adsurl = {https://ui.adsabs.harvard.edu/abs/2022ApJ...928....8F}
}

@ARTICLE{2019AsBio..19...64T,
       author = {{Tilley}, Matt A. and {Segura}, Ant{\'\i}gona and {Meadows}, Victoria and {Hawley}, Suzanne and {Davenport}, James},
        title = "{Modeling Repeated M Dwarf Flaring at an Earth-like Planet in the Habitable Zone: Atmospheric Effects for an Unmagnetized Planet}",
      journal = {Astrobiology},
         year = 2019,
        month = jan,
       volume = {19},
       number = {1},
        pages = {64-86},
          doi = {10.1089/ast.2017.1794},
archivePrefix = {arXiv},
       eprint = {1711.08484},
 primaryClass = {astro-ph.EP},
       adsurl = {https://ui.adsabs.harvard.edu/abs/2019AsBio..19...64T}
}

@ARTICLE{2018ApJS..235...38T,
       author = {{Thompson}, Susan E. and {Coughlin}, Jeffrey L. and {Hoffman}, Kelsey and {Mullally}, Fergal and {Christiansen}, Jessie L. and {Burke}, Christopher J. and {Bryson}, Steve and {Batalha}, Natalie and {Haas}, Michael R. and {Catanzarite}, Joseph and {Rowe}, Jason F. and {Barentsen}, Geert and {Caldwell}, Douglas A. and {Clarke}, Bruce D. and {Jenkins}, Jon M. and {Li}, Jie and {Latham}, David W. and {Lissauer}, Jack J. and {Mathur}, Savita and {Morris}, Robert L. and {Seader}, Shawn E. and {Smith}, Jeffrey C. and {Klaus}, Todd C. and {Twicken}, Joseph D. and {Van Cleve}, Jeffrey E. and {Wohler}, Bill and {Akeson}, Rachel and {Ciardi}, David R. and {Cochran}, William D. and {Henze}, Christopher E. and {Howell}, Steve B. and {Huber}, Daniel and {Pr{\v{s}}a}, Andrej and {Ram{\'\i}rez}, Solange V. and {Morton}, Timothy D. and {Barclay}, Thomas and {Campbell}, Jennifer R. and {Chaplin}, William J. and {Charbonneau}, David and {Christensen-Dalsgaard}, J{\o}rgen and {Dotson}, Jessie L. and {Doyle}, Laurance and {Dunham}, Edward W. and {Dupree}, Andrea K. and {Ford}, Eric B. and {Geary}, John C. and {Girouard}, Forrest R. and {Isaacson}, Howard and {Kjeldsen}, Hans and {Quintana}, Elisa V. and {Ragozzine}, Darin and {Shabram}, Megan and {Shporer}, Avi and {Silva Aguirre}, Victor and {Steffen}, Jason H. and {Still}, Martin and {Tenenbaum}, Peter and {Welsh}, William F. and {Wolfgang}, Angie and {Zamudio}, Khadeejah A. and {Koch}, David G. and {Borucki}, William J.},
        title = "{Planetary Candidates Observed by Kepler. VIII. A Fully Automated Catalog with Measured Completeness and Reliability Based on Data Release 25}",
      journal = apjs,
         year = 2018,
        month = apr,
       volume = {235},
       number = {2},
          eid = {38},
        pages = {38},
          doi = {10.3847/1538-4365/aab4f9},
archivePrefix = {arXiv},
       eprint = {1710.06758},
 primaryClass = {astro-ph.EP},
       adsurl = {https://ui.adsabs.harvard.edu/abs/2018ApJS..235...38T}
}

@ARTICLE{2015ApJS..217...31M,
       author = {{Mullally}, F. and {Coughlin}, Jeffrey L. and {Thompson}, Susan E. and {Rowe}, Jason and {Burke}, Christopher and {Latham}, David W. and {Batalha}, Natalie M. and {Bryson}, Stephen T. and {Christiansen}, Jessie and {Henze}, Christopher E. and {Ofir}, Aviv and {Quarles}, Billy and {Shporer}, Avi and {Van Eylen}, Vincent and {Van Laerhoven}, Christa and {Shah}, Yash and {Wolfgang}, Angie and {Chaplin}, W.~J. and {Xie}, Ji-Wei and {Akeson}, Rachel and {Argabright}, Vic and {Bachtell}, Eric and {Barclay}, Thomas and {Borucki}, William J. and {Caldwell}, Douglas A. and {Campbell}, Jennifer R. and {Catanzarite}, Joseph H. and {Cochran}, William D. and {Duren}, Riley M. and {Fleming}, Scott W. and {Fraquelli}, Dorothy and {Girouard}, Forrest R. and {Haas}, Michael R. and {He{\l}miniak}, Krzysztof G. and {Howell}, Steve B. and {Huber}, Daniel and {Larson}, Kipp and {Gautier}, Thomas N., III and {Jenkins}, Jon M. and {Li}, Jie and {Lissauer}, Jack J. and {McArthur}, Scot and {Miller}, Chris and {Morris}, Robert L. and {Patil-Sabale}, Anima and {Plavchan}, Peter and {Putnam}, Dustin and {Quintana}, Elisa V. and {Ramirez}, Solange and {Silva Aguirre}, V. and {Seader}, Shawn and {Smith}, Jeffrey C. and {Steffen}, Jason H. and {Stewart}, Chris and {Stober}, Jeremy and {Still}, Martin and {Tenenbaum}, Peter and {Troeltzsch}, John and {Twicken}, Joseph D. and {Zamudio}, Khadeejah A.},
        title = "{Planetary Candidates Observed by Kepler. VI. Planet Sample from Q1--Q16 (47 Months)}",
      journal = apjs,
         year = 2015,
        month = apr,
       volume = {217},
       number = {2},
          eid = {31},
        pages = {31},
          doi = {10.1088/0067-0049/217/2/31},
archivePrefix = {arXiv},
       eprint = {1502.02038},
 primaryClass = {astro-ph.EP},
       adsurl = {https://ui.adsabs.harvard.edu/abs/2015ApJS..217...31M}
}

@ARTICLE{2022yCat.1355....0G,
       author = {{Gaia Collaboration}},
        title = "{VizieR Online Data Catalog: Gaia DR3 Part 1. Main source (Gaia Collaboration, 2022)}",
      journal = {VizieR Online Data Catalog},
         year = 2022,
        month = may,
          eid = {I/355},
        pages = {I/355},
          doi = {10.26093/cds/vizier.1355},
       adsurl = {https://ui.adsabs.harvard.edu/abs/2022yCat.1355....0G}
}

@ARTICLE{2021ESC.....5..239T,
       author = {{Todd}, Zoe R. and {Szostak}, Jack W. and {Sasselov}, Dimitar D.},
        title = "{Shielding from UV Photodamage: Implications for Surficial Origins of Life Chemistry on the Early Earth}",
      journal = {ACS Earth and Space Chemistry},
         year = 2021,
        month = feb,
       volume = {5},
       number = {2},
        pages = {239-246},
          doi = {10.1021/acsearthspacechem.0c00270},
       adsurl = {https://ui.adsabs.harvard.edu/abs/2021ESC.....5..239T}
}

@ARTICLE{2023CmChe...6..259S,
       author = {{Sajeev}, Y.},
        title = "{Prebiotic chemical origin of biomolecular complementarity}",
      journal = {Communications Chemistry},
         year = 2023,
        month = nov,
       volume = {6},
       number = {1},
          eid = {259},
        pages = {259},
          doi = {10.1038/s42004-023-01060-8},
       adsurl = {https://ui.adsabs.harvard.edu/abs/2023CmChe...6..259S}
}

@ARTICLE{2022AsBio..22..242R,
       author = {{Ranjan}, Sukrit and {Kufner}, Corinna L. and {Lozano}, Gabriella G. and {Todd}, Zoe R. and {Haseki}, Azra and {Sasselov}, Dimitar D.},
        title = "{UV Transmission in Natural Waters on Prebiotic Earth}",
      journal = {Astrobiology},
         year = 2022,
        month = mar,
       volume = {22},
       number = {3},
        pages = {242-262},
          doi = {10.1089/ast.2020.2422},
archivePrefix = {arXiv},
       eprint = {2110.00432},
 primaryClass = {astro-ph.EP},
       adsurl = {https://ui.adsabs.harvard.edu/abs/2022AsBio..22..242R}
}

@article{Mojzsis1996,
	author = {Mojzsis, S. J. and Arrhenius, G. and McKeegan, K. D. and Harrison, T. M. and Nutman, A. P. and Friend, C. R. L.},
	date = {1996/11/01},
	doi = {10.1038/384055a0},
	id = {Mojzsis1996},
	isbn = {1476-4687},
	journal = {Nature},
	number = {6604},
	pages = {55--59},
	title = {Evidence for life on Earth before 3,800 million years ago},
	url = {https://doi.org/10.1038/384055a0},
	volume = {384},
	year = {1996}}

@ARTICLE{2014AJ....148...64S,
       author = {{Shkolnik}, Evgenya L. and {Barman}, Travis S.},
        title = "{HAZMAT. I. The Evolution of Far-UV and Near-UV Emission from Early M Stars}",
      journal = aj,
         year = 2014,
        month = oct,
       volume = {148},
       number = {4},
          eid = {64},
        pages = {64},
          doi = {10.1088/0004-6256/148/4/64},
archivePrefix = {arXiv},
       eprint = {1407.1344},
 primaryClass = {astro-ph.SR},
       adsurl = {https://ui.adsabs.harvard.edu/abs/2014AJ....148...64S}
}

@ARTICLE{2016ApJ...824..102L,
       author = {{Loyd}, R.~O.~P. and {France}, Kevin and {Youngblood}, Allison and {Schneider}, Christian and {Brown}, Alexander and {Hu}, Renyu and {Linsky}, Jeffrey and {Froning}, Cynthia S. and {Redfield}, Seth and {Rugheimer}, Sarah and {Tian}, Feng},
        title = "{The MUSCLES Treasury Survey. III. X-Ray to Infrared Spectra of 11 M and K Stars Hosting Planets}",
      journal = apj,
         year = 2016,
        month = jun,
       volume = {824},
       number = {2},
          eid = {102},
        pages = {102},
          doi = {10.3847/0004-637X/824/2/102},
archivePrefix = {arXiv},
       eprint = {1604.04776},
 primaryClass = {astro-ph.SR},
       adsurl = {https://ui.adsabs.harvard.edu/abs/2016ApJ...824..102L}
}

@ARTICLE{2024ApJ...976...43W,
       author = {{Wang}, Song and {Li}, Xue and {Han}, Henggeng and {Liu}, Jifeng},
        title = "{Predicting Photospheric Ultraviolet Emission from Stellar Evolutionary Models}",
      journal = apj,
         year = 2024,
        month = nov,
       volume = {976},
       number = {1},
          eid = {43},
        pages = {43},
          doi = {10.3847/1538-4357/ad87d0},
archivePrefix = {arXiv},
       eprint = {2410.11611},
 primaryClass = {astro-ph.SR},
       adsurl = {https://ui.adsabs.harvard.edu/abs/2024ApJ...976...43W}
}

@ARTICLE{2018ApJS..236...47W,
       author = {{Willmer}, Christopher N.~A.},
        title = "{The Absolute Magnitude of the Sun in Several Filters}",
      journal = apjs,
         year = 2018,
        month = jun,
       volume = {236},
       number = {2},
          eid = {47},
        pages = {47},
          doi = {10.3847/1538-4365/aabfdf},
archivePrefix = {arXiv},
       eprint = {1804.07788},
 primaryClass = {astro-ph.SR},
       adsurl = {https://ui.adsabs.harvard.edu/abs/2018ApJS..236...47W}
}

@ARTICLE{2025ApJS..280...13W,
       author = {{Wang}, Jia-Hui and {Xiang}, Maosheng and {Zhang}, Meng and {Xie}, Ji-Wei and {Ge}, Jian and {Zhang}, Jinghua and {Mou}, Lanya and {Liu}, Ji-Feng},
        title = "{Spectroscopic Ages for 4 Million Main-sequence Dwarf Stars from LAMOST DR10 Estimated with a Data-driven Approach}",
      journal = apjs,
         year = 2025,
        month = sep,
       volume = {280},
       number = {1},
          eid = {13},
        pages = {13},
          doi = {10.3847/1538-4365/aded16},
archivePrefix = {arXiv},
       eprint = {2508.03019},
 primaryClass = {astro-ph.SR},
       adsurl = {https://ui.adsabs.harvard.edu/abs/2025ApJS..280...13W}
}

@ARTICLE{2016AsBio..16...68R,
       author = {{Ranjan}, Sukrit and {Sasselov}, Dimitar D.},
        title = "{Influence of the UV Environment on the Synthesis of Prebiotic Molecules}",
      journal = {Astrobiology},
         year = 2016,
        month = jan,
       volume = {16},
       number = {1},
        pages = {68-88},
          doi = {10.1089/ast.2015.1359},
archivePrefix = {arXiv},
       eprint = {1511.00698},
 primaryClass = {astro-ph.EP},
       adsurl = {https://ui.adsabs.harvard.edu/abs/2016AsBio..16...68R}
}

@ARTICLE{2025SoSyR..59...72A,
       author = {{Abrevaya}, X.~C. and {Odert}, P. and {Leitzinger}, M. and {Oppezzo}, O. and {Luna}, G.~J.~M. and {Patel}, M.~R. and {Hanslmeier}, A.},
        title = "{The EXO-UV Program: Latest Advances of Experimental Studies to Investigate the Biological Impact of UV Radiation on Exoplanets}",
      journal = {Solar System Research},
         year = 2025,
        month = jul,
       volume = {59},
       number = {6},
          eid = {72},
        pages = {72},
          doi = {10.1134/S0038094624602019},
       adsurl = {https://ui.adsabs.harvard.edu/abs/2025SoSyR..59...72A}
}

@ARTICLE{2016ApJ...826..195K,
       author = {{Kay}, C. and {Opher}, M. and {Kornbleuth}, M.},
        title = "{Probability of CME Impact on Exoplanets Orbiting M Dwarfs and Solar-like Stars}",
      journal = apj,
         year = 2016,
        month = aug,
       volume = {826},
       number = {2},
          eid = {195},
        pages = {195},
          doi = {10.3847/0004-637X/826/2/195},
archivePrefix = {arXiv},
       eprint = {1605.02683},
 primaryClass = {astro-ph.SR},
       adsurl = {https://ui.adsabs.harvard.edu/abs/2016ApJ...826..195K}
}

@ARTICLE{2024A&A...688A.138P,
       author = {{Pe{\~n}a-Mo{\~n}ino}, L. and {P{\'e}rez-Torres}, M. and {Varela}, J. and {Zarka}, P.},
        title = "{Magnetohydrodynamic simulations of the space weather in Proxima b: Habitability conditions and radio emission}",
      journal = aap,
         year = 2024,
        month = aug,
       volume = {688},
          eid = {A138},
        pages = {A138},
          doi = {10.1051/0004-6361/202349042},
archivePrefix = {arXiv},
       eprint = {2405.19116},
 primaryClass = {astro-ph.EP},
       adsurl = {https://ui.adsabs.harvard.edu/abs/2024A&A...688A.138P}
}

@ARTICLE{2019ApJ...881..114Y,
       author = {{Yamashiki}, Yosuke A. and {Maehara}, Hiroyuki and {Airapetian}, Vladimir and {Notsu}, Yuta and {Sato}, Tatsuhiko and {Notsu}, Shota and {Kuroki}, Ryusuke and {Murashima}, Keiya and {Sato}, Hiroaki and {Namekata}, Kosuke and {Sasaki}, Takanori and {Scott}, Thomas B. and {Bando}, Hina and {Nashimoto}, Subaru and {Takagi}, Fuka and {Ling}, Cassandra and {Nogami}, Daisaku and {Shibata}, Kazunari},
        title = "{Impact of Stellar Superflares on Planetary Habitability}",
      journal = apj,
         year = 2019,
        month = aug,
       volume = {881},
       number = {2},
          eid = {114},
        pages = {114},
          doi = {10.3847/1538-4357/ab2a71},
archivePrefix = {arXiv},
       eprint = {1906.06797},
 primaryClass = {astro-ph.SR},
       adsurl = {https://ui.adsabs.harvard.edu/abs/2019ApJ...881..114Y}
}

@ARTICLE{2022SciA....8I9743H,
       author = {{Hu}, Junxiang and {Airapetian}, Vladimir S. and {Li}, Gang and {Zank}, Gary and {Jin}, Meng},
        title = "{Extreme energetic particle events by superflare-asssociated CMEs from solar-like stars}",
      journal = {Science Advances},
         year = 2022,
        month = mar,
       volume = {8},
       number = {12},
          eid = {eabi9743},
        pages = {eabi9743},
          doi = {10.1126/sciadv.abi9743},
       adsurl = {https://ui.adsabs.harvard.edu/abs/2022SciA....8I9743H}
}
    \end{singlespace}
    \clearpage
    \end{refsection}


\end{document}